\documentclass[fleqn]{2020SCGE}
\usepackage{graphics}
\usepackage{textalpha}
\usepackage{xspace}
\newcommand{\HUBS}{\textit{HUBS}\xspace}
\newcommand{\Chandra}{\textit{Chandra}\xspace}
\newcommand{\ROSAT}{\textit{ROSAT}\xspace}
\newcommand{\eROSITA}{\textit{eROSITA}\xspace}
\newcommand{\XMMNewton}{\textit{XMM-Newton}\xspace}

\begin{document}

\ensubject{subject}

\ArticleType{Article}
\SpecialTopic{SPECIAL TOPIC: }
\Year{2023}
\Month{January}
\Vol{0}
\No{0}
\DOI{??}
\ArtNo{000000}
\ReceiveDate{\today}
\AcceptDate{January 00, 0000}
\OnlineDate{\today}

\title{Scientific Objectives of\\ the \textit{Hot Universe Baryon Surveyor} (\HUBS) Mission}{Scientific Objectives of the \HUBS Mission}

\author[1]{Joel Bregman}{}
\author[2]{Renyue Cen}{}
\author[3,4]{Yang Chen}{}
\author[5]{Wei Cui}{cui@tsinghua.edu.cn}
\author[6]{Taotao Fang}{}
\author[7,8]{Fulai Guo}{}
\author[9]{\\Edmund Hodges-Kluck}{}
\author[5]{Rui Huang}{}
\author[10,11]{Luis C. Ho}{}
\author[12]{Li Ji}{}
\author[7,8]{Suoqing Ji}{suoqing@shao.ac.cn}
\author[2,12]{Xi Kang}{}
\author[13]{Xiaoyu Lai}{}
\author[5]{\\Hui Li}{}
\author[12,1]{Jiangtao Li}{}
\author[2]{Miao Li}{}
\author[3,4]{Xiangdong Li}{}
\author[14]{Yuan Li}{}
\author[15]{Zhaosheng Li}{}
\author[16]{Guiyun Liang}{}
\author[17]{\\Helei Liu}{}
\author[12]{Wenhao Liu}{}
\author[18]{Fangjun Lu}{}
\author[5]{Junjie Mao}{}
\author[19]{Gabriele Ponti}{}
\author[20]{Zhijie Qu}{}
\author[21]{Chenxi Shan}{}
\author[10]{\\Lijing Shao}{}
\author[7]{Fangzheng Shi}{}
\author[22]{Xinwen Shu}{}
\author[3,4]{Lei Sun}{}
\author[6]{Mouyuan Sun}{}
\author[23]{Hao Tong}{}
\author[6]{Junfeng Wang}{}
\author[24]{\\Junxian Wang}{}
\author[25]{Q. Daniel Wang}{}
\author[26]{Song Wang}{}
\author[24]{Tinggui Wang}{}
\author[27,10]{Weiyang Wang}{}
\author[28]{\\Zhongxiang Wang}{}
\author[5]{Dandan Xu}{dandanxu@mail.tsinghua.edu.cn}
\author[21]{Haiguang Xu}{hgxu@sjtu.edu.cn}
\author[29]{Heng Xu}{}
\author[27,10]{Renxin Xu}{}
\author[3,4]{Xiaojie Xu}{}
\author[24]{\\Yongquan Xue}{}
\author[12]{Hang Yang}{}
\author[7,8]{Feng Yuan}{fyuan@shao.ac.cn}
\author[12]{Shuinai Zhang}{}
\author[5]{Yuning Zhang}{}
\author[30]{\\Zhongli Zhang}{}
\author[21]{Yuanyuan Zhao}{}
\author[31]{Enping Zhou}{}
\author[3,4]{Ping Zhou}{}


\AuthorCitation{J Bregman et al.}

\address[1]{Department of Astronomy, University of Michigan, Ann Arbor, MI, 48109-1107, USA}
\address[2]{Institute for Astronomy, School of Physics, Zhejiang University, Hangzhou 310027, China}
\address[3]{School of Astronomy and Space Science, Nanjing University, Nanjing 210023, China}
\address[4]{Key Laboratory of Modern Astronomy and Astrophysics, Nanjing University, Ministry of Education, Nanjing 210023, China}
\address[5]{Department of Astronomy, Tsinghua University, Beijing 100084, China}
\address[6]{Department of Astronomy, Xiamen University, Xiamen, Fujian 361005, China}
\address[7]{Astrophysics Division, Shanghai Astronomical Observatory, Chinese Academy of Sciences, Shanghai 200030, China}
\address[8]{Key Laboratory for Research in Galaxies and Cosmology, Shanghai Astronomical Observatory, Chinese Academy of Sciences, Shanghai 200030, China}
\address[9]{NASA Goddard Space Flight Center, Greenbelt, MD 20771, USA}
\address[10]{Kavli Institute for Astronomy and Astrophysics, Peking University, Beijing 100871, China}
\address[11]{Department of Astronomy, School of Physics, Peking University, Beijing 100871, China}
\address[12]{Purple Mountain Observatory, Chinese Academy of Sciences, Nanjing 210023, China}
\address[13]{Department of Physics and Astronomy, Hubei University of Education, Wuhan 430205, China}
\address[14]{Department of Physics, University of North Texas, Denton, TX 76203, USA}
\address[15]{Key Laboratory of Stars and Interstellar Medium, Xiangtan University, Xiangtan 411105, China}
\address[16]{CAS Key Laboratory of Optical Astronomy, National Astronomical Observatories, Chinese Academy of Sciences, Beijing 100101, China}
\address[17]{School of Physical Science and Technology, Xinjiang University, Urumuqi 830046, China}
\address[18]{Key Laboratory for Particle Astrophysics, Institute of High Energy Physics, Chinese Academy of Sciences, Beijing 100049, China}
\address[19]{INAF-Osservatorio Astronomico di Brera, I-23807 Merate (LC), Italy}
\address[20]{Department of Astronomy \& Astrophysics, the University of Chicago, Chicago, IL 60637, USA}
\address[21]{School of Physics and Astronomy, Shanghai Jiao Tong University, Shanghai 200240, China}
\address[22]{Department of Physics, Anhui Normal University, Wuhu 241002, China}
\address[23]{School of Physics and Materials Science, Guangzhou University, Guangzhou 510006, China}
\address[24]{Department of Astronomy, University of Science and Technology of China, Hefei 230026, China}
\address[25]{Department of Astronomy, University of Massachusetts, Amherst, MA 01003, USA}
\address[26]{Key Laboratory of Optical Astronomy, National Astronomical Observatories, Chinese Academy of Sciences, Beijing 100101, China}
\address[27]{School of Physics and State Key Laboratory of Nuclear Physics and Technology, Peking University, Beijing 100871, China}
\address[28]{Department of Astronomy, School of Physics and Astronomy, Yunnan University, Kunming 650091, China}
\address[29]{National Astronomical Observatories, Chinese Academy of Sciences, Beijing 100101, China}
\address[30]{Shanghai Astronomical Observatory, Key Laboratory of Radio Astronomy, Chinese Academy of Sciences, Shanghai 200030, China}
\address[31]{Huazhong University of Science and Technology, Wuhan 430074, China}


\abstract{The \textit{Hot Universe Baryon Surveyor} (\HUBS) is a
 proposed space-based X-ray telescope for detecting X-ray emissions from the hot
 gas content in our universe. With its unprecedented spatially-resolved
 high-resolution spectroscopy and large field of view, the \HUBS mission will be
 uniquely qualified to measure the physical and chemical properties of the hot
 gas in the interstellar medium, the circumgalactic medium, the intergalactic
 medium, and the intracluster medium. These measurements will be valuable for
 two key scientific goals of \HUBS, namely to unravel the AGN and stellar
 feedback physics that governs the formation and evolution of galaxies, and to
 probe the baryon budget and multi-phase states from galactic to cosmological
 scales. In addition to these two goals, the \HUBS mission will also help us
 solve some problems in the fields of galaxy clusters, AGNs, diffuse X-ray
 backgrounds, supernova remnants, and compact objects. This paper discusses the
 perspective of advancing these fields using the \HUBS telescope.}

\keywords{X-ray telescopes, Galactic halo, X-ray sources}

\PACS{95.55.Ka, 98.35.Gi, 98.70.Qy}

\thispagestyle{empty}
\maketitle


\begin{multicols}{2}

\section{Introduction}\label{sec:intro}

Over 99\% of the baryonic matter in the Universe is in the form of ionized
\noindent plasma, among which diffuse gas is the most prevalent form that spans
over a wide range of scales from the interstellar medium (ISM) to the
circumgalactic medium (CGM), the intracluster medium (ICM), and the
intergalactic medium (IGM). The diffuse gas is a key component in the cosmic
baryon budget and cycle, as it is the main reservoir of mass, metals and energy,
and thus regulates the formation of stars and galaxies (e.g.,
\cite{rees1977cooling,naab2017theoretical}). In particular, galaxies may embrace
pristine cosmological inflows as fuel for star formation, and, in the meanwhile,
eject a significant fraction of mass and metals back to the surrounding diffuse gas
via so-called feedback processes, e.g., supernova feedback and active galactic
nucleus (AGN) feedback. The interplay between the inflow and outflow of baryons
in galaxies is a key process in the formation and evolution of galaxies.
Compared with dark matter which only interacts gravitationally, baryonic
matter is more richly and sensitively imprinted with the history of galactic
feedback and galaxy evolution, due to the more complicated and less understood
baryonic physics, including gas (magneto) hydrodynamics, ionization, cooling and
heating, and chemical enrichment. Therefore, understanding the physical and
chemical properties of diffused gas residing in ``galactic ecosystems'' is a
fundamental problem in modern astrophysics \cite{Faucher2023}.

Unlike the ionized plasma in the form of stars which emit%
\Authorfootnote \noindent photons powered by nuclear reactions and thus are
easily observable, the ionized diffuse gas within and between galaxies is more
difficult to probe. Fortunately, atomic physics enables the possible detection of diffuse gas via emission and absorption across a wide range of the
electromagnetic spectrum, among which the X-ray band is highly prominent. X-rays
are ubiquitous in astrophysical environments: at large (galaxy scale and above)
scales, gas is usually virialized at the virial temperature that increases with
the greater enclosed mass of the system, and thus hot diffuse gas is commonly
expected especially in massive astrophysical systems. For instance, galaxy
clusters are filled with hot ICM with temperatures up to $T\sim10^{8}$ K. At
smaller scales, the gas can be efficiently heated by the feedback processes from
stars and AGNs. Due to the nature of the emission mechanism, the gas thermal
properties, i.e., temperatures and densities of the X-ray-emitting gas, can be
directly obtained from the X-ray spectra. In addition, through the detection of
emission and absorption lines, the X-rays can also be used to determine the
kinematic properties and chemical compositions of the hot gas, which contain a
tremendous amount of important information for the physical processes such as
AGN and stellar feedback and interaction between galaxies. The X-ray observation
is thus a powerful tool to probe the thermal states and kinematics of the
diffuse gas, and thus to understand the baryon budget and cycle in the Universe.

The \textit{Hot Universe Baryon Surveyor} (\HUBS) mission \cite{Cuietal2020JLTP}
is a timely effort to unravel the mystery of the baryon budget and cycle. The
design of \HUBS is highly optimized for detecting extended X-ray emission from
the diffuse hot gas in and around galaxies \cite{2020SPIE11444E..2SC}. Compared
with other X-ray missions such as \XMMNewton and \Chandra, two features of \HUBS
stand out: (1) large field-of-view (of 1$^\circ$ half-power diameter), and (2)
high spectral resolution (of $<$1 eV for the central sub-array and 2 eV for the
main array at 1 keV). The hybrid design of the detector is adopted to enhance
absorption-line studies through observations of point-like background sources
like active galactic nuclei (AGN) or gamma-ray bursts (GRBs), while staying
within the capability of current readout technologies, which limit the number of
pixels in the detector array.  Other technical trade-offs have also been made.
For instance, the spectral range is capped at 2 keV, to make it easier to
realize high spectral resolution while maintaining good quantum efficiency of
the detector. Because CGM and IGM are of very low density, the X-ray emission
from them is expected to be extremely weak but, fortunately, be dominated by
spectral lines. \HUBS is designed to make full use of the unprecedented spectral
capabilities of the transition edge sensor-based microcalorimeters in detecting
hot CGM and IGM, through narrow-band imaging around strong emission lines, and
in deriving their physical and chemical properties, through high-resolution
X-ray spectroscopy. Moreover, for detecting very extended emissions, the larger
the field of view, the more photons are let in, and the higher the signal-to-noise
ratio. Furthermore, a large field of view provides high efficiency in covering  
nearby galaxies and galaxy groups or clusters.
We refer the readers to Table~1 in \cite{Cuietal2020JLTP} and Table~1 in
\cite{2020SPIE11444E..2SC} for detailed specifications of \HUBS.

The high-resolution X-ray spectroscopic observations with \HUBS are also
expected to enable advancement in many other areas, including inflows and
outflows in active galactic nuclei (AGN), elemental abundances and distribution
in supernova remnants (SNRs), the origin of diffuse X-ray background, flaring
activities, relativistic effects in compact objects, as well as lunar or
planetary X-ray emission associated with solar wind charge exchange, which
produces foreground X-ray emissions and bears high relevance to active research
in the fields of laboratory astrophysics and atomic physics.

This paper discusses the scientific objectives of \HUBS. As already mentioned, the design and optimization of the \HUBS payload are driven by a set of core scientific objectives. Here, we present the core science at two levels, which cover two marginally overlapped spatial scales, with one focusing on feedback processes and baryon/metal cycling in the galactic ecosystem, and the other on the hot baryon budget and multi-phase status at larger scales. The former is discussed in \S\ref{sec:feedback} while the latter in \S\ref{sec:BaryonsOnLargeScale}, respectively.  In  \S\ref{sec:Galactic-Sciences}, Galactic sciences of \HUBS are discussed. In \S\ref{sec:unlocking}, we explore the capability of \HUBS by reviewing adopted atomic models and analysis techniques. We finally report the current status of \HUBS in \S\ref{sec:status} and conclude in \S\ref{sec:summary}.

\section{Galactic ecosystem: feedback and baryon cycles}
\label{sec:feedback}

AGN and stellar feedback is a bottleneck in the study of galaxy formation and
evolution, and it is also a cutting-edge topic in astrophysics in recent years. AGN feedback is believed to be dominant in relatively massive galaxies while stellar feedback is believed to be more important in less massive ones \cite{Liyp2018}. Both of them are closely
related to star formation activity in galaxies and are responsible for the production of galaxy wind, generating an X-ray halo around galaxies, called circumgalactic medium \cite{Tumlinson2017,Faucher2023}. 

With the help of \HUBS, we will be able to answer a few key questions driving our understanding of the feedback physics and its relationship to galaxy evolution, including:

$\bullet$ \emph{How does AGN feedback suppress star formation and even quench the galaxy, and what are the respective roles of different feedback modes and different AGN outputs (radiation, wind, and jet) in these processes?}

$\bullet$ \emph{How are different phases of the ISM/CGM regulated by the AGN and stellar
 feedback?}
 
$\bullet$ \emph{How much energy is released, and how much matter is heated/accelerated during the interaction of feedback with the ISM/CGM?}

$\bullet$ \emph{How can \HUBS help to constrain important non-thermal physics in
the CGM?}

The hot circumgalactic medium (CGM) may contain
considerable mass and metals, and is crucial for us to understand the mechanism of feedback. High energy resolution imaging spectroscopy X-ray observations of galaxies and their surrounding CGM/IGM play an important role in our understanding of the role of the co-evolution and interplay between galaxies and their environments. As highlighted by the Decadal Survey on Astronomy and Astrophysics 2020 (Astro2020) \cite{Astro2020}, studying various forms of stellar and AGN feedback on the galaxy ecosystem over a large physical scale is critical in unveiling the hidden drivers of galaxy growth, including the connection between star formation and the ISM, the cycling of gas and metal in and out of the galactic disk and halo, as well as the ionizing sources of the Universe.

\begin{figure}[H]
  \centering
  \includegraphics[width=0.47\textwidth]{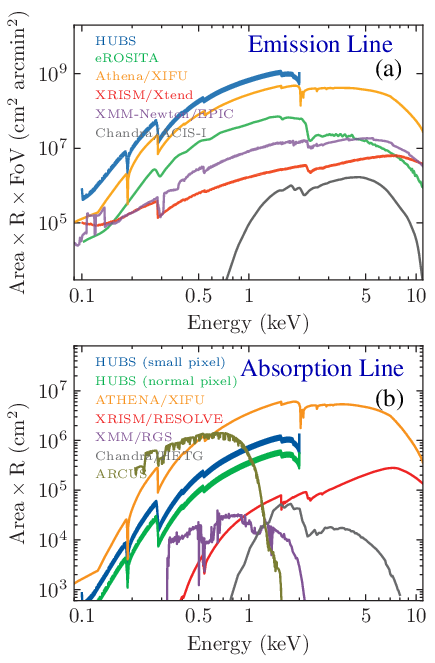}
  \caption{Comparison of the FoM of a few X-ray missions in the detection of emission lines from an extended source (a; $FoM_{\rm em}$) and absorption lines from a point-like source (b; $FoM_{\rm ab}$). The response files of the $60\times60$ normal array of \HUBS are used in (a), while those of both the normal array and the $12\times12$ central sub-array are used in (b). For instruments with an X-ray grating (\XMMNewton/RGS, \Chandra/HETG, and ARCUS), all available grating orders are combined together. \HUBS has the highest $FoM_{\rm em}$ at $\lesssim2\rm~keV$ among all the X-ray telescopes either currently in use or planned in the future, while it is also outstanding in $FoM_{\rm ab}$, although not as powerful as some future X-ray missions with grating (\textit{Athena} and \textit{Arcus}).
  }
  \label{fig:FoM}
\end{figure}

With \HUBS, the CGM and the associated physical processes can be probed through direct detection of emission lines, or absorption line studies with the observations of bright background sources, taking advantage of the superior spectral resolution of the central sub-array. In addition, observations of the CGM are also crucial to understand the baryon/metal cycling and budget. With its breakthrough technology of combining the high energy resolution micro-calorimeter detector and a large-FOV X-ray telescope \cite{2020SPIE11444E..2SC}, \HUBS has an unprecedented capability in resolving and detecting individual X-ray emission lines from the hot plasma either in thermal equilibrium or not (Fig.~\ref{fig:FoM}a). It is also outstanding in absorption line studies of X-ray bright background sources (Fig.~\ref{fig:FoM}b). The former is extremely important in the study of how the stellar and AGN feedbacks affect the galactic ecosystem. Within $\sim5\rm~years$ of science operation, \HUBS will conduct milestone studies of stellar and AGN feedback mainly in two ways: either by moderately deep surveys of nearby objects with large angular sizes, resolving hot gas features with a physical size from star clusters to massive galaxy clusters ($\sim10$--$10^{6}\rm~pc$); or by deep enough observations of individual galaxy halos at moderate distances to measure the physical and chemical properties of the hot CGM.

\subsection{AGN feedback}
\label{subsec:AGN_feedback}

Active galactic nuclei (AGN) are the observational manifestation of matter inflow toward supermassive black holes, which can be found at the centers of almost all massive galaxies \cite{Netzer2015}. Compared to their host galaxies, AGNs are miniature in both size and mass. However, they are expected to provide substantial feedback (in the form of outflows) to their host galaxies and beyond \cite{Magorrian1998,Ferrarese2000,Fabian2012,Kormendy2013}. 

Generally speaking, AGN-driven outflows have two forms: radio jets and ionized winds. Highly collimated relativistic jets, mainly observed in the radio band, can be found in some AGN accreting at relatively low efficiencies \cite{Fabian2012}. Radio jets are also known as the kinetic or maintenance mode of AGN outflow. Ionized winds with large solid angles are prevalent in AGN accreting at high efficiencies \cite{Laha2021}. Ionized winds are often referred to as the radiative mode of AGN outflow. Presently, grating spectrometers in the X-ray and UV bands are the main working horses to probe these ionized winds via absorption spectroscopy since these winds are too small to be resolved via direct imaging.

In the X-ray band, there are mainly three types of ionized winds: warm absorbers, ultrafast outflows, and obscuring winds. The classical warm absorbers are identified with multiple narrow absorption lines with a typical outflow velocity of $\lesssim 10^3~{\rm km~s^{-1}}$ \cite{Crenshaw2003,Costantini2007,Kaastra2011,Mao2017b,Mehdipour2018,Mao2019b,WangYJ2022}. Ultrafast outflows are mainly inferred from the absorption features of highly ionized Fe {\sc xxvi} and/or Fe {\sc xxv} in the hard X-ray band \cite{Reeves2003,Tombesi2010b,Nardini2015,Tombesi2015,Parker2017}. The outflow velocity of ultrafast outflows can reach up to about a third of the speed of light ($\sim10^{4-5}~{\rm km~s^{-1}}$). Ultrafast outflows occupy the high column density ($N_{\rm H}$), ionization parameter ($\xi$), and outflow velocity ($v_{\rm out}$) part of the parameter space, while warm absorbers occupy the other side of the parameter space. Transient obscuring winds are currently identified mainly in coordinated multi-wavelength observations \cite{Kaastra2014,Ebrero2016a,Mehdipour2017,Longinotti2019,Kara2021,Mao2022b}.
In the $N_{\rm H}-\xi-v_{\rm out}$ parameter space, obscuring winds are in between warm absorbers and ultrafast outflows, overlapping more with the former.  

To study weak absorption lines of AGN winds, two key instrument parameters are
critical: energy resolution ($R$) and effective area ($A_{\rm eff}$). The larger
the product $R\times A_{\rm eff}$ (as the figure of merit), the better we can
constrain weak absorption lines. As discussed at the beginning of
\S\ref{sec:feedback}, compared to existing grating spectrometers aboard
\XMMNewton and \Chandra, \HUBS will greatly advance our knowledge of AGN winds. 

Among the three types of AGN winds in the X-ray band, the classical warm absorbers are the most frequent to be detected \cite{Reynolds1997,McKernan2007,Laha2014}. Nonetheless, we still have gaps in our understanding of the warm absorber, e.g., its number density and distance to the black hole remain largely uncertain. These two parameters, linked with each other via the measurable ionization parameter, are essential to infer the origin of the warm absorber as well as its impact on the circumnuclear media and beyond. 

The number density of ionized winds can be constrained with density-sensitive metastable absorption lines. In theory, such diagnostics can cover more than ten orders of magnitude in number density \cite{Mao2017b}. In practice, successful applications are rather scarce: NGC\,4151 \cite{King2012,Ogorzalek2022} and GRO\,J1655-40 \cite{Miller2008,Tomaru2023}. The latter is an X-ray binary. In addition, upper or lower limits were obtained for Mrk\,279 \cite{Kaastra2004} and NGC\,5548 \cite{Mao2017b}. This is partly due to the insufficient figure of merit ($R\times A_{\rm eff}$) of current instruments. The situation can be significantly improved with future missions like \HUBS.

\begin{figure}[H]
\centering
\includegraphics[width=0.50\textwidth, trim={0.cm 0.cm 1.0cm 0.cm}, clip]{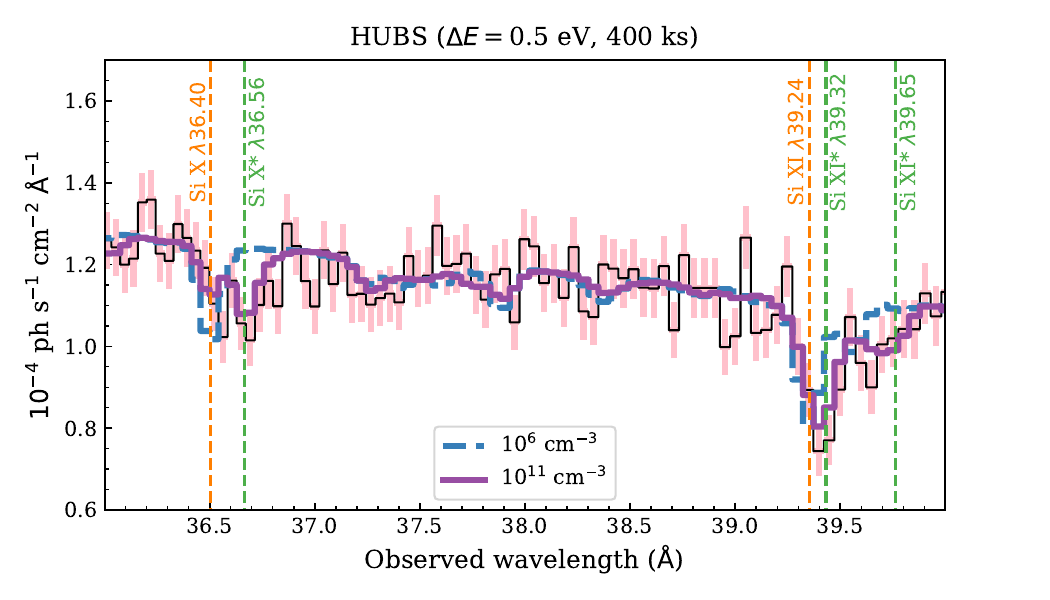}
\caption{\HUBS simulation of density diagnostics with metastable absorption lines. The central grid with 0.5 eV energy resolution is used for the simulation, which has an exposure time of 400 ks. The simulated data are shown in black with $1\sigma$ uncertainties shown in pink. The redshift of the AGN is 0.00386, while the warm absorber has a column density of $10^{21}~{\rm cm^{-2}}$ and an outflow velocity of $-300~{\rm km~s^{-1}}$. Warm absorber models with two different number densities are shown: $10^6~{\rm cm^{-3}}$ (dashed curve in blue) and $10^{11}~{\rm cm^{-3}}$ (solid curve in purple). In the high-density model, absorption lines (in orange) from the ground level are weaker, while absorption lines (in green) from the metastable levels are stronger. This is mainly due to collisional excitation from the ground to the metastable level in a high-density wind. }
\label{fig:plot_dnma_msal}
\end{figure}

Figure~\ref{fig:plot_dnma_msal} illustrates such diagnostics with \HUBS. In this simulation, the $0.2-2$~keV observed continuum flux is $\sim7\times10^{-12}~{\rm erg~s^{-1}~cm^{-2}}$. This flux level is low when compared to those of the well-studied targets like NGC\,5548 \cite{Kaastra2000,Mao2017b} and NGC\,3783 \cite{Kaspi2000,Mao2019b}. Even for this low continuum flux, with 400 ks exposure, the central grid of \HUBS (with the $<1$ eV energy resolution) can well constrain the density of the wind with multiple key diagnostic lines. For targets with higher continuum flux, the required exposure time might be further reduced. In addition, the long exposure observations of \HUBS can help us to better understand the wind density and in turn location via spectral timing analyses \cite{Kaastra2012,Silva2016,Juranova2022,Rogantini2022}. Moreover, potential eclipse events of the clumpy absorber could also provide useful constraints on its location and thus density \cite{Gallo2021}.  
  
Another fundamental parameter of the warm absorber that is not well constrained is its opening angle. This can be indirectly inferred from the warm absorber occurrence fraction in a sample of type 1 AGN. In the era of ASCA, the occurrence fraction of the warm absorber is $\sim50$~\% \cite{Reynolds1997}. The occurrence fraction was revised to $\sim65$~\% in the era of \XMMNewton and \Chandra \cite{McKernan2007,Laha2014}. However, limited by the capability of current instruments, these sample studies are limited to bright targets. The largest sample size among the three studies is merely 26. 

Such sample studies can be improved with \HUBS. On one hand, its energy resolution and effective area are suitable for weak absorption line studies. On the other hand, its large field of view enables us to accumulate a large number of high-quality spectra in an efficient way. Figure~\ref{fig:plot_dnma_msal} shows the \HUBS simulation of warm absorber features of a serendipitous AGN when observing other core science or observatory science targets. If not filled with extended sources, the 1-square-degree field of view (with 3600 integral field units) might have a few serendipitous AGN with detectable warm absorber features \cite{LuoB2017,XueYQ2017}. Even for a rather low continuum $0.2-2$~keV flux of $1.0\times10^{-13}~{\rm erg~s^{-1}~cm^{-2}}$ (nearly two orders of magnitude lower than any of the sample targets in \cite{Reynolds1997,McKernan2007,Laha2014}), warm absorber features such as the Fe M-shell unresolved transition array (UTA) feature can be well detected with the 2 eV energy resolution of the normal grid. That is to say, without requesting dedicated observations, we can still accumulate a good sample of AGN warm absorbers. 

\begin{figure}[H]
\centering
\includegraphics[width=0.50\textwidth, trim={0.cm 0.cm 1.0cm 0.cm}, clip]{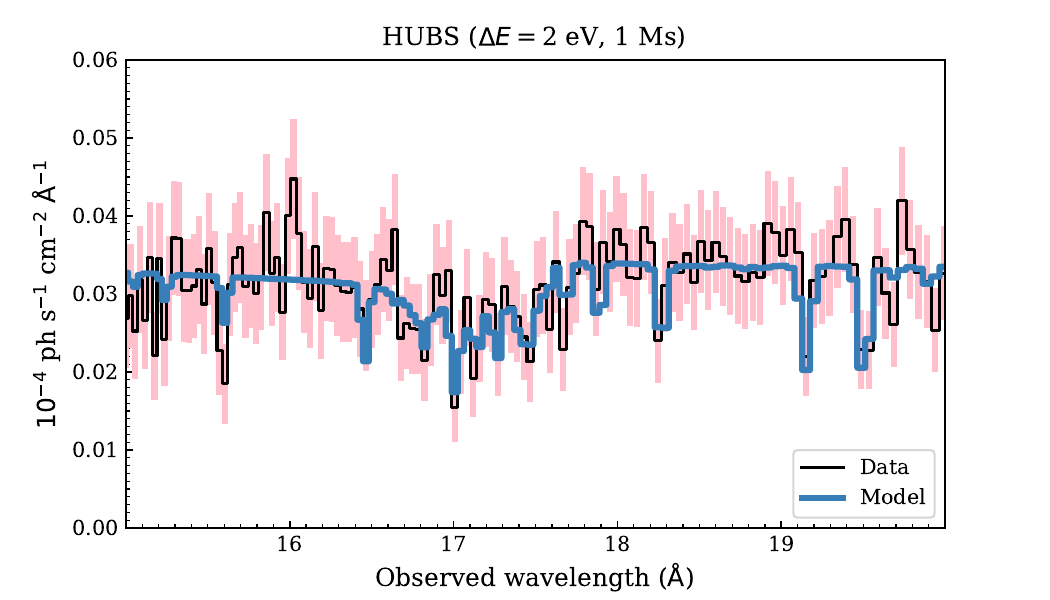}
\caption{\HUBS simulation of warm absorber features of a serendipitous AGN when observing other core science or observatory science targets. The normal grid with 2 eV energy resolution is used for the simulation, which has an exposure time of 1 Ms. The simulated data are shown in black with $1\sigma$ uncertainties shown in pink. The warm absorber model is shown in blue. The $0.2-2$~keV observed continuum flux of the AGN ($z=0.03$) is $1.0\times10^{-13}~{\rm erg~s^{-1}~cm^{-2}}$ and the warm absorber has a column density of $10^{21}~{\rm cm^{-2}}$. The characteristic Fe M-shell unresolved transition array (UTA) feature around $17$~\AA\ (in the observed frame) and other warm absorber lines are clearly detected. }
\label{fig:plot_dnma_wa}
\end{figure}

By the same token, we can search for ultrafast outflows in the soft X-ray band in these serendipitous AGNs. The occurrence fraction of ultrafast outflows is at least 30~\% in the sample studies by \cite{Tombesi2010b,Tombesi2013a,Tombesi2014b}. These fast winds are rarely detected in the soft X-ray band though \cite{Pounds2003,Gupta2013,Longinotti2015,Gupta2015,Parker2017,Reeves2020}. Moreover, ultrafast outflows can be quite variable \cite{Tombesi2015,Matzeu2017,Parker2017,Midooka2022}. The long exposure observations of \HUBS can also help us to better understand the evolution of these powerful winds. 

Some AGN winds are probably powered by accretion disks. As a result, the disk winds may take a significant fraction of the accretion power away from the accretion disk, which significantly alters the disk temperature profiles \cite{cheng2019modelling}, energy density distributions \cite{Slone2012} and disk sizes \cite{Li2019, Sun2019}. 

Transient accretion onto supermassive black holes (SMBH) in galactic centers can also lead to winds and outflows. The most interesting class is the stellar tidal disruption event (TDE). A TDE occurs when a star passes too close to a SMBH and gets tidally 
disrupted and accreted, producing a flare of radiation peaking in the
UV and soft X-rays \cite{Rees1988}. TDE provides a unique probe of quiescent SMBHs in normal galaxies, especially the physical processes associated with various accretion states on a practically observable timescale of several years. The super-Eddington fall-back rates and potentially super-Eddington accretion rates in TDEs have been predicted in several theoretical works \cite{Guillochon2014, Dai2015}, which can lead to fast, radiation-driven outflows, as seen in MHD simulations of such systems \cite{Coughlin2014, Jiang2014}. 

X-ray outflows in TDEs are crucial in constraining the super-Eddington accretion flows, yet observations are still sparse \cite{Lin2017}. This is possibly due to the X-ray faintness of most TDEs, particularly those discovered in the optical bands. An ionized X-ray outflow (from the blueshifted {\sc O viii} absorption trough) with a velocity of 0.2 c has been reported in a nearby TDE ASASSN-14li, but it disappeared in the late-time observations \cite{Kara2018}. If a super-Eddington radiatively driven outflow were observed, it would have ``turned off''  only one year after the TDE luminosity peak. Interestingly, with dedicated \Chandra grating spectroscopy observations, X-ray outflows with a much lower velocity ($\sim$300 km s$^{-1}$) in ASASSN-14li are detected, which are also variable in physical conditions \cite{Miller2015}. Such a low-velocity outflow component would be common among TDEs, which could be interpreted as absorption through a super-Eddington wind or through a filament of stellar debris. Since TDEs are mainly peaking in the soft X-ray bands \cite{Saxton2020}, with its superior sensitivity and spectral resolution, \HUBS will allow for measuring precisely the soft X-ray outflow properties at different accretion states. 
For instance, \HUBS is able to measure the evolution of ionized absorption features as a function of luminosity and/or accretion rate, especially in the rising and decaying phase of the TDE flares, establishing the connection between the outflows and the super-Eddington accretion process, which is still poorly constrained. 

In the current model of galaxy formation and evolution, feedback is the most
uncertain physical process.  Almost all cosmological simulations adopt
subgrid models of AGN feedback which are highly uncertain and distinctively different between
one and another. In this sense, simulations of the evolution of a single galaxy have a significant advantage because a much higher resolution can be achieved and the inner boundary of the AGN accretion flow, which is the Bondi radius of black hole accretion, can even be resolved numerically. This ensures that by self-consistently evolving the gas flow from hundreds of kpc scales down to the inner Bondi radius, we can reliably trace the mass flux crossing the Bondi radius, i.e., the mass accretion rate of the AGN, which is crucial for the determination of the magnitude of AGN activity. Moreover, in this case, the state-of-the-art AGN physics \cite{yuan2014hot} can be incorporated into the simulations  (e.g., \cite{yuan2018active}). Fig.~\ref{fig:AGN_sims} shows such an example of simulation results, which describes the evolution of a single elliptical galaxy when the effects of AGN feedback are taken into account.  
Fig.~\ref{fig:HUBSmockspectra} shows the mock spectra of an elliptical galaxy produced by the gas from 0.1-10 {\rm kpc} using the response of \HUBS telescope with four different models of AGN feedback, together with the fitting results of the one-temperature model. This result shows the ability of the high spectral resolution spectra obtained by \HUBS to discriminate different feedback models.   
Next step, it is crucial to develop the model further by incorporating more physics into the model and to study in detail the roles of different AGN outputs, such as radiation, wind, and jet in cold and hot feedback modes, in AGN feedback. Moreover, we also need to incorporate the simulation results into cosmological simulations.  

\subsection{Stellar feedback}
\label{subsec:stellar_feedback}

Though there is no doubt that supernova feedback is crucial to the suppression of star formation and the launch of galactic winds, people now realize that early feedback from massive stars (e.g. fast OB winds and ionizing radiation) is equally important since the total energy output from them is comparable or even larger than that of the supernovae (SNe; e.g. \cite{agertz_etal13, geen_etal15, grudic_etal21}). More importantly, early feedback operates as soon as massive stars are formed, much earlier than the SN explosion at the death of the stars, which takes at least ten Myr. Given the fact that most star-forming regions are short-lived, early feedback processes determine the evolution and fate of individual star-forming regions. Modern galaxy formation simulations have already shown that early feedback disrupts local density concentrate, shuts off star formation (e.g. \cite{hopkins_etal20}), clears out material and enhances the effects of subsequent SN feedback, reduces the clustering of stars (e.g. \cite{smith_etal21}), enhances chemical enrichment in galactic environments \cite{mao2019nitrogen,mao2021elemental}, and changes the mass and spatial distribution of GMCs (e.g. \cite{li_etal20}). 

However, there exists huge uncertainty regarding the intensity of these processes and the coupling to the ambient multi-phase turbulence-dominated medium. Diffuse X-ray observations of nearby star-forming regions, therefore, provide a direct laboratory to confront the theoretical expectation and observations on the effects of the feedback from the smallest scales. Previous observations have already demonstrated that diffuse X-ray emission is ubiquitous in star-forming regions (e.g. \cite{strickland2007iron, richings_etal10, LiJ13a}). One famous example is the 30 Doradus, the most well-studied massive star-forming region in LMC.

\begin{figure}[H]
  \centering
  \includegraphics[width=\columnwidth]{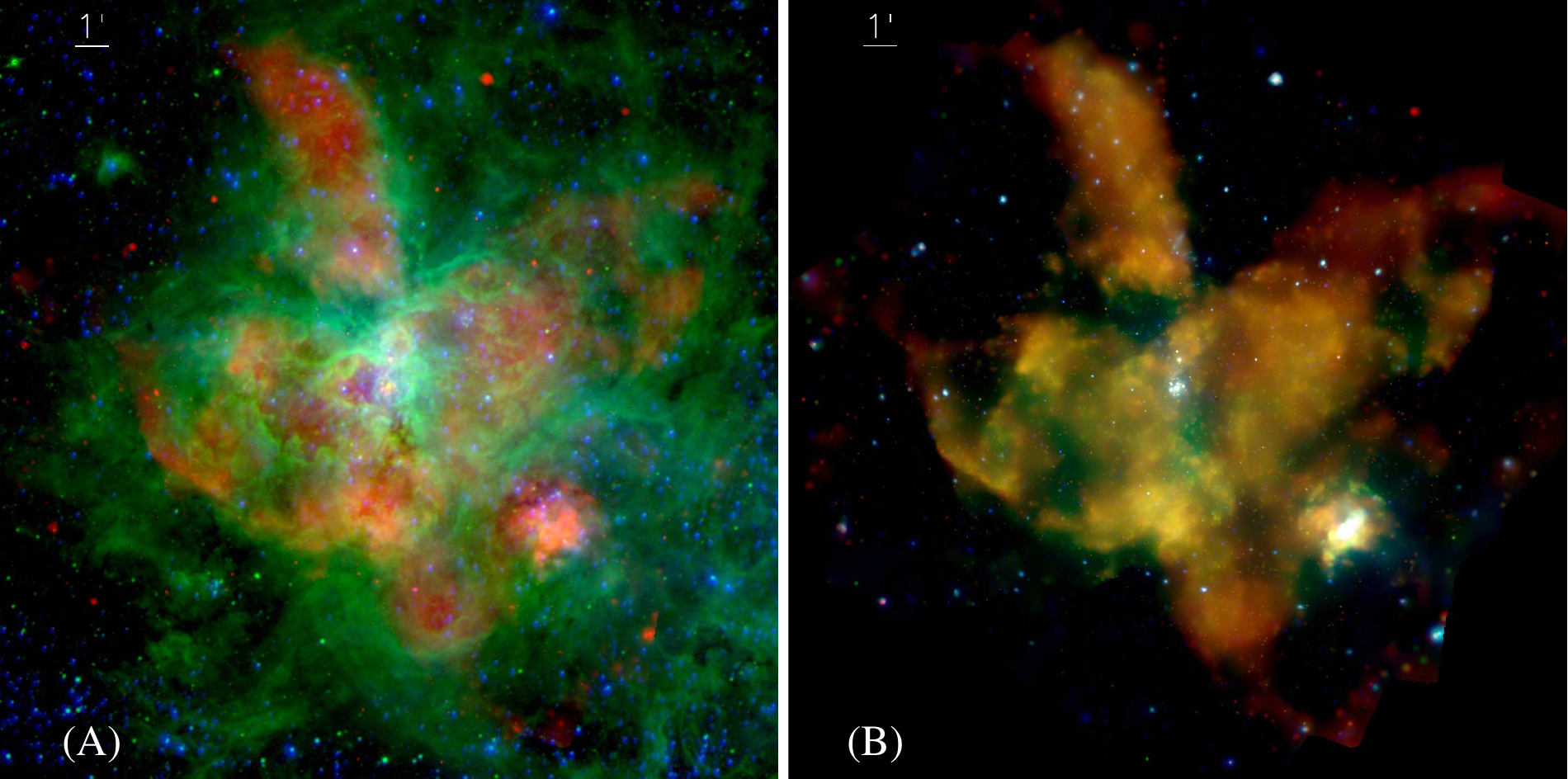}
  \caption{(A) Multi-wavelength image of 30 Doradus, one of the largest star-forming regions located in Large Magellanic Cloud. Million-degree hot gas emitting X-rays (red) is surrounded by warm ionized and neutral gas traced by UV (blue) and $H_\alpha$ (green). (B) \Chandra images in 0.5-1 keV (red), 1-2 keV (green), and 2-8 keV (blue) bands (\cite{cheng_etal21}).}
  \label{fig:30dor_img}
\end{figure}

\begin{figure}[H]
  \centering
  \includegraphics[width=0.45\columnwidth]{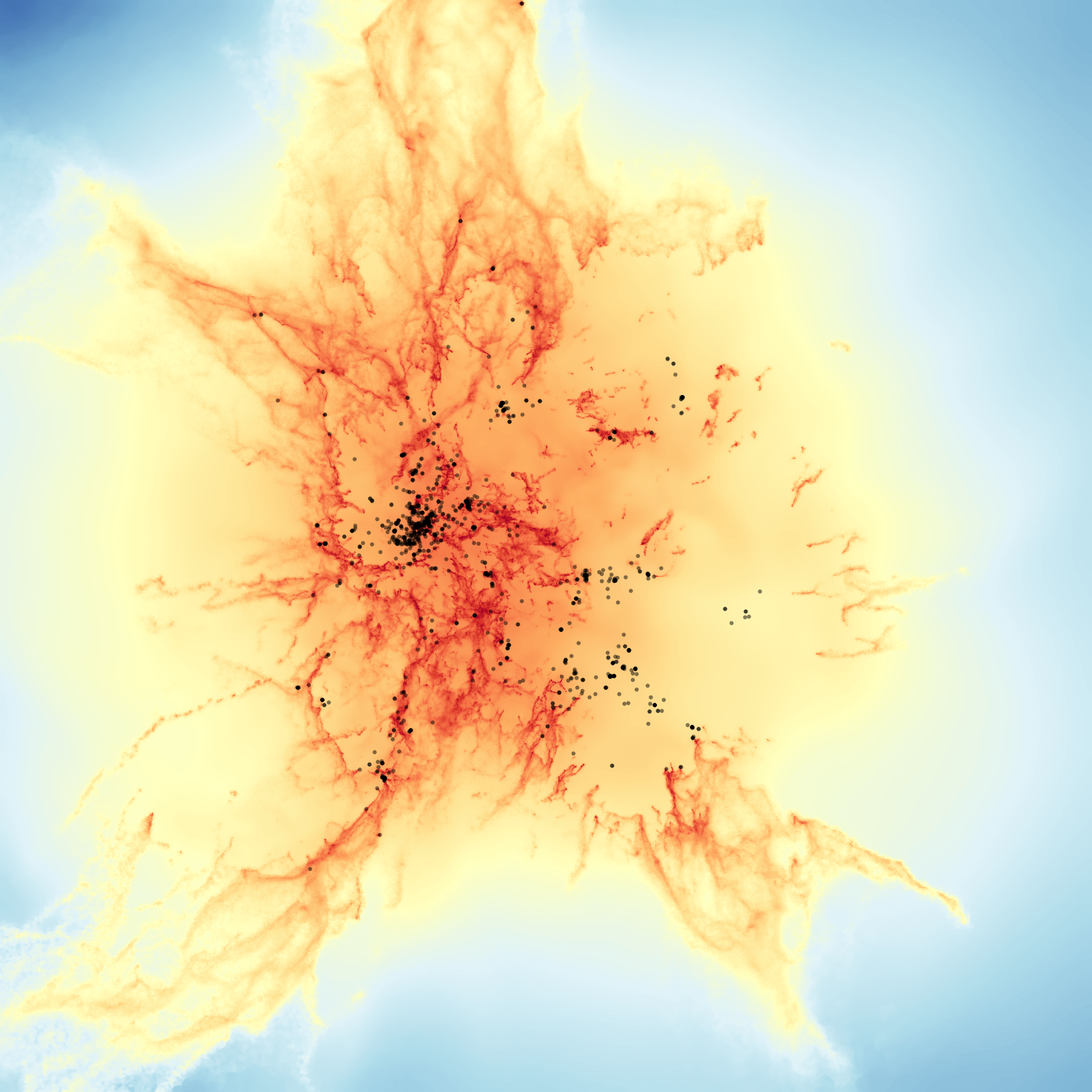}
  \includegraphics[width=0.45\columnwidth]{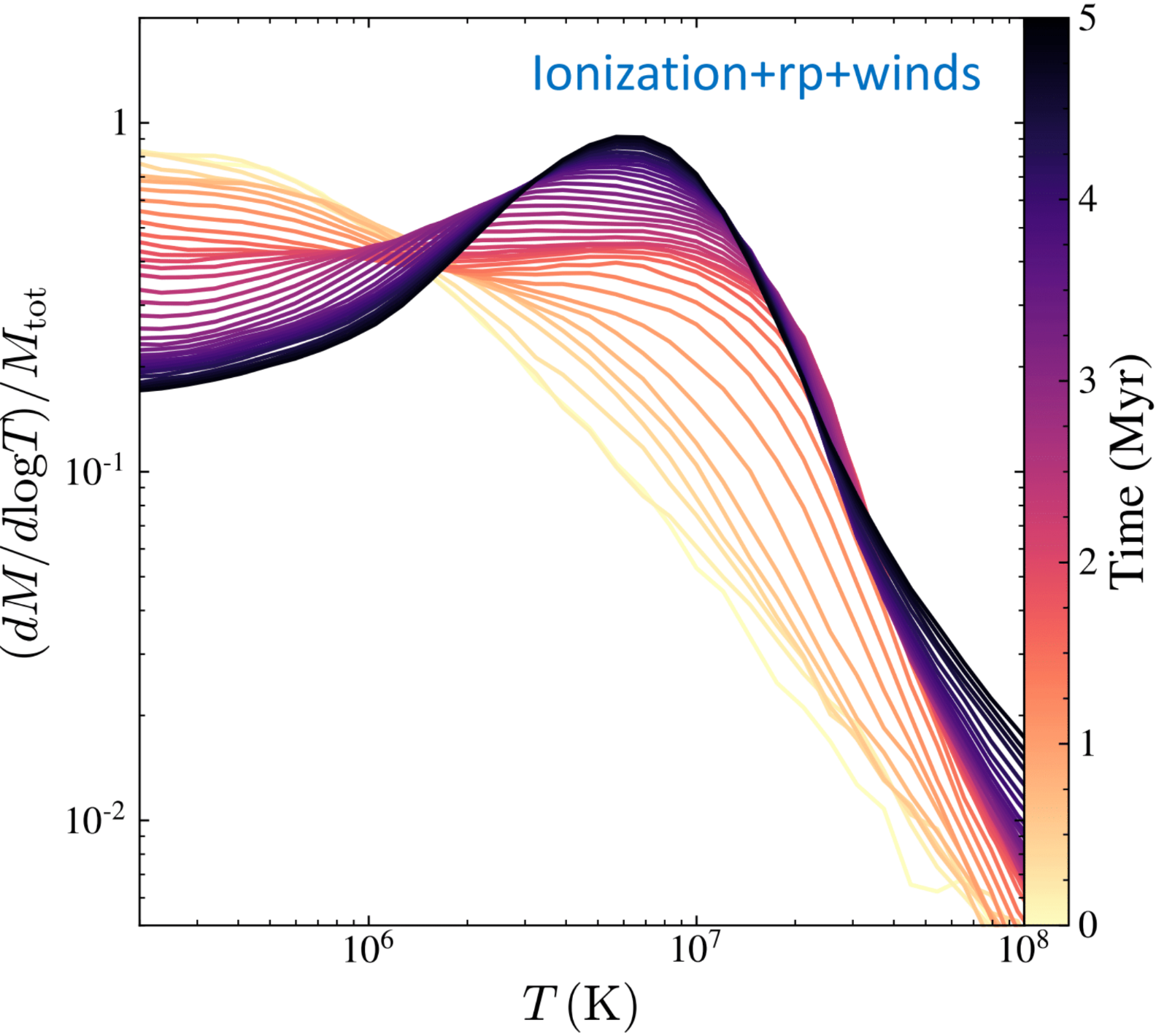}
  \caption{Left: Gas density projection plot of a simulated GMC in the late stage of its evolution. Fast stellar winds and ionizing radiation from massive stars disrupt the cloud and produce high-velocity high-temperature gas. The location of the massive stars formed in the cloud is shown as black dots. Right: Mass-weighted temperature distribution in the same simulation. The different color shows the distribution at different epoch. The exact shape of the distribution depends strongly on the stellar feedback processes included in the simulations (Li et al. in prep.).}
  \label{fig:gmc-sim}
\end{figure}

\begin{figure}[H]
    \centering
    \includegraphics[width=\columnwidth]{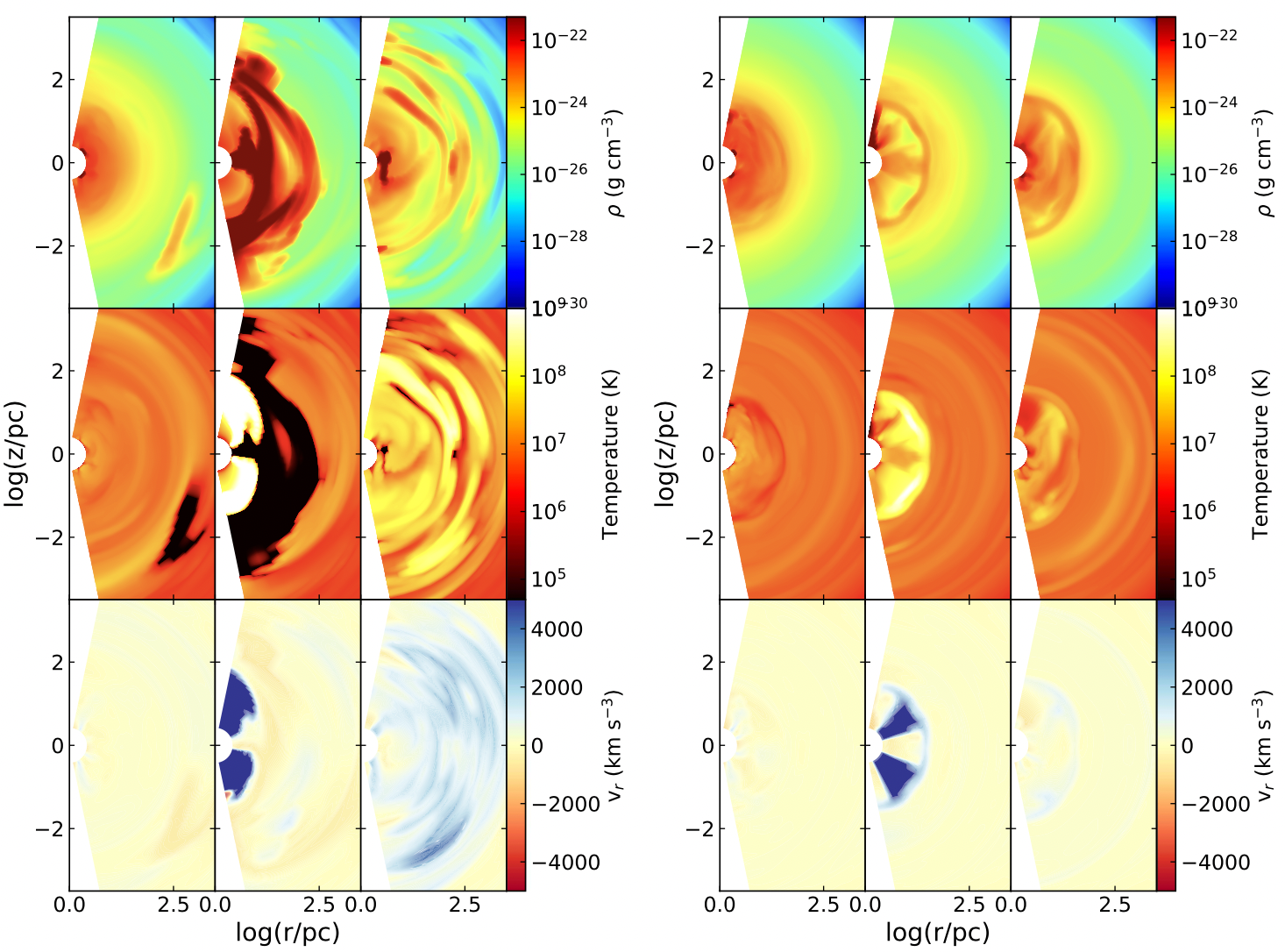}
    \caption{Spatial distribution of density, temperature, and radial velocity
    at three different times obtained from high-resolution simulations of the evolution of an elliptical galaxy when AGN feedback is taken into account. The figure is adapted from \cite{yuan2018active}.}
    \label{fig:AGN_sims}
  \end{figure}

\begin{figure}[H]
  \centering
  \includegraphics[width=\columnwidth]{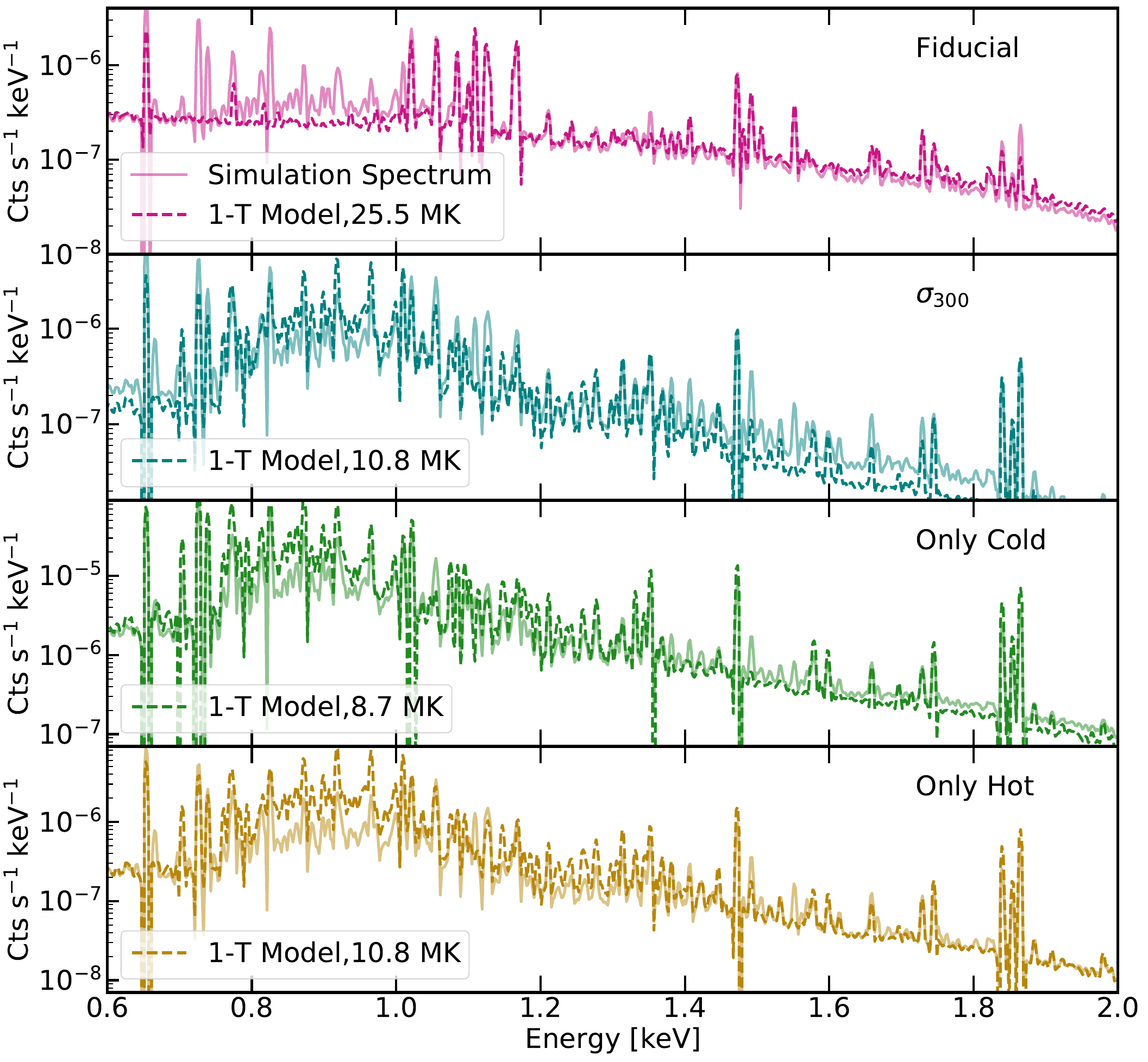}
  \caption{The simulated spectra of an elliptical galaxy emitted by the gas from 0.1-10 {\rm kpc} with four different AGN feedback models. The \HUBS response is used when producing the spectra. The best-fit spectra based on the one-temperature model are shown by the dashed lines. Taken from \cite{2022arXiv221012886V}.}
  \label{fig:HUBSmockspectra}
\end{figure}

Fig.~\ref{fig:30dor_img} shows a composite image of X-ray, H$\alpha$, and UV bands. The complex exhibits many blisters and bubbles filled with hot plasma traced by soft X-rays. These bubbles are surrounded by warm gas traced by UV and H$\alpha$ emission, demonstrating the importance of the interface between gas of different phases. The 30 Doradus region has been investigated extensively in X-rays. Back in the 90s, Wang \& Helfand 1991 (\cite{wang_helfand91}) first revealed the diffuse X-ray emission using the Einstein Observatory and fitted the spectra with an isothermal plasma model of the temperature of $5 \times 10^6\,\mathrm{K}$. Most recently, using \textit{Suzaku} observations of the 30 Doradus, Cheng et al. 2021 (\cite{cheng_etal21}) showed that the spectra are better modeled with a log-normal temperature distribution, which leads to a higher estimate of the total thermal energy and gas pressure. However, due to the limited spectral resolution of the existing X-ray telescopes, the detailed temperature distribution and total thermal energy of the hot gas are not well determined, therefore limiting our understanding of how feedback modulates hot gas around young star clusters. 
From a theoretical perspective, recently numerical simulations start to be able to model various key physical processes in massive star-forming regions. In Fig.~\ref{fig:gmc-sim}, we show the results from a radiation-hydrodynamic that takes into account various stellar feedback processes such as fast stellar winds and ionizing radiation from massive stars (Li et al. in prep.). The combined effects of feedback disrupt the cloud very quickly in only a few Myr and generate a huge amount of hot gas with a complicated temperature distribution, which is a sensitive probe of the detailed feedback implementation. 

\HUBS will provide a major improvement on this matter, thanks to its large field of view and excellent spectral resolution. The 2 eV spectral resolution will reveal individual metal lines with different ionization states and determine the accurate temperature and electron number density at each pixel of the observation. The high spectral resolution will also allow us to determine the shift of the line center and in turn the line-of-sight velocity of the hot gas. With \HUBS's huge field of view, with only a single pointing, we can obtain all the above key physical quantities in a spatially-resolved fashion across the whole star-forming region. These physical quantities together with the derived total energy budgets are powerful measures to directly constrain the dynamics of the regions caused by early stellar feedback processes from a single stellar population.

\begin{figure*}
  \centering
  \includegraphics[width=0.7\textwidth]{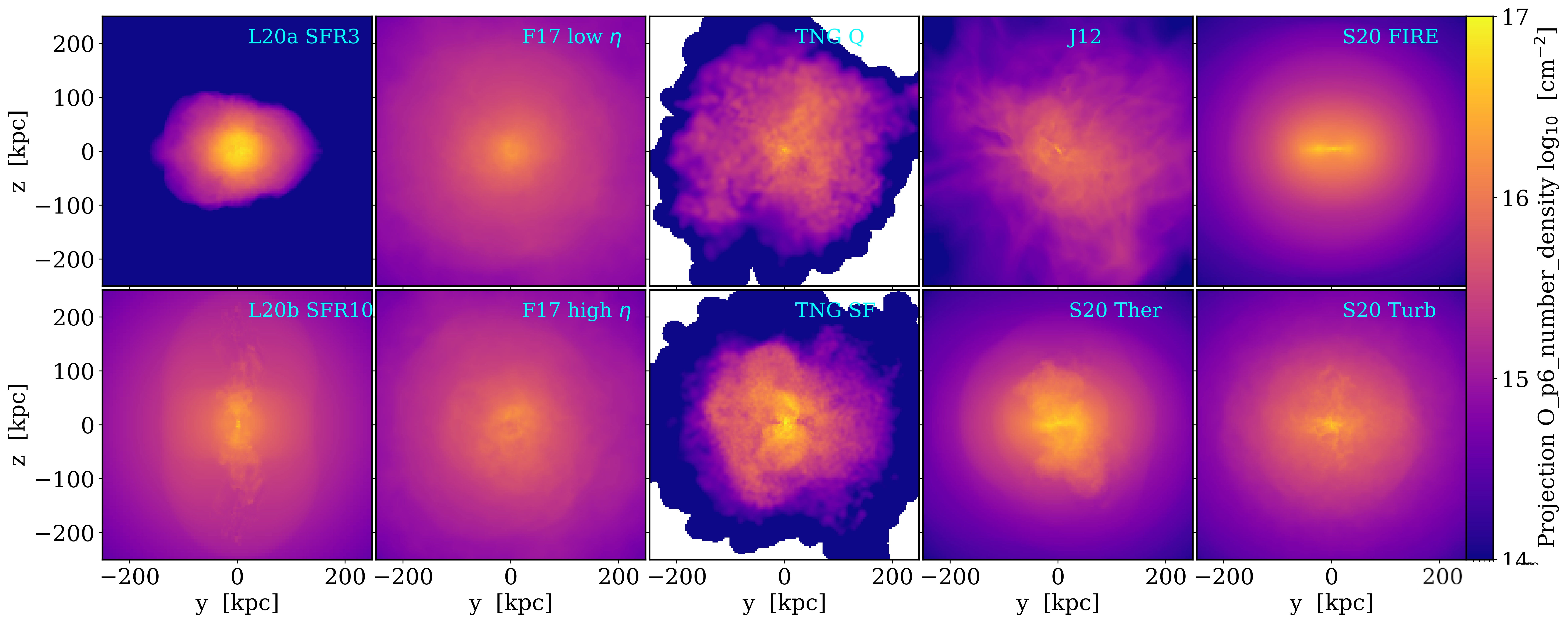}
  \includegraphics[width=0.7\textwidth]{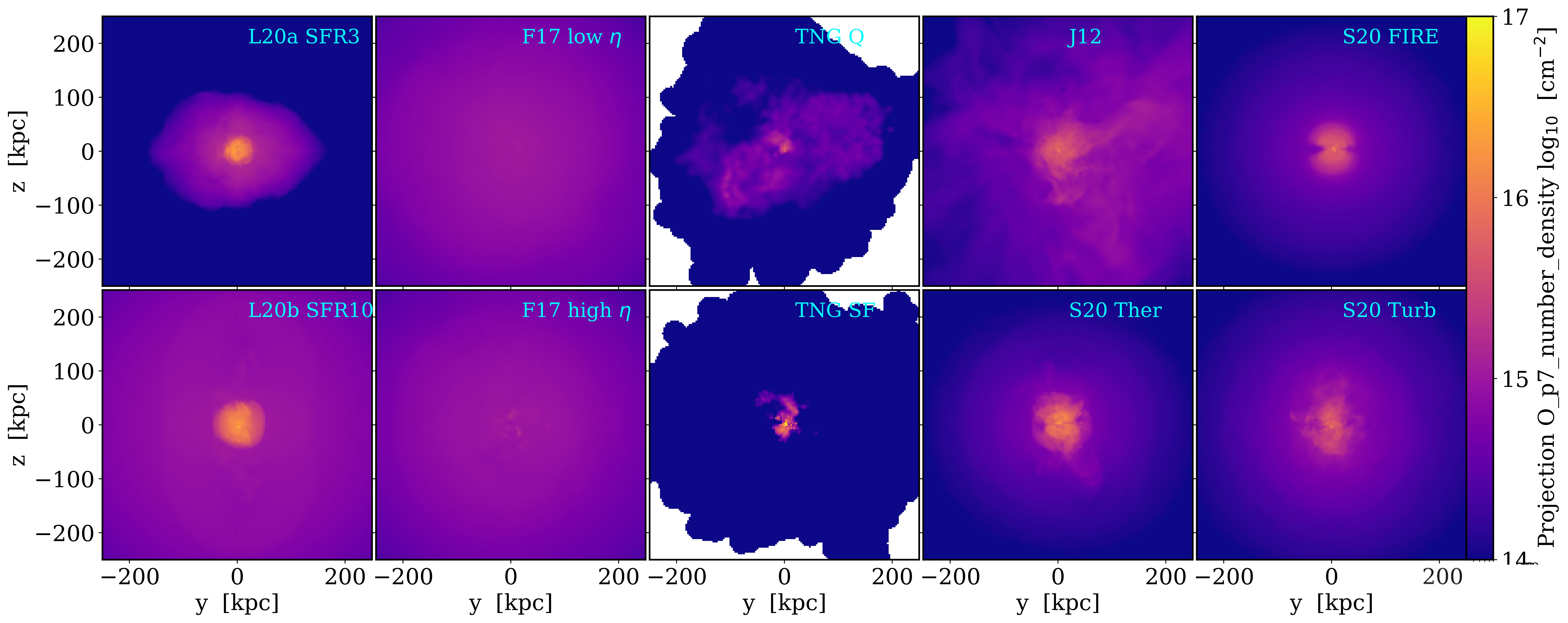}
  \caption{CGM O{~\scriptsize VII} (top) and O{~\scriptsize VIII} (bottom)
  column density projection. Different simulation uses different feedback
  models, resulting in different density distributions of these columns (Li et
  al., in prep).}
  \label{fig:OVII_OVIII}
\end{figure*}

Star-forming regions are highly turbulent, multi-phase media. It has been
demonstrated that wind feedback naturally creates fractal hot bubble surfaces,
where cooling becomes extremely efficient in the interface between different
phases of the gas (\cite{lancaster_etal21}). The interface is an ideal site to
trigger charge exchange between ionized and neutral gas. The charge exchange
(CX) X-ray emission is therefore a powerful tool to quantify the physical
conditions between hot plasmas and cold neutral gas within the star-forming
regions. Significantly different from collisional thermal emission, the X-ray
spectra from CX present enhanced forbidden and intercombination lines relative
to the resonance lines, such as the He$\alpha$ triplet of O {\sc vii}, N {\sc
vi}, and Ne {\sc ix} (\cite{gu_etal16}). However, clear signatures of CX in
X-rays are extremely difficult to detect (\cite{montmerle_townsley12}) because
it requires high spectral resolutions and S/N ratios to resolve the triplets.
\HUBS will be a game changer for the detection of CX spectra signature due to
its large effective area which enhances the S/N rations, and will provide solid
evidence and quantification of the fractal nature of the turbulent medium in
star-forming regions.

Another advantage of the \HUBS high spectral resolution is that the spectra can
be used to determine the accurate abundance of many elements, such as O, Mg, Ne,
and Fe. These measurements will provide stringent constraints on the metal yield
from massive stars from the stellar population synthesis, which is still very
uncertain due to the different models of binary evolution. Besides the total
metal yields, the spatially-resolved metallicity distribution will also be used
to study the time-dependent metal loading and the process of metal diffusion,
both of which are key physical ingredients of the large-scale galaxy formation
models (e.g. \cite{escala_etal18}). Besides D30 Doradus, other interesting
sources include Carina Nebula, NGC 3603, and M17. A compilation of star-forming
regions of various evolutionary stages will provide us with a great opportunity
to confront the theoretical understanding of the stellar feedback in different
stellar populations.

\subsection{CGM: constraining feedback physics and baryon cycle}
\label{subsec:CGM_physics}

At scales of a few hundred $\mathrm{kpc}$s, the fraction of baryons in the universe
contributing to the circumgalactic medium of galaxies is crucial to answering
the question of ``missing baryon''. Part of the missing baryons is likely to
exist in the hot medium outside the galaxy \cite{cen1999baryons}, but the exact
location is unclear: It may be mainly in the dark matter halo near the galaxy,
or around a large-scale structure other than the dark matter halo
\cite{tanimura2017probing}. The amount of baryons contained in the thermal
medium is determined by the accretion driven by dark matter halos and the feedback of
galaxies. It is now widely accepted that feedback in less massive and massive
galaxies are dominated by supernovae and AGNs respectively
\cite{efstathiou2000model,Kormendy2013}, while how they affect the baryon
content of the circumgalactic medium remains under active investigation.
Thankfully, since the density of the CGM is not forbiddingly low, the emitted
X-rays of the circumgalactic medium are still above the detection limit.
Observations of the X-ray system of the galactic peripheral thermal medium can
constrain the mass of the galactic medium, which will help us understand the
distribution of baryons and feedback physics.

Because of its low density, CGM is very sensitive to the
feedback effect, and the properties of CGM generated by different simulations
are very different (e.g., \cite{2022arXiv221012886V}). Therefore, comparison with
the observed CGM can be used to identify the credibility of the feedback model.
Fig.~\ref{fig:OVII_OVIII} is a comparison of the distribution map of CGM
O{~\scriptsize VII}, O{~\scriptsize VIII} column density (Li et al. in prep),
including the mainstream cosmological simulation IllustrisTNG and zoom-in
simulation \cite{joung2012gas}, and the simulation of a single galaxy
\cite{fielding2017supernovae,li2020supernovae,su2020cosmic}, each simulation
adopts different feedback models. It can be seen that the O{~\scriptsize VII}
and O{~\scriptsize VIII} in different simulations are very different. \HUBS's
observations of CGM in neighboring galaxies can be compared with the results of
numerical simulations to evaluate the reduction of CGM by different simulations
and limit galaxy feedback models.

To better constrain the galaxy formation models, a large sample of the hot CGM
is needed and systematically compared with the numerical simulations. At
present, there are only dozens of disk galaxies that have detected halo,
basically within 30 Mpc. Generally speaking, in less massive galaxies, the
thermal CGM is fainter and less extended. \HUBS can use its large field of view
to significantly improve observation efficiency and see fainter, farther and
less massive galaxies, increasing the number of samples by at least an order 
of magnitude. In this way, there will be a statistically complete sample in
various physical parameter intervals of galaxies, such as galaxy mass, stellar
activity in the galaxy, and the large-scale environment in which the galaxy is
located. This provides a comprehensive picture of how the properties of the hot
CGM change with the physical parameters of the galaxy. Numerical simulations
that have developed rapidly in recent years will also cover these physical
parameters and systematically predict the outflow of these galaxies and the
thermal medium around galaxies. Comparing the total luminosity, metal abundance,
spectrum and other information of the observed X-rays with the results of
numerical simulation, we can (1) constrain the feedback processes in galaxies under
different conditions and give quantitative limits on the mass cycle of galaxies;
(2) constrain the mass of hot baryons contained in the CGM from the statistical
point of view, and answer the “missing baryon” problem. 

On galactic scales, how the galactic winds from SN and AGN feedback including
the AGN jets \cite{wang2022magnetically,yang2022modeling} interact with
cosmological inflows is the basis for the formation of galaxies. A key question
in galaxy formation is how galaxies obtain mass from their CGM and when to
stop growing \cite{somerville2015physical}. Competition between inflows and
feedback directly determines the growth rate of galaxies. Inflow will increase
the mass of galaxies, and the galaxy wind can not only take away mass and metals
from galaxies, but also reduce the accretion of new gas from the periphery. If
this effect is significant, the inflow will be blocked and the star formation
will be suppressed \cite{yuan2018active}. The converging location of inflow and
outflow is the CGM. Therefore, the observation and study of the galactic media
will provide important clues to the formation and evolution of galaxies.

\begin{figure}[H]
  \centering
  \includegraphics[width=\columnwidth]{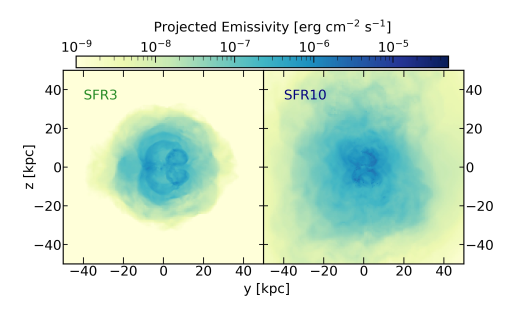}
  \caption{The cross-section of the MW-like thermal halo and the X-ray surface
  brightness. The left panel shows the case with the star formation rate of
  $3 \mathrm{M_\odot/yr}$ (current level), and right panel
  the star formation rate of 10 $\mathrm{M_\odot/yr}$. (Vijayan \& Li 2022)}
  \label{fig:xray_sfr}
\end{figure}

To quantitatively constrain the strength of feedback and inflow, a number of
attempts have been made. The Numerical Investigation of a Hundred Astrophysical
Objects (NIHAO) project \cite{wang2015nihao} simulated a total of 100 galaxies
with masses ranging from dwarfs to the Milky Way (MW)-like galaxies. Compared
with other simulations (e.g., Eagle, Illustris), the NIHAO simulation has some
advantages in the number of samples and resolutions. The NIHAO simulation
includes pre-stellar feedback, so it can reproduce the stellar-halo mass
relationship and is suitable for studying the baryon cycle. The impact of SN
feedback on the baryon cycle is further studied \cite{tollet2019nihao}, where
the percentage of baryons ejected out of the dark matter halo and returned to the galaxy
are predicted respectively. Because the brightness of the thermal halo decreases
with the increase of the radius, existing observations generally only see
radiation within a few kpcs from the galaxy. 

It is expected that \HUBS can increase the detectable physical scale by more
than one order of magnitude, so that the current research on the inner region of
the thermal halo can be extended to the entire CGM (even the IGM), thereby
constraining radial profiles of various properties (e.g., density and
temperature). Theoretically, inflow and outflow have very different effects on
the properties of the thermal CGM (metal abundance, temperature, etc.). A
remarkable feature is that the metal abundance in the outflow-dominated area
will be much higher than that in the inflow. The outflow of different
intensities will also produce very different effects. More energetic outflows
launched by intensive star formation could enrich metals and heat up the gaseous
halo at larger radii, which leads to a higher X-ray luminosity (see
Fig.~\ref{fig:xray_sfr}, numerical simulation results \cite{vijayan2022x}).
\HUBS's good spectral resolution can provide important clues to the properties of
the thermal gas around the galaxy, such as temperature and metal abundance.
Combined with its better spatial resolution, it can limit the inflow, outflow
and the area where they interact, from which the feedback strength and the
intensity of the inflow can be inferred. In addition, due to the invisibility of
large-scale environments, there is no constraint on how many thermal gases there
are in the media around the galaxy, while \HUBS's observation of large-scale
thermal gas will provide an important constraint on this.

It was a long mystery that the expected amount of baryons around galaxies are not detected in existing multi-wavelength observations (e.g., \cite{Planck13b, Werk14, Prochaska2017, Eckert15, LiJ18, Bregman18, Bregman22}). For example, with the NIHAO galaxy formation simulations, \cite{2017MNRAS.466.4858W} made predictions for the baryonic budget in present-day Milky-Way ($M_{200} \sim 10^{12}~M_{\odot}$) type galaxies. They found that, compared to a universal cosmic baryon fraction of $f_b = \Omega_b / \Omega_m = 0.15$, haloes of this mass scale are typically ``missing'' 30\% of the expected baryons, which are relocated to beyond two times the virial radii and are dominated by a diffuse warm-hot gas.  

\begin{figure}[H]
\centering
\includegraphics[width=0.47\textwidth,trim=22mm 18mm 18mm 15mm, clip]{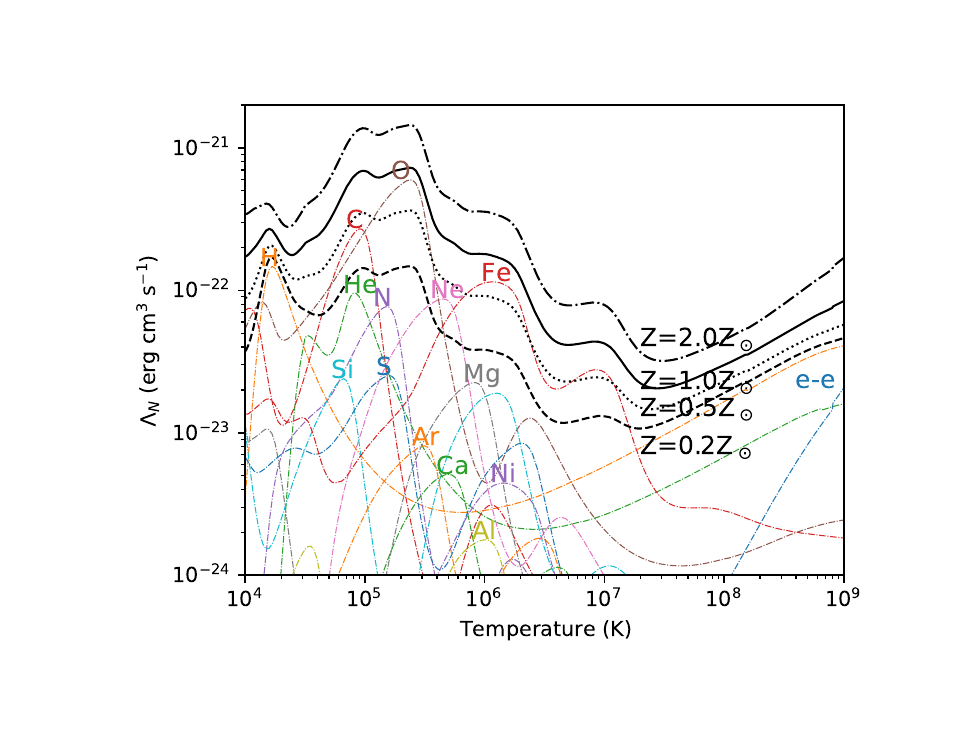}
\caption{Radiative cooling curve of the thermal plasma under collisional ionization equilibrium (CIE) based on the AtomDB database (\url{http://www.atomdb.org}). The vertical axis is the normalized radiative cooling rate of the plasma defined as $\Lambda_N\equiv\frac{U}{\tau_{\rm cool}n_{\rm e}n_{\rm t}}$ (in a unit of $\rm erg~cm^{3}~s^{-1}$), where $U$ is the internal energy of the gas [$U=\frac{3}{2}(n_{\rm e}+n_{\rm t})kT$], $\tau_{\rm cool}$ is the radiative cooling timescale, while $n_{\rm t}$ and $n_{\rm e}$ are the total ion and electron number densities, respectively. The horizontal axis is the temperature of the plasma. Different colored dash-dotted curves are the contribution by different elements (we only consider 14 elements with the strongest emissions) assuming an abundance of $Z=1.0~Z_\odot$, with ``e-e'' denoting the electron-electron bremsstrahlung emission. The thick black curves are the sum of all these components under different abundances (the dashed, dotted, solid, and dash-dotted curves correspond to $Z=0.2, 0.5, 1.0, 2.0~Z_\odot$, respectively).}
\label{fig:CoolingCurve}
\end{figure}

The game is to find this gas and map its distribution. As can be seen in Fig.~\ref{fig:CoolingCurve}, the temperature at the peak of the radiative cooling curve ($T\sim10^{5.5}\rm~K$) is close to the virial temperature of $L^\star$ galaxies. For $L^\star$ or super-$L^\star$ galaxies with more massive halos ($M_{\rm halo}\gtrsim10^{12-13}~M_{\odot}$), the virial temperature could fall in the X-ray emitting range where the radiative cooling efficiency is relatively low compared to lower mass halos, as the latter has higher metal cooling efficiency. In this case, there could exist an extended and stable X-ray-emitting hot gaseous halo that potentially contains a significant fraction of the ``missing baryons'' (e.g., \cite{LiJ17, LiJ18, Bregman22}).

\begin{figure}[H]
\centering
\includegraphics[width=0.47\textwidth]{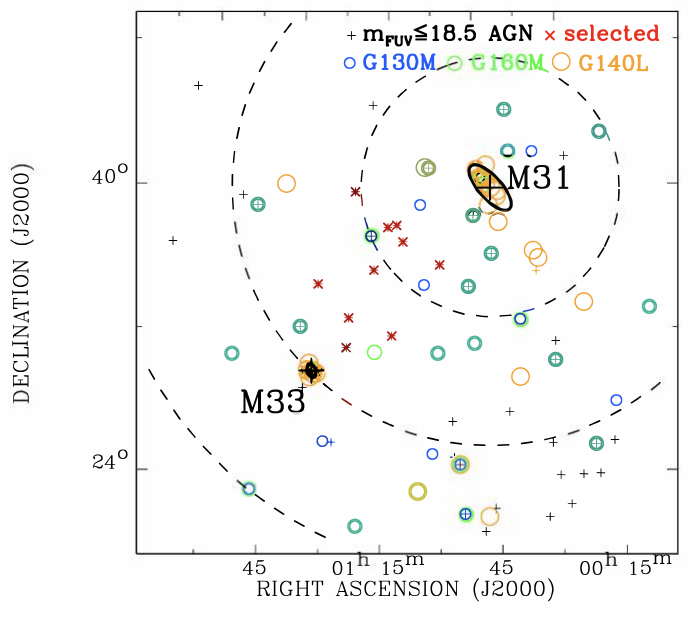}
\caption{
Location of UV bright (with GALEX FUV magnitude $m_{\rm FUV} \leq 18.5$) AGNs (black plus) around M31 and M33. The large plus sign and the solid ellipse mark the location and extension of the optical disk of the two galaxies. The three large dashed circles have a radius of $r = 100, 200, 300\rm~kpc$ from the center of M31. Small colored circles as denoted on the top right are HST/COS observations of objects in the surrounding area \cite{Rao13}, or are MW halo stars which allow for a determination of the absorption from the foreground MW halo \cite{Lehner15}.}
\label{fig:M31BckUVAGN}
\end{figure}

The first science case in this regard is to observe nearby objects. In this case, the hot CGM could be studied in unprecedented detail even with moderate exposures (e.g., \cite{LiJ13a,Qu21}). The Andromeda galaxy (M31), with a stellar mass of $M_*=(1-1.5)\times10^{11}~M_\odot$, a dark matter halo mass of $M_{\rm 200}=(8-11)\times10^{11}~M_\odot$ \cite{Tamm12}, ${\rm SFR}\approx0.4~M_\odot~yr^{-1}$ \cite{Barmby06}, and $d\approx0.78\rm~Mpc$ ($1^\prime\approx230\rm~pc$), is thus the best case to search for the large-scale accreted hot CGM. It is the external galaxy with the largest angular size of the virial radius ($r_{\rm vir} \approx 300{\rm~kpc} \approx 23^\circ$), while the companion galaxy M33 locates at a projected distance of $\sim200\rm~kpc$ from M31, within its dark matter halo (see Fig.~\ref{fig:M31BckUVAGN} for the configuration). Such a large angular size of the dark matter halo makes M31 unique for the most detailed study of the multi-phase CGM. In particular, there are a lot of UV-bright background AGN projected within $r_{\rm vir}$ of M31 (Fig.~\ref{fig:M31BckUVAGN}), allowing for UV absorption line studies of the cool and warm gases from the CGM \cite{Lehner20}. Furthermore, as our closest massive neighbor in the local group, M31 also received a lot of observations in many other bands, which help us to study both its dark matter halo and the multi-phase CGM (e.g., \cite{Braun04,Patel17}). These multi-wavelength observations, together with the proposed large sky area survey with \HUBS will need $\sim(200-300)~\times$ 15 ks = $(3.0-4.5)$ Ms
observations to cover the entire area of interest, such as the M31-M33 stellar and gas stream (\cite{Richardson11}). This will provide us with a unique panchromatic view of the baryon budget among the stars and the multi-phase CGM.

\begin{figure}[H]
\centering
\includegraphics[width=0.35\textwidth,angle=270]{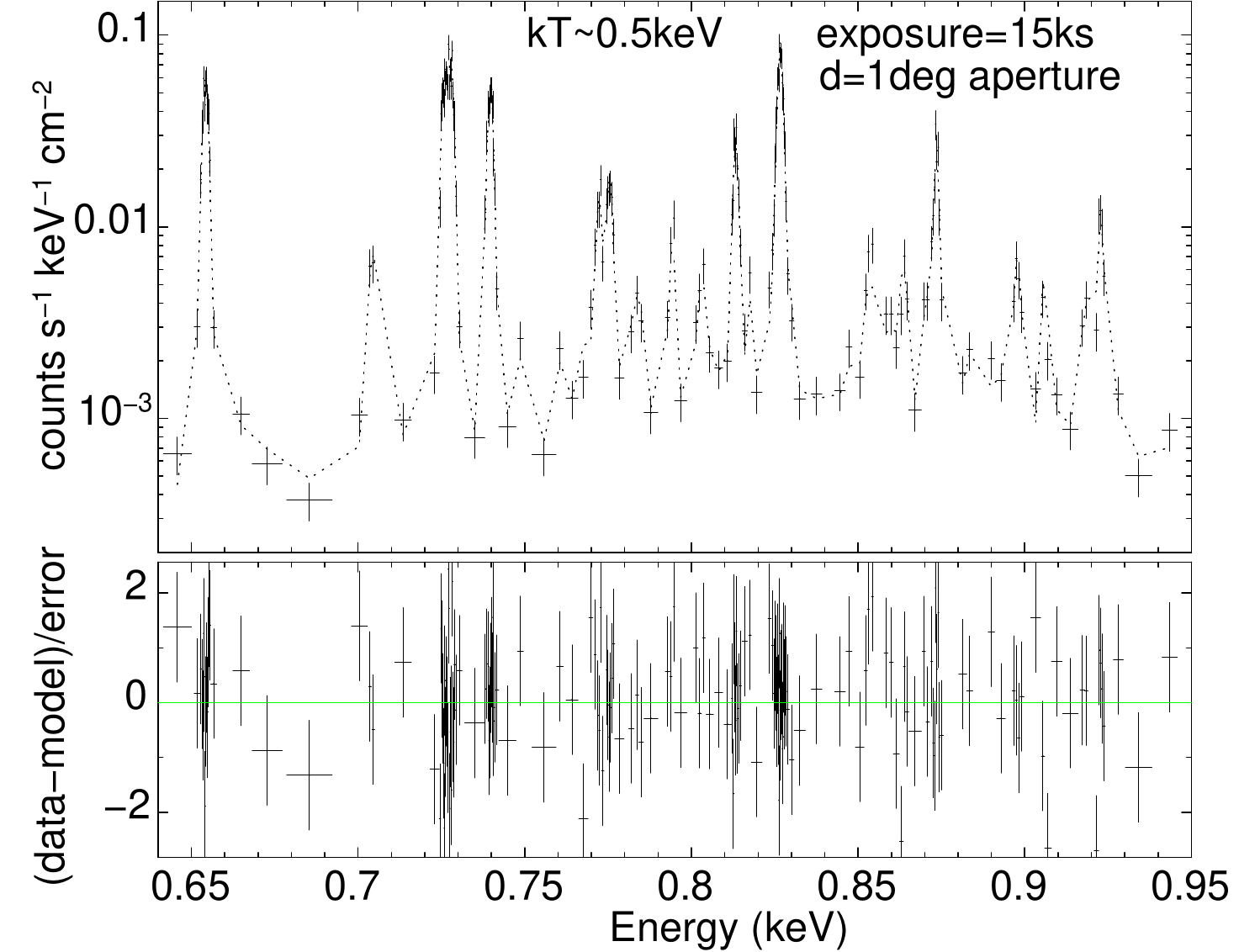}
\caption{Simulated 15~ks \HUBS spectrum
extracted from the $1^\circ \times 1^{\circ}$ FOV coverage of the M31 halo. The dotted curve shows the $kT\sim0.5\rm~keV$ 
model subjected to Galactic foreground absorption.}
\label{fig:HUBSSimuSpecM31}
\end{figure}

Fig.~\ref{fig:HUBSSimuSpecM31} shows the simulated \HUBS spectrum extracted from the entire FOV toward the direction of M31, with an exposure of only $\sim15\rm~ks$. 
The real selection of the spectral extraction aperture depends on both the brightness of the feature of interest and the scientific goal. Here the $1^\circ \times 1^{\circ}$ aperture is still enough to separate the M31-M33 stream from the surrounding medium \cite{Richardson11}, which is a large-scale structure with gaseous counterparts \cite{Braun04}. In many cases when we do not need such a high signal-to-noise ratio, a higher angular resolution down to the instrument limit ($1^\prime\approx230\rm~pc$) could be adopted, which is impossible for more distant galaxies. Due to the large angular size of the object, such \HUBS observations of local galaxies are still very time-consuming and typically require a few mega-seconds.

We also expect to collect a sample of $\lesssim10$ massive galaxies at moderate distances for \HUBS. The X-ray emission of the CGM could arise not only from the feedback of AGN and stellar sources (e.g., \cite{Strickland04,LiJ13b,Wang16}), but also from the accretion shock heating and gravitational compression of the IGM (e.g., \cite{Benson00}). The relative importance of these two potentially interrelated mechanisms likely depends on a galaxy's mass, as well as other properties such as the SFR and the environment. The extended hot CGM could potentially contain a large fraction of a galaxy's ``missing baryons''. However, due to its low density and metallicity, the X-ray emissivity of this extended hot CGM is extremely low \cite{Crain13}. In order to detect it and characterize its spatial distribution, we need a galaxy sample that is massive enough so the virialized gas has a temperature falling in the X-ray emitting band. These galaxies also need to be quiescent in star formation to avoid disproportionately strong X-ray emission from metal-enriched feedback material, as well as in a non-cluster environment such that the ICM would not contaminate the measurement of the CGM in the galaxy vicinity \cite{LiJ16,LiJ17}. Furthermore, it will also be better if the galaxies are located at a moderate distance of $d\sim(50-100)\rm~Mpc$, so the \HUBS FOV will cover at least a significant fraction of the virial radius, and the redshifted soft X-ray emission lines could be separated from the MW foreground emission (\cite{LiJ20, Zhang22}). The best cases are thus super-$L^\star$ quiescent galaxies (e.g., \cite{Dai12, Bogdan13, Bogdan15, Anderson16}). With the high energy resolution and low background of \HUBS, we can extract narrow-band images covering individual emission lines with significantly suppressed MW foreground emission, and probe its radial distribution out to almost the virial radius.

\begin{figure}[H]
\centering
\includegraphics[width=0.47\textwidth]{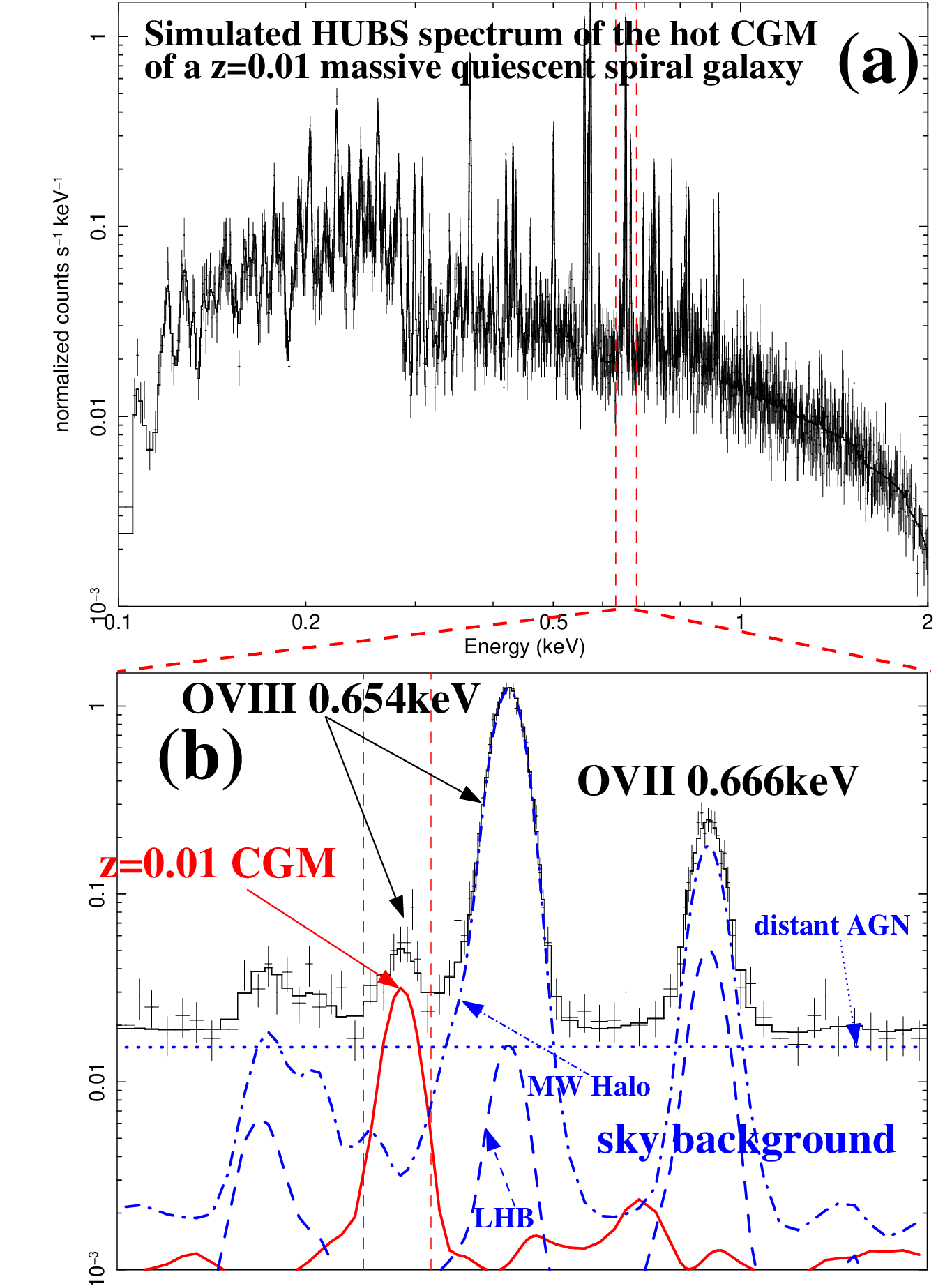}
\caption{Simulated 1~Ms \HUBS spectrum of the hot CGM at $r \leq 0.2r_{\rm 200}$ around a $z=0.01$ ($d\sim50\rm~Mpc$) galaxy \cite{LiJ16,LiJ20}. Panel~(b) 
is the zoom-in in the energy range of 0.63–0.68~keV of panel~(a). The red curve is the hot CGM ($kT=0.6\rm~keV$ plasma), while the blue curves are various sky background components (local hot bubble LHB; MW halo; distant AGN). The redshifted 
O{\scriptsize~VIII} line from the CGM can be separated from the MW halo component.}
\label{fig:CGMMASSSimuHUBSSpec}
\end{figure}

When probing X-ray emission from low surface brightness features such as the extended CGM, the most important thing is not only the photon statistic, but also the level and fluctuation of the sky background. With broadband X-ray imaging observations, we can typically detect the hot CGM only within $r\lesssim(20-30)\rm~kpc$ or $r\lesssim 0.1r_{\rm 200}$ (e.g., \cite{LiJ16,LiJ17,LiJ18,LiJ20}).
We present a simulated $\sim1\rm~Ms$ \HUBS spectrum of a $z=0.01$ ($d\approx50\rm~Mpc$) massive quiescent galaxy in Fig.~\ref{fig:CGMMASSSimuHUBSSpec}, using the spectral model from \cite{LiJ16}. It is clear that some key diagnostic emission lines of the hot gas, such as the redshifted O{\sc viii} line at the rest-frame energy of $0.654\rm~keV$, could be separated from the same emission line arising from the MW halo. This will significantly increase the signal-to-noise ratio of the redshifted hot gas emission lines in narrow-band imaging observations.

We must note that the objects to be included in such observations need to be carefully selected according to their redshifts. A shorter distance will be helpful to collect more photons to study the physical and chemical properties of the brightest part of the hot CGM, but the contamination from the MW foreground makes it difficult to detect the faint extended hot CGM which potentially contains a larger fraction of the baryons \cite{Bregman18,LiJ18}. On the other hand, a too-large distance will significantly reduce the flux of the object and makes the project unfeasible. The best choice will be objects at $d\sim(50-100)\rm~Mpc$, such as the CGM-MASS sample studied in \cite{LiJ16,LiJ17,LiJ18}. We also would like to emphasize that the galaxies in the mass range of the CGM-MASS galaxies often have a large discrepancy in the measured hot CGM mass based on X-ray or Sunyaev-Zel'dovich (SZ) observations \cite{Planck13b, LiJ17, Qu18, Bregman22},
which could be partially caused by the poorly constrained hot gas density profile \cite{Bogdan15,LiJ18}. This is another reason to have deep X-ray observations probing the hot CGM from a large fraction of the dark matter halo. The total \HUBS observation time needed to complete such a survey will be a few mega-seconds, depending on the real sample size and the adjustment of the exposure time for individual galaxies based on existing \Chandra and \XMMNewton observations \cite{Bogdan15,LiJ17}. 

\subsection{Additional feedback physics}
\label{subsec:CGM_physics_additional}

Due to the high level of complexity of galactic environments, the feedback
processes might involve more physics than hydrodynamics and gravity. The impact
of non-thermal physics on galaxy formation, such as magnetic fields and cosmic
rays, has been long overlooked until recently the importance of these additional
feedback physics starts to be investigated and recognized \cite{Faucher2023}. A
number of theoretical models have been developed, predicting distinctively
different CGM properties with a wide range of physical parameter spaces poorly
constrained by existing observations. \HUBS will be uniquely qualified to test
these models and constrain the feedback physics in the CGM, which is discussed
in detail as follows.

\emph{Magnetic fields:} The magnetic field strength in the CGM is expected to be much weaker than that in the ISM, as the CGM is expected to be
more diffuse and less dense. In the MW halo, the best-fitting $B$-field values
are $\sim 1$--$10\,\mathrm{\mu G}$ \cite{haverkorn15,han2017observing}. 
In recent years, the strength and topology of galactic scale magnetic field in the CGM started to be well constrained in radio observations, via either polarization or Faraday rotation measure (RM) synthesis (\cite{Irwin12a,Krause20}). 
The observed magnetic energy density in the CGM could be either higher or lower than the hot gas pressure \cite{li2008chandra,fang2012hot,Irwin12b,MoraPartiarroyo19,Stein19}, indicating a variety of roles the magnetic field plays in the global gas flows.
On the simulation side, a variety of magnetic field
strengths and topologies in the CGM are predicted by different sets of
simulations, such as SURGE \cite{vandevoort2021the} and FIRE
\cite{ponnada2022magnetic}, which is still under active investigation. For
instance, \cite{vandevoort2021the} found that the magnetic fields in the
simulations even become dominant in the bi-conical regions. Therefore, the
impact of magnetic fields in the CGM might not be negligible.

\begin{figure}[H]
\centering
\includegraphics[width=\columnwidth]{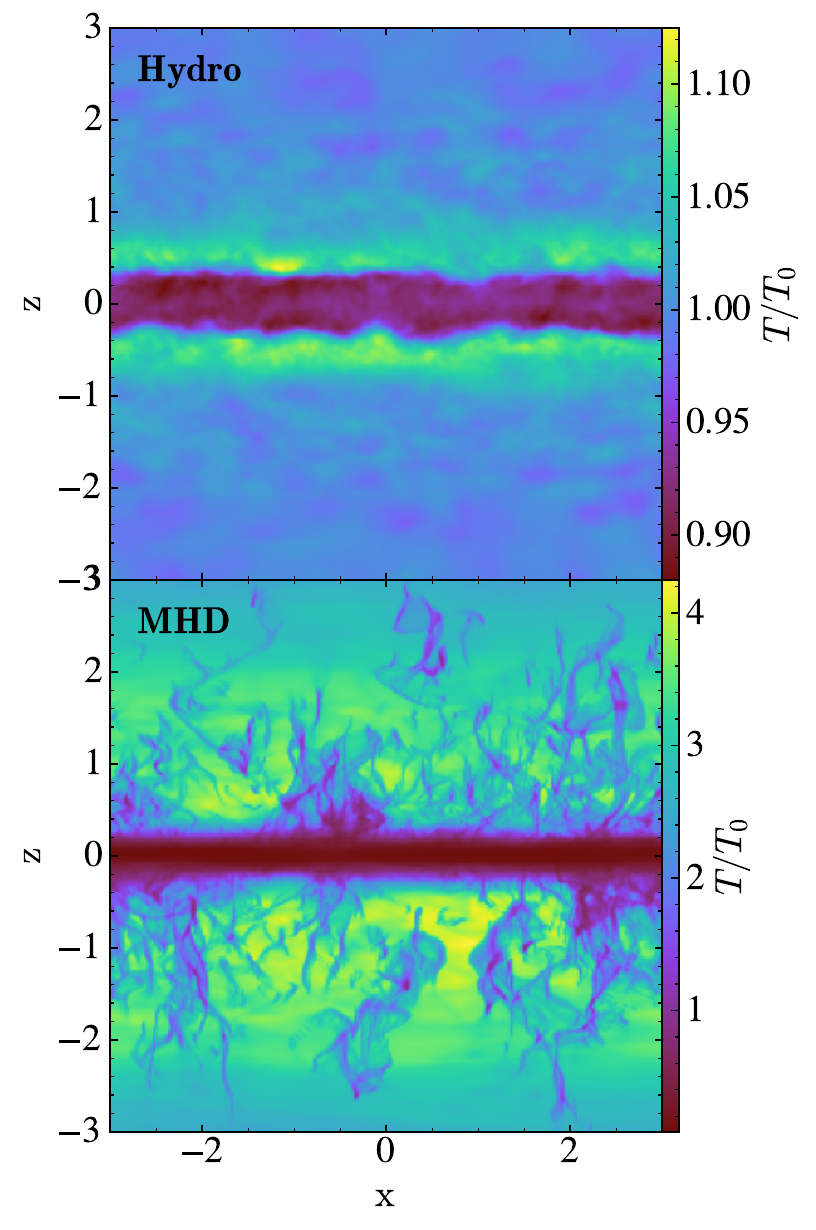}
\caption{Projections of the gas temperature $T$ (normalized to the initial
temperature $T _0$) of simulated patches of the CGM in simulations without (top)
and with (bottom) magnetic fields. Magnetic fields dramatically change the
thermal states of the CGM by enhancing the thermal instability as well as the
production of the cool gas, while the CGM remains single-phase in the case
without magnetic fields. Datasets used in this figure are from \cite{ji2018}.}
\label{fig:thermal_instab}
\end{figure}

Magnetic fields affect CGM in a few ways. First, magnetic fields can provide
non-thermal magnetic pressure to the CGM, which could even be comparable to the
local thermal pressure of the field strengths reaches a few $\mathrm{\mu G}$. In
this case, the halo gas is partially supported by the magnetic pressure
$P_\mathrm{mag} = |\bm{B}|^2 / 4\pi$, and can stay at a lower thermal pressure/temperature \cite{vandevoort2021the}. Second, magnetic fields can facilitate the
production of cool gas in the CGM by enhancing thermal instability
\cite{ji2018}. As shown in Fig.~\ref{fig:thermal_instab}, at the presence of
magnetic fields (where the magnetic energy and gas thermal energy are
comparable), a significant amount of cool filaments arise in the CGM via
enhanced thermal instabilities (bottom), in contrast to the case without
magnetic fields where the CGM remains a single phase (top). Finally, magnetic
fields help the survival of the cool gas by suppressing turbulent mixing with the
hot phase via magnetic tension
\cite{dursi2008draping,mccourt2015magnetized,Ji2019,yang2023radiative}, or reduce thermal
conduction between the cool and hot phases via anisotropic conduction
\cite{parrish2009anisotropic}. Therefore, the overall impact of magnetic fields
is to increase the fraction of the cool gas in the CGM, and thus potentially
alter the CGM thermal status which can be tested by \HUBS.

\emph{Cosmic rays:} Cosmic rays (CRs) are ultra-relativistic protons/electrons
coupled with local plasma magnetic fields via Lorentz forces
\cite{zweibel2013the}. At the galactic/ISM scale, CRs at GeV energies (which
dominants the CR energy spectrum) are produced by supernovae and AGN shock
acceleration, and are transported by turbulence and magnetic fields in ISM and
CGM. CRs are believed to be in roughly energy equipartition with the magnetic
fields and thermal pressure in the ISM \cite{draine2010physics}. 
In recent years, the CR energy density and transport mechanisms have been better constrained via spatial analysis of the synchrotron radio continuum emissions detected above the galactic disks (e.g., \cite{Irwin12a,Krause18,Heald22,Stein22}).

\begin{figure}[H]
  \centering
  \includegraphics[width=1.05\columnwidth]{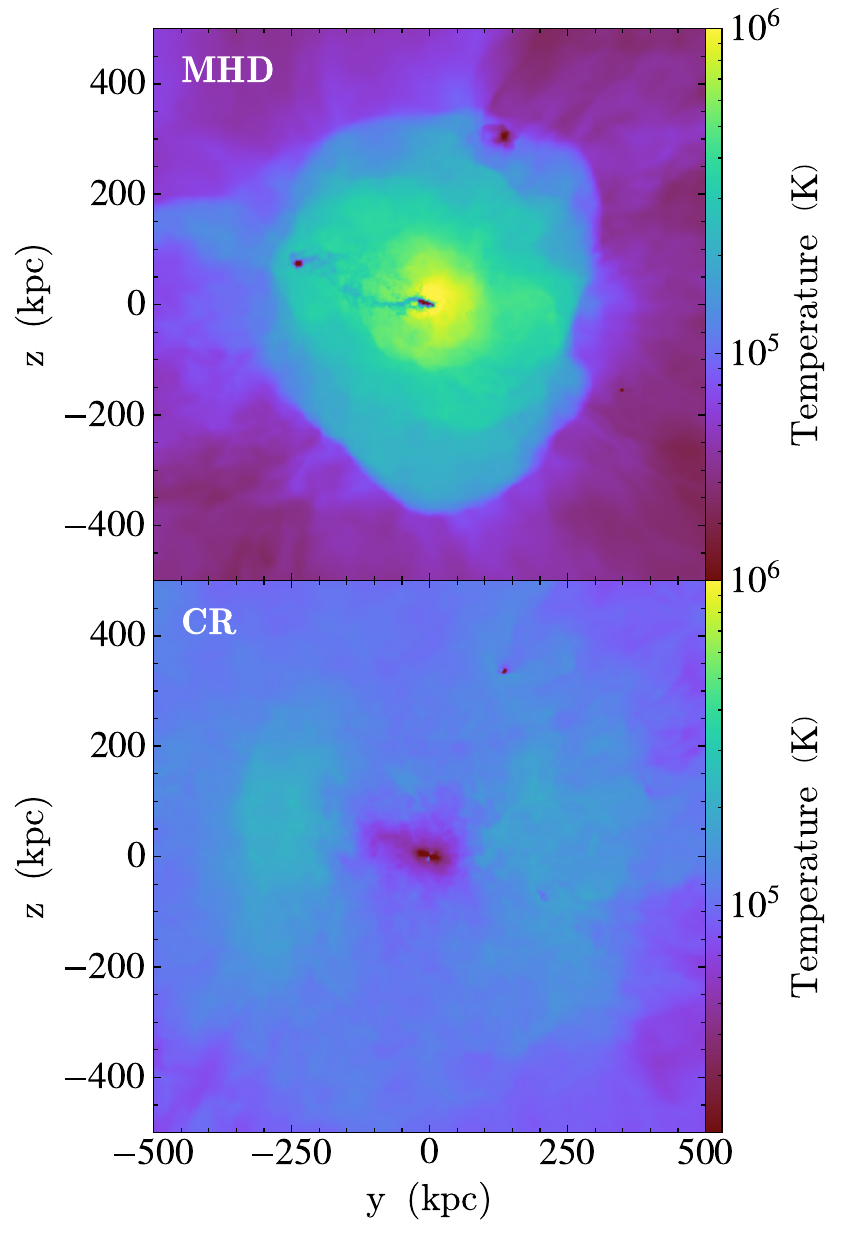}
  \caption{Projections of the gas temperature in MW-mass galaxy halos without
  (left) and with (right) cosmic rays. CGM is at the virial temperature of $T\sim
  10^{5-6}\,\mathrm{K}$ in the case without cosmic rays, with clear boundaries
  between the hot and cool phases around the virial radius $\sim
  250\,\mathrm{kpc}$ which traces the location of the virial shock fronts. In
  contrast, with cosmic rays, the galaxy halo is filled with much cooler
  ($T\sim$ a few $10^4\,\mathrm{K}$) CGM without the signature of the virial
  shock.  
  Figures are adapted from Fig. 1 in \cite{ji2021virial}.}
  \label{fig:cosmic_ray}
  \end{figure}

Recent theoretical studies suggest that with reasonable CR injection rates and
transport coefficients, the CR energy density in the CGM can be comparable, or
even significantly exceed, the thermal pressure in the CGM (e.g.,
\cite{salem2016role,farber2018impact,butsky2018role,ji2020properties,buck2020effects,ji2021virial}).
Ji et al. 2020 (\cite{ji2020properties}) found that the CR pressure in the CGM can be
one order of magnitude larger than the thermal pressure in the CGM, leading to a
CR pressure-dominated galaxy halo where the halo gas is primarily supported by
CR pressure rather than gas thermal pressure, and the temperature of the halo
gas is much lower than the virial temperature of $\sim
10^{5\text{--}6}\,\mathrm{K}$, as shown in Fig.~\ref{fig:cosmic_ray} from the
FIRE-2 simulations\footnote{The Feedback in Realistic Environments (FIRE)
Collaboration:
\href{http://fire.northwestern.edu}{\url{http://fire.northwestern.edu}}}. In the
meanwhile, virial shocks expected in massive ($M_\mathrm{halo}\gtrsim
10^{11.5}M_\odot$) galaxy halos are also absent from the CR pressure-dominated
CGM. Although this scenario is roughly consistent with the observed CGM
properties via quasar absorption lines such as H{\scriptsize~I} and
O{\scriptsize~VI} column densities \cite{Werk14, Prochaska2017}, two-dimensional
CGM emission maps which are expected from future \HUBS observations can provide a
more direct test of the CR pressure-dominated CGM. In particular, the
morphologies and intensities of the soft X-ray emission from the CGM can be used
to distinguish between the CR pressure-dominated and the thermal
pressure-dominated CGM. In addition, the \HUBS kinematics resolution
 can reach up to $1000\,\mathrm{km/s}$ in absorption and $300\,\mathrm{km/s}$ in emission, both of which are sufficient to probe
the structures of virial shocks in the CGM. 

\subsection{Potential case studies on feedback physics tailored for \HUBS}
\label{subsec:potential_case_studies}

\begin{figure}[H]
 \centering
 \includegraphics[width=0.47\textwidth]{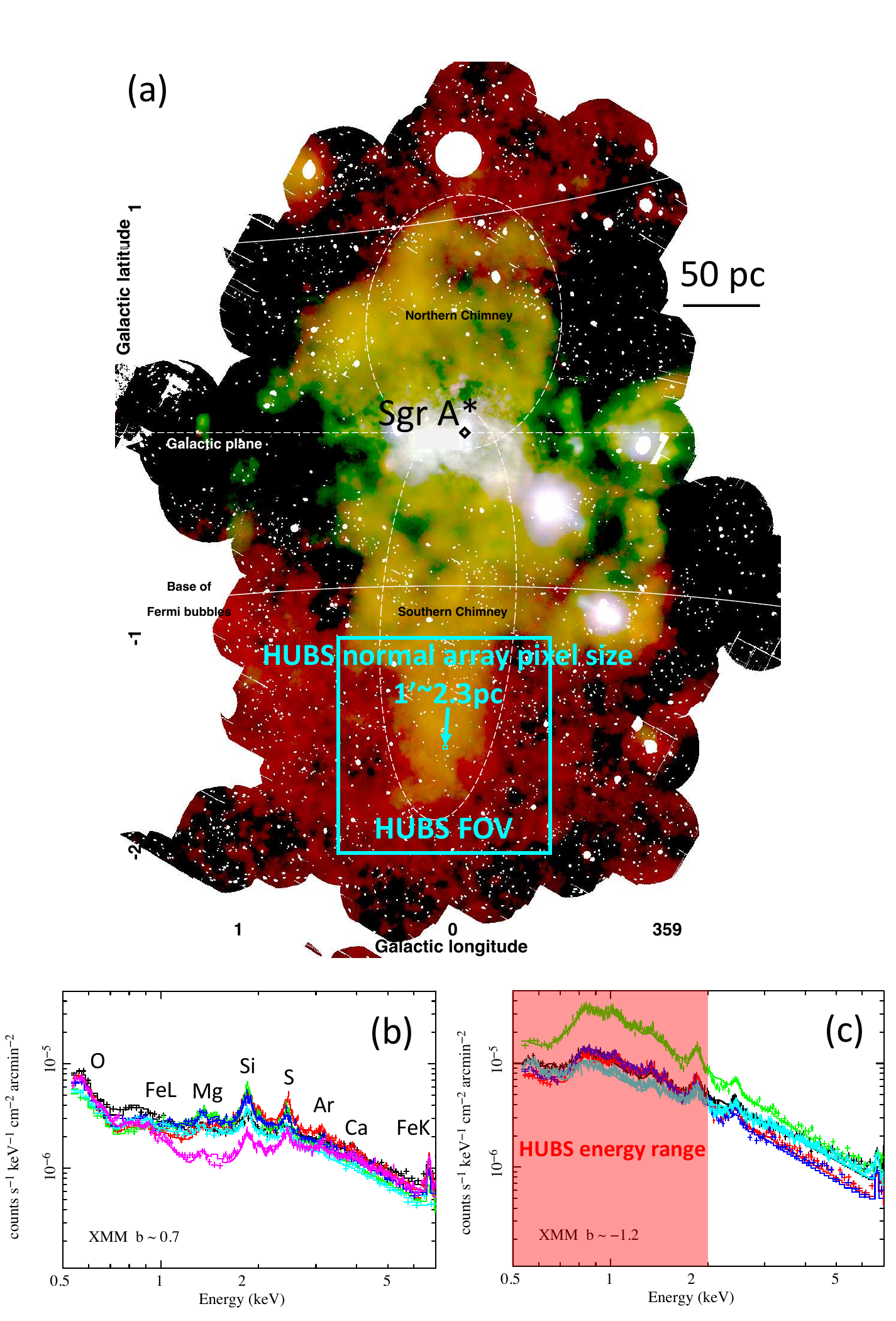}
 \caption{\XMMNewton observations of the ``chimney'' in the Galactic Center area \cite{Ponti19}. (a) Tri-color X-ray images of the central $300\times500\rm~pc$ of the Milky Way: red: 1.5-2.6~keV; green: 2.35-2.56 keV covering the S{\scriptsize~XV} lines; blue: continuum emission in the 2.7-2.97~keV band. The diamond at the center marks Sgr~A$^\star$. The two dashed eclipses are the two chimneys indicating collimated bipolar outflows. The large cyan box shows the FOV of \HUBS ($1^\circ\times1^\circ$), while the tiny box marked with the arrow is the pixel size of the \HUBS normal array ($1^\prime\times1^\prime$). At the distance to the Galactic center ($\sim8\rm~kpc$), a $1^\prime$ \HUBS pixel corresponds to $\sim2.3\rm~pc$. (b) and (c) are the \XMMNewton spectra extracted from a few regions at a Galactic latitude of $b\sim+0.7^\circ$ (b) and $b\sim-1.2^\circ$ (c), respectively. At $|b|\gtrsim1^\circ$, the foreground extinction is moderate, and many soft X-ray emission lines can be detected [emission line bumps of different elements are marked in (b)]. \HUBS is sensitive to the soft X-ray line features at $\lesssim2\rm~keV$ [shaded area in (c)]. }
 \label{fig:Ponti19GC}
\end{figure}

In order to probe the feedback physics mentioned above, we herein propose a few well-studied objects for some possible \HUBS follow-up observations, which may lead to breakthrough scientific output in our understanding of stellar and AGN feedback.

Sagittarius~A$^\star$ (Sgr~A$^\star$) located at the center of the Milky Way (MW) is the nearest supermassive black hole (SMBH), so provides us with a unique opportunity to witness the details of AGN or stellar (if more active star formation exists in the past) feedback close to its launching site. There are increasing lines of evidence that outflows of energy and metal-enriched materials from the central tens of parsecs of galaxies have shaped the observed structures on a variety of larger scales \cite{Ponti19,Wang21,Yang22}. Fig.~\ref{fig:Ponti19GC} shows the \XMMNewton observations of the Galactic center area \cite{Ponti19}. The two ``chimneys'' suggest collimated bi-conical outflows, which further connect to the larger-scale coherent structures such as the ``Fermi bubbles'' in $\gamma$-ray \cite{Su10}, the ``\eROSITA bubbles'' in X-ray \cite{Predehl20}, or the ``WMAP Haze'' in microwave \cite{Finkbeiner04,Planck13a}, with typical sizes roughly on the order of the galaxy itself (more than one order larger than the ``chimneys''). Existing X-ray observations of the Galactic center area already show interesting fine structures highlighted in the emission from special ions (e.g., the bipolar ``chimneys'' revealed in the S{\small~XV} emission in Fig.~\ref{fig:Ponti19GC}a), many of them also have multi-wavelength coherent structures or counterparts (also see \cite{Wang21}). However, the energy resolution of the X-ray CCD spectrum is insufficient to separate individual emission lines (Fig.~\ref{fig:Ponti19GC}b,c), which limits the constraint of the physical and chemical properties of the outflows. \HUBS will for the first time resolve fine spectral structures of the hot gas in a large area above the Galactic plane close to Sgr~A$^\star$. Since the foreground extinction is very strong toward the Galactic center direction and \HUBS is only sensitive at $\lesssim2\rm~keV$ (Fig.~\ref{fig:Ponti19GC}b,c), the future \HUBS survey will most likely focus on the area with the Galactic latitude $|b|\gtrsim1^\circ$ (e.g., the cyan box shown in Fig.~\ref{fig:Ponti19GC}a). We can either map the ``chimneys'' area shown in Fig.~\ref{fig:Ponti19GC}a or a larger sky area covering a significant fraction of the ``\eROSITA bubbles'', depending on the desired depth and the available observing time, or the required ``effective angular resolution''.

\begin{figure}[H]
\centering
\includegraphics[width=0.35\textwidth]{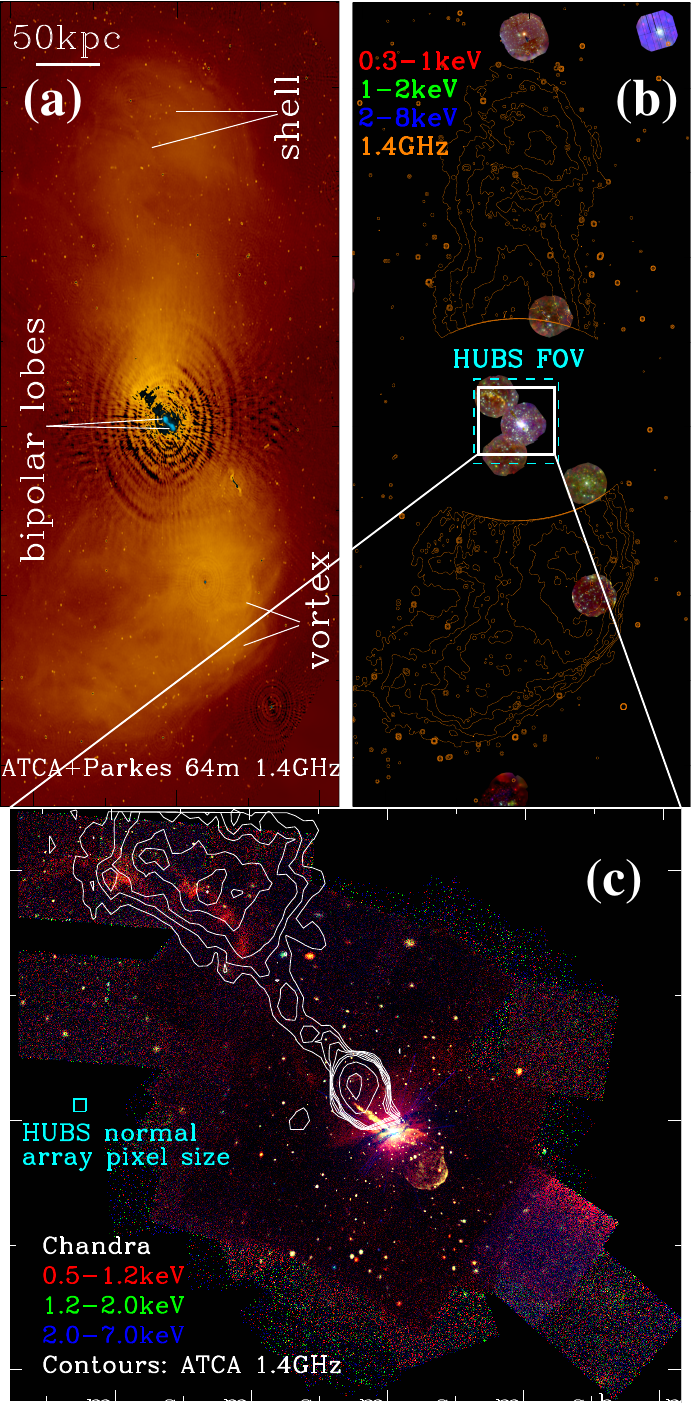}
\caption{(a) ATCA+Parkes 64m 1.4GHz radio continuum image of the $\gtrsim500\rm~kpc$ scale giant radio lobes in Cen~A \cite{Feain11}. Some prominent radio continuum features identified by \cite{Feain11} are marked on the image (``shell'' and ``vortex''). The contrast of the image may not be high enough to clearly show these features as the dynamic range of the image is very large. The $\sim10\rm~kpc$ scale bipolar radio lobes are plotted in blue in the very nuclear region. We also plot a scale bar of $50\rm~kpc$ assuming a distance to Cen~A of $d\approx3.8\rm~Mpc$. (b) \XMMNewton tricolor images of Cen~A and the surrounding area with the same FOV as (a). The orange contours are the 1.4~GHz image in (a) with the central region masked. The cyan dashed box in the center is the \HUBS FOV, while the white solid box is the FOV of the \Chandra image in (c). (c) \Chandra tricolor images of Cen~A and the Northern Middle Radio Lobe. The contours are the high-resolution ATCA 1.4GHz radio continuum image from \cite{Neff15} (only covering the northern half). The small cyan box is the pixel size of the $60\times60$ \HUBS normal array.}
\label{fig:CenA}
\end{figure}

Centaurus~A (Cen~A; NGC~5128) is the nearest FR-I radio galaxy located at a distance of $d\approx3.8\rm~Mpc$ ($1^\prime\sim1.1\rm~kpc$; \cite{Harris10}). It is the central galaxy of one of the two subgroups comprising the Cen~A/M83 group, and the 4th nearest galaxy group only after the local group, IC342, and the M81 group. Cen~A is the 5th brightest external galaxy in optical, after LMC, SMC, M31, and M33; and 2nd brightest extragalactic radio source, only after Cygnus~A. Fig.~\ref{fig:CenA} shows the multi-scale radio and X-ray structures of Cen~A, indicating complex interactions between the AGN jet and the galaxy environment. The small distance and plenty of multi-scale structures related to the AGN-ISM/CGM/IGM interaction make Cen~A unique for detailed analysis of AGN feedback. Combined with the study of the Galactic center region as described above, the proposed \HUBS observations can be used to study the feedback processes over more than five orders of physical scales (from $\lesssim10\rm~pc$ to $\sim1\rm~Mpc$). The large FOV of \HUBS makes it ideal to observe a large object such as Cen~A. We will need $\sim30$ \HUBS observations to cover the entire area of interest surrounding Cen~A, which is much more efficient than the \Chandra or \XMMNewton (Fig.~\ref{fig:CenA}b). The angular resolution of \HUBS is still sufficient to resolve some fine structures such as the chain of knots in the northern jet (Fig.~\ref{fig:CenA}c). The energy resolution ($E/\Delta E\approx500$ $@$1~keV for the $60\times60$ normal array; $E/\Delta E\approx1000$ $@$0.6~keV for the $12\times12$ central sub-array) is sufficient to measure the physical and chemical properties of the hot gas, and is also typically marginally sufficient to measure the shift or broadening of the soft X-ray emission lines from a normal galactic outflow, especially for those from the gaseous medium strongly turbulated by the AGN (e.g., the Perseus cluster as observed by \textit{Hitomi} \cite{Hitomi16}; the \HUBS central sub-array has an energy resolution comparable to \textit{Hitomi} at the Fe~K lines).

As one of the nearest nuclear starburst galaxies with an edge-on orientation, M82 provides us with the best view of the multi-phase galactic superwind driven by starburst feedback \cite{Strickland04,LiJ13a}. Existing high-resolution X-ray grating spectroscopy observations of the halo of M82 indicate complicated emission line spectra, with contributions from both the thermal plasma and some non-thermal components such as the charge exchange (CX; see the \XMMNewton/RGS spectra from \cite{ZhangS14}). However, such grating observations are limited to relatively compact objects. In most of the nearby galaxies like M82, we still rely on X-ray CCD imaging spectroscopy observations, with which the decomposition of different emission components, so the measurement of the physical and chemical properties of the hot gas, can be very uncertain (e.g., \cite{LiJ13a,Lopez20}). Although located in the M81 group which has giant tidal tails detected in colder gas \cite{Yun94}, most of the interesting features related to the galactic superwind (such as the northern ``cap'' $\sim11\rm~kpc$ above the galactic plane; e.g., \cite{Hoopes05}) could be covered with a single \HUBS observation. At a distance of $d\approx3.53\rm~Mpc$, the $1^\prime$ \HUBS normal pixel corresponds to $\sim1\rm~kpc$, which is still helpful to perform some spatially resolved analysis (e.g., \cite{Lopez20}), although the fine structures of the superwind cannot be resolved. The $15^{\prime\prime}$ pixel size of the central sub-array could better sample the PSF, but cannot significantly increase the angular resolution. The velocity of the hot wind constrained in different ways should be $>10^3\rm~km~s^{-1}$ (e.g., \cite{Strickland09,Melioli13}), significantly exceeds the outflow velocity of the cold gas (typically $\sim500\rm~km~s^{-1}$; e.g., \cite{ZhangS14}). Measuring the velocity difference of different gas phases is not only important in measuring the energy content of the outflow, but also in quantitatively estimating the CX contribution. Furthermore, we can also use the high-resolution spectra taken with the micro-calorimeter on board \HUBS to better constrain the gradients of the intrinsic absorption column density, temperature, and metallicity, as well as some other derived parameters (electron number density, thermal pressure, cooling timescale, etc.), of the hot gas outflow \cite{Lopez20}. These measurements can be compared to numerical simulations to determine the thermalization efficiency of supernovae (SNe) energy and the mass loading factor of the cool gas, which are key parameters of stellar feedback models (e.g., \cite{Strickland09,Lopez20}).

\section{Galaxy cluster and large-scale structure}
\label{sec:BaryonsOnLargeScale}

Quantifying the cosmic baryon budget, its multi-phase status, as well as its cooling and accretion activities, over a variety of physical scales from galaxies to groups/clusters or even the cosmic webs (e.g., \cite{Eckert15,Tanimura20,Tanimura22}) can help us to understand various hidden drives of galaxy growth in their larger-scale environment \cite{Astro2020, 2022MNRAS.509.3148W}. The baryonic matter abundance and distribution inside the cluster and group halos (e.g., \cite{2004ApJ...617..879L, 2009ApJ...703..982G, 2012ApJ...744..159L, 2013ApJ...778...14G, 2013A&A...551A..23E}), as well as in the CGM of galaxies (e.g., \cite{2008AJ....135..922K, 2012ApJ...760L...7K, 2013ApJ...776..115N, 2014ApJ...796..140P, Werk14, Prochaska2017}) have been extensively studied through multi-wavelength observations. On larger scales, a good fraction of the cosmological baryons at $z>1$ have been detected inside the cosmic web with their abundance measured mainly through Lyman-$\alpha$ observations (e.g., \cite{1997ApJ...489....7R, 2014ApJ...795L..12L}). 
At lower redshifts, however, baryons inside the cosmic web are much more difficult to be probed due to the very low gas column density. 

Many efforts have been made to search for cosmic baryons at such redshifts and it has been shown that approximately half of the total baryon budget at these epochs are locked up in the CGM of galaxies, the intragroup medium (IGrM), the ICM of clusters and the IGM in neutral and diffuse phases (\cite{1998ApJ...503..518F}). The other half remain ``missing'' observationally and cosmological simulations have shown that they are locked up in a warm-hot ($10^5\,\mathrm{K} < T < 10^7,\mathrm{K}$) phase in the IGM -- referred to as the WHIM, as a result of significant heat-up and removal by star forming and feedback processes (e.g., \cite{2005Natur.433..495N, 2005ApJ...620...21K, 2006ApJ...650..560C, 2007MNRAS.374..427D, 2008Sci...319...55N, 2010MNRAS.408.2051D, 2016MNRAS.457.3024H, 2019MNRAS.486.3766M}). Determining the ``missing'' baryon budget is expected to be most promising through the next generation X-ray spectroscopy (e.g., \cite{2001ApJ...554L...9P, 2003PASJ...55..879Y, 2013arXiv1306.2324K, 2018Natur.558..406N}). In this regard, \HUBS will play a breakthrough role in probing the hot baryons in their multi-phase states on a variety of physical scales in the Universe. We herein present a few \HUBS core science projects which could potentially greatly advance our understanding of the hot baryon budget of the local Universe. 

\subsection{Multi-phase hot gas in galaxy groups and clusters}
\label{subsec:Multi-phaseHotGas}

The hot gas inside the dark matter halo of galaxies, groups and clusters, referred to as the CGM, IGrM, and ICM, respectively, is an optically thin, collisionally ionized, multi-phase medium. The physical states and their spatial distribution are modulated by both external and internal factors (e.g., \cite{2022MNRAS.509.3148W}). On larger scales, pristine gas is accreted from the connected cosmic web to the halo, together with relatively low metallicity gas stripped from the infalling satellite galaxy halos. On smaller scales, AGN and stellar feedback eject material, metal, energy/heat and momentum back to the halo environment, some of them can reach about several tens or even hundreds of kiloparsecs, where this hot polluted ejection meets the cold accretion from the outside and falls back when it sufficiently cools down. Overall the halo gas is experiencing gravitational accretion and heating, impact compression, collisional ionization and excitation, radiative cooling, etc.. Internally driven processes generally cause gas to move in the outward direction, although the angular distribution of inflows and outflows may be different and complex. Different spatial and time scales of the various processes involved therefore naturally lead to the multi-phased nature of this hot halo plasma. Observationally resolving the spatial structures in temperature, density, metallicity, and ionization state will be crucial to probe the contributions and strengths of individual processes that together modulate the hot halo gas across hundreds of kiloparsecs. 

Regarding such a multi-phase IGrM or ICM gas, a cooler component is often detected in the central few tens of parsecs of relaxed (or nearly relaxed) groups and clusters, which accounts for up to several tens of percent of the total X-ray luminosity in $0.5-2$ keV, after the projection effect is corrected (see \cite{2001PASJ...53..401M} for a review). In the X-ray imaging spectroscopic analysis of the \Chandra (\cite{Vikhlinin09}), \XMMNewton (\cite{Pierre16}) and \textit{Suzaku} (\cite{2007PASJ...59S..23K}) data, this cooler component is routinely modeled either as a single-phase, with a temperature decreasing inward monotonously, or as a cool spectral component co-existing with a hot component (i.e., the ICM defined in the ordinary sense). In this latter case, corresponding to a two-phase scenario, a relative ``volume filling factor'' that varies with radius is introduced to characterize the spatial distribution of the cooler gas. In these two scenarios, the formation and evolution of the cold gas are believed to be intrinsically different, which unfortunately cannot be distinguished by current data. In some cases (e.g., Abell 1795, \cite{2012ApJ...749..186G}) a third weak gas component with an even lower temperature is necessary to improve the spectral fitting. The cooler components in the Virgo cluster and Abell 1795 are found to be more metal-enriched than their hotter counterparts. However, in many other cases, the temperature range and metal abundances of the cooler component are poorly constrained, and are often fixed to certain values in the multi-component spectral fitting. Since the amount of gas cannot be well determined, a certain form of emission measure distribution as a function of temperature (e.g., a power-law form) has to be imposed. These uncertainties may have a considerable impact on the accuracy of measurements of metallicity (e.g., the so-called ``Fe-bias'', see \cite{2000MNRAS.311..176B}) and other gas properties. For example, \cite{2012ApJ...749..186G} reported that a systematic bias up to about $\sim 10\%$ can arise for the dynamical mass of the central region, which approximately equals the typical deviation between masses measured with the X-ray and the gravitational lensing techniques. Mounting evidence shows that the coexistence of cold and hot phases cannot be interpreted simply in terms of the hot bubble(s) inflated by the central AGN in the circumstance of a cooler environment. Although it has been proposed that the cold-phase gas may be the cD coronal gas confined by magnetic loops surrounded by the intruding hot ICM, or simply it may be a consequence of radiative energy loss in part of the ICM (\cite{2006PhR...427....1P}), direct observational evidence is still absent. 

\begin{figure}[H]
\centering
\includegraphics[width=1\columnwidth]{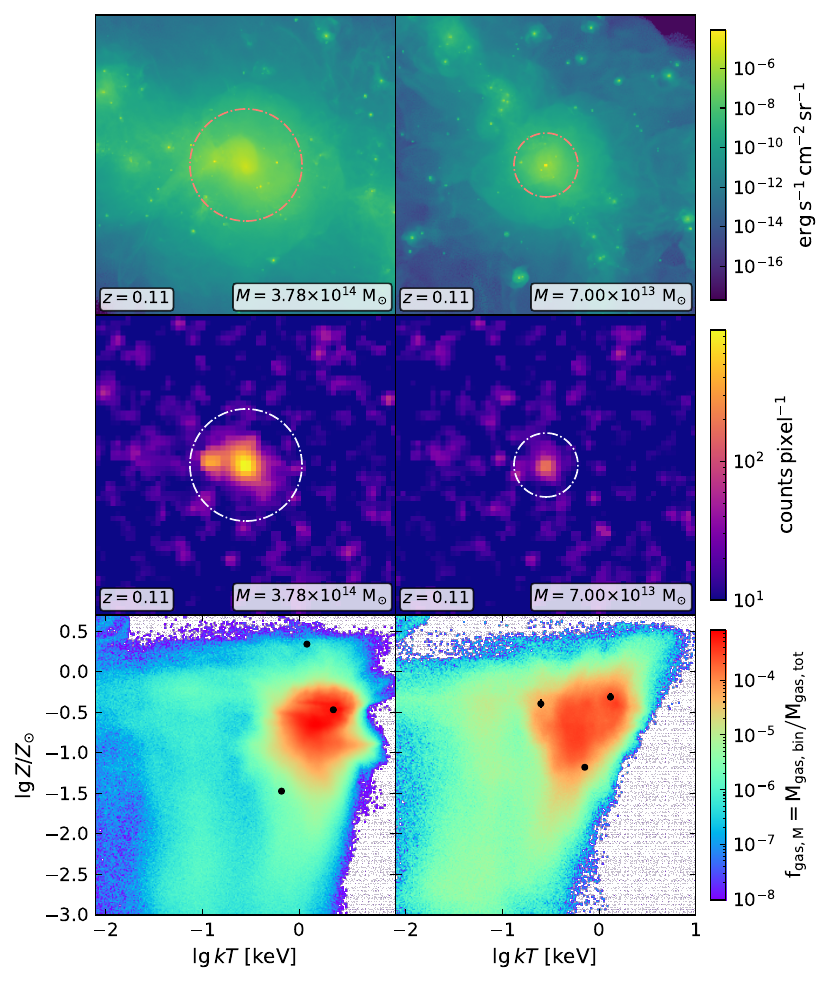}
\caption{Top panel: X-ray emissivity maps of the hot gas in a cluster-sized halo (left) and a group-sized halo (right) selected at $z = 0.11$ from the IllustrisTNG simulation. The images are produced with $256 \times 256$ pixels for each, which covers a sky region of 1 square degree (the field of view of \HUBS). The dash-dotted circles represent the virial radius ($r_{200}$). Middle panel: the mock \HUBS O {\sc vii} intensity maps for the two gas halos (i.e., the narrow-band images extracted at the position of (redshifted) O {\sc vii} emission line) with $60 \times 60$ pixels in 1 square degree (1 arcmin/pixel, i.e., the resolution of \HUBS). Bottom panel: the metallicity-temperature ($Z-kT$) phase diagrams of the two gas halos, where the black points represent the bestfits of the three-APEC spectral model. See \cite{Zhang22} for more details.} 
\label{fig:TNGMockGChalos2}
\end{figure}

The large FOV and low instrumental background of \HUBS make it best optimized to detect large-scale low surface brightness features, such as the extended multi-phase medium in cluster halos or even the cosmic web \cite{2020SPIE11444E..2SC}. To detect the extragalactic hot gas emission with low-surface brightness, the foreground emission due to the MW hot CGM should be securely removed.
With the high energy resolution of HUBS, a lower limit of redshift ensures the separation between the emission lines from the MW and targeted features (Fig.~\ref{fig:CGMMASSSimuHUBSSpec}; also see \cite{LiJ20, Zhang22}). This will strongly help to remove the sky background, enabling the detection of extremely low surface brightness features. As demonstrated in the recent work of \cite{Zhang22}, which is implemented based on the IllustrisTNG simulation (\cite{TNG2018Marinacci, TNG2018Naiman, TNG2018VS, TNG2018Pillepich, TNG2018Nelson}), \HUBS is capable of detecting the soft X-ray emission of the IGrM in group halos out to $z=0.3$, or that of the ICM in cluster halos located at slightly higher redshifts, when operating in either imaging or spectral mode for 1 Ms. Fig.\,\ref{fig:TNGMockGChalos2} presents the X-ray emissivity maps (top) and the \HUBS-observed O {\sc vii} intensity maps (middle) of the hot gas in a cluster-sized halo (left column) and a group-sized halo (right column) simulated at $z = 0.11$. The metallicity-temperature ($Z-kT$) distributions of the gas particles in the two gas halos are plotted in the bottom panel, where two sets of black points are used to mark the bestfits of the adopted spectral model consisting of three APEC components. In particular, the mock images are made with the \HUBS field of view and spatial resolution, i.e., $60 \times 60$ pixels in 1 square degree. The results also show that, although it is possible to pick out the primary emission components by applying a simple spectral model (the three-APEC model in this case), more advanced tools designed for the analysis of high-resolution spectral data are needed to describe the gas properties more accurately.

\subsection{Searching for hot baryons in the cosmic web} \label{subsec:StackCosmicWeb}

Recent studies have shown that it is possible to identify the baryonic filamentary structures of the cosmic web through stacking Lyman-$\alpha$ emissions (\cite{2018MNRAS.475.3854G}), or the thermal Sunyaev-Zel'dovich signals (\cite{2019A&A...624A..48D, 2019MNRAS.483..223T}). An ongoing effort is, with the aid of cosmological hydrodynamic simulations, to find out the feasibility of detecting the X-ray emission of the hot baryons inside the filaments through the stacking technique \cite{2020A&A...643L...2T,2022A&A...667A.161T}. Due to the extremely low gas column density, direct observations are nearly impossible. However, this may be achieved with the help of optical tracers, because galaxies that live inside filaments can be employed as a natural indicator of the cosmic web location. Many galaxy surveys, such as SDSS (\cite{SDSSDR11-12}), GAMA (\cite{GAMA2011GroupCatalogue}), 2dFGRS (\cite{2001MNRAS.328.1039C}), WiggleZ (\cite{2010MNRAS.401.1429D}) as well as the Millennium Galaxy Catalogue Survey (\cite{2005MNRAS.360...81D}), have provided a variety of galaxy catalogs that together cover a good fraction of the full sky. Using the sky position and redshift information of these galaxies as inputs, edge-extracting software can readily reveal the large-scale structure of the cosmic web. With the optical tracers, we now have a great base for detecting the hot baryons hidden inside the filaments. However, the situation is still tricky, because a large spatial coverage and sufficient spatial resolutions in both transverse and sightline directions are necessary. \HUBS is exactly suited for this purpose. The filaments have typically widths in the range of several hundred kiloparsecs to a few megaparsecs, our spectral resolution will be able to resolve them at a distance of up to a few hundred megaparsecs. This capability may be important to enhance the signal strength with or without stacking. The large field of view of \HUBS can effectively cover a significant patch of the sky. With several tens of pointings, 
\HUBS can collect X-ray emissions within a 
sufficiently large sky patch to allow mapping of the cosmic web structure (see Figure 3 of \cite{Barsanti_GAMA_Web22} for a demonstration of the cosmic web structure within a slice of a 10-degree cone out to $z\sim 0.1$). As shown in \cite{Zhang22}, out to redshift $z=0.1$ \HUBS will be able to detect hot gas in galaxies, groups, and clusters in both imaging and spectral modes given suitable exposure times. With an energy resolution of 2 eV, corresponding to a redshift resolution of $\delta_z \sim 0.003$ at such distances, the strong O {\sc vii} and O {\sc viii} emission lines will essentially act as tomography tracers,  indicating hot gas distributions at different redshifts. Selected patches in these line intensity maps \emph{in both spatial and redshift dimensions} shall then be stacked according to the pre-identified cosmic web features as probed by optical tracers. The stacked signal will then be compared with signals derived from random stackings. Through such comparisons, not only will we perceive the existence of the hidden baryons inside the filaments, but also may we learn about the temperature and metallicity distributions of the hot ionized plasma inside the cosmic web.

\subsection{Cluster observations for cosmology} \label{subsec:cosmology}

\HUBS will provide unprecedented opportunities for cosmology study with galaxy clusters, based on its large FoV, low instrumental background and superior spectral resolution. Cosmological constraints will mainly come from the following two aspects in the context of \HUBS observations.

\noindent{\bf A. Constraints from cluster mass function}

Galaxy cluster and group (``cluster'' hereafter) population can be used as an important probe to constrain cosmological models and to investigate the properties of dark matter and dark energy\cite{Bohringer04,Vikhlinin09}, the latter of which dominates within $z \lesssim 0.5$ \cite{Frieman08}. In order to achieve cluster samples as complete as possible, wide-field surveys have been performed through the observations of the Sunyaev-Zel'dovich (SZ) effect, the weak gravitational lensing, and the X-ray emission of the ICM (e.g., \cite{Adami18,Liu22}). By obtaining cluster mass function, we can infer important parameters of the dark matter model, e.g., the matter density as $\Omega_{\rm M}$ and the amplitude of linear matter density fluctuations as $\sigma_8$ when a flat universe is assumed. Furthermore, by combining X-ray and Sunyaev-Zel'dovich effect observations the absolute distances of galaxy clusters can be calculated, which allows the measurement of the Hubble constant \cite{Frieman08}. 

The completeness of the detected cluster population, which is crucial to the constraints on cosmological parameters, is directly determined by the survey area and depth. Deep large surveys are time-consuming and expensive, thus the cluster population is comprehensively investigated only until $z\sim0.1-0.2$ as of today. In the meantime, deficiency still exists in the faint end with a lower mass cut of about $10^{14} M_{\odot}$, making it very difficult to obtain a complete cluster mass function. Moreover, limited by the current FoV and instrumental background of detectors, very few clusters have been observed out to their virial radii, especially the low-$z$ ones which are crucial to constrain dark energy models \cite{Ettori06}.

Several leading X-ray surveys were conducted in the last decade, among which the \ROSAT All-Sky Survey (RASS) \cite{Voges1999} is the first full-sky survey in soft X-rays. The studies of the few thousand RASS clusters, which are detected above the flux limit of $\sim10^{-12}$ erg s$^{-1}$ cm$^{-2}$ and are mostly high mass systems located at the low redshift space, have offered a fundamental basis to cluster cosmology (e.g., \cite{Mantz2010,Mantz15}). Recently, \eROSITA All-Sky Survey (eRASS), the successor of RASS, is the most promising project for cosmological constraints in X-rays \cite{Predehl21}. Upon completion\footnote{However, the eRASS completion is currently uncertain, as \eROSITA on board the German-Russian Spectrum-Roentgen-Gamma mission has been switched off since 26th February 2022, with only four of the eight planned all-sky survey passes finished \cite{eROSITAwebsite}. The resumption of the telescope's operation has not yet been determined.}, eRASS is expected to detect $\sim10^5$ clusters, most of which are bright sources, in eight all-sky survey scans \cite{Pillepich12}. The depth of eRASS (an average exposure time of 2.5 ks per field), however, is relatively shallow and will limit its application in cluster cosmology (e.g., measurement of the gas fraction within the virial radius). In fact, although it is estimated that the eRASS survey will provide constraints of $\Delta\Omega_{\rm M}=0.012$ and $\Delta\sigma_8=0.036$ with combined probes from cluster number counts and angular clustering \cite{Pillepich12}, $\Delta\Omega_{\rm M}\sim0.05$ and $\Delta\sigma_8\sim0.07$ has just been achieved based on the first results from the proof-of-concept mini-survey with cluster number counts only, i.e., \eROSITA Final Equatorial Depth Survey \cite{Chiu2023}; these are at similar levels of the constraints provided by the \Chandra and \ROSAT cluster archives \cite{Mantz15} or by the XXL survey of \XMMNewton \cite{Pierre16,Pacaud18}. Deeper investigations with a considerably wide field are desired in order to achieve a complete cluster sample out to $z \lesssim 0.5$, including those low-mass systems.

\HUBS, due to its large FoV and low instrumental background, has great potential to collect a complete sample of clusters extending to larger redshifts, by carrying out a deep survey covering an area of $\sim$15 deg$^2$ with an average exposure of 300 ks. This surveying field, hence being named \HUBS-DF (\HUBS deep field), will be sited in the Galaxy And Mass Assembly (GAMA) survey footprint, particularly the GAMA02 field \cite{Baldry18}, therein abundant multi-band survey data have been archived, which is very important to assist cluster identification and study.

\begin{figure}[H]
\centering
\includegraphics[scale=0.17]{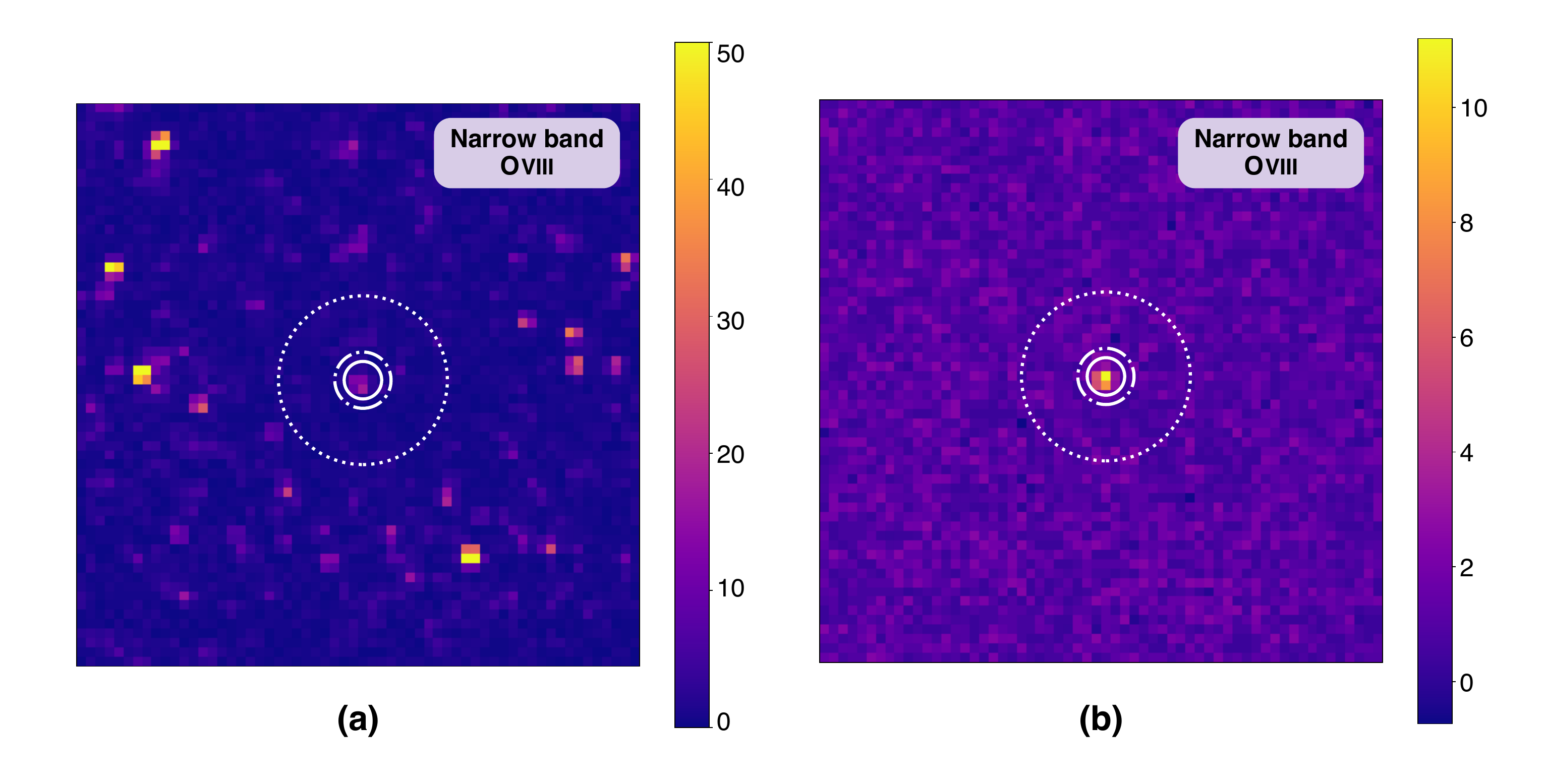}
\caption{Mock X-ray images of a cluster with $M_{500}=5\times10^{13} M_{\odot}$ at $z\sim0.5$. (a) is the 18 eV narrow band image around the redshifted O {\sc viii} emission line; (b) same as (a) with the point sources above $5\sigma$ of the background removed. See \cite{Zhang22} for details of the simulation. The dashed, dot-dashed, and dotted circles denote the radii $r_{500}$, $r_{200}$, and $3r_{200}$, respectively.} 
\label{fig:OVIII}
\end{figure}

With a low instrumental background and superior 2 eV energy resolution \cite{2020SPIE11444E..2SC}, \HUBS has the superb capability to resolve galaxy groups (or the faintest clusters) among crowded foreground/background AGNs and normal galaxies in narrow-band (vicinity of O {\sc vii} and O {\sc viii} lines) images based on a recently published work of the \HUBS team \cite{Zhang22}. By extrapolating this result, it can be easily seen that a cluster with $M_{500}=5\times10^{13} M_{\odot}$ at $z\sim0.5$ can be resolved in the O {\sc viii} line with an exposure of 300 ks (Figure \ref{fig:OVIII}). We consider this quasi-monochromatic imaging of \HUBS a novel method to identify faint and/or distant clusters, and believe that \HUBS has excellent feasibility to measure ICM in extended redshift range (see also\cite{Zhang22}). However, to achieve a complete cluster sample within $z \sim 0.5$, strong background source confusion of \HUBS with flux limit of $\sim5\times10^{-15}$ erg s$^{-1}$ cm$^{-2}$ must be removed. A drift scan or an assembly of 4500 stacked shallow observations (1 ks for each) with each pointing shifted by 3$^{\prime}$ is proposed, to improve the angular resolution of the central 11.3 deg$^2$ of \HUBS-DF to $15^{\prime\prime}$ with the core $12\times12$ detector array \cite{2020SPIE11444E..2SC}, so as to lower down source confusion limit by one magnitude in this area.

Since one of the most important issues is the synergies between \HUBS and multi-wavelength facilities for source identifications, the survey will be performed within the 25 deg$^2$ north portion of the XMM-XXL survey in GAMA02, which also overlaps with the VIPERS redshift survey \cite{Garilli14}. In the narrow-band image of \HUBS, background AGNs above $10^{-15}$ erg s$^{-1}$ cm$^{-2}$ (in 0.5-2.0 keV) can be efficiently removed according to the existing XMM-XXL source catalog, and meanwhile, the candidate clusters can be confirmed by utilizing exiting cluster catalogs (especially XXL cluster catalog \cite{Adami18}, the Atacama Cosmology Telescope (ACT) SZ cluster catalog \cite{Hilton21}, and Wen-Han (WH2022) cluster catalog \cite{Wen22}) and ancillary multi-frequency observations, such as the Sloan Digital Sky Survey (SDSS) \cite{Blanton17}, the Wide-field Infrared Survey Explorer (WISE) \cite{Wright2010}, the Galaxy Evolution Explorer (GALEX) \cite{Bianchi2017}, and the Canada-France-Hawaii Telescope Lensing Survey (CFHTLenS) \cite{Erben2013}. 

\begin{figure}[H]
\centering
\includegraphics[scale=0.35]{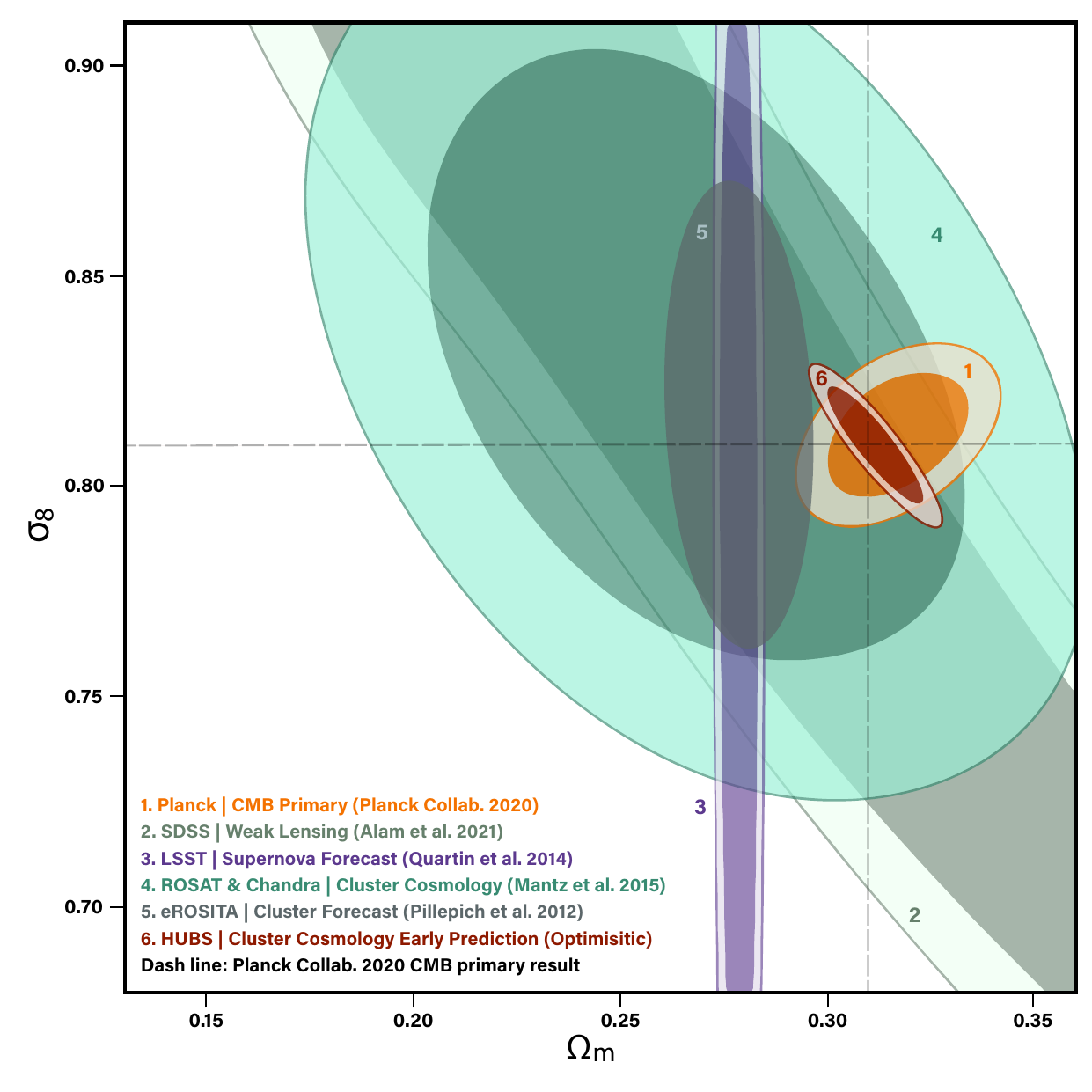}
\caption{Predicted \HUBS constraints on $\Omega_{\rm M}$ and $\sigma_8$, which is compared to the results derived from \textit{Planck} CMB study, SDSS weak lensing study, LSST supernova forecast, \ROSAT and \Chandra cluster survey, and \eROSITA cluster forecast (references in the picture). Note the last three constraints all used the cluster number counts method.} The \eROSITA cluster forecast marks the 1$\sigma$ distribution of the parameter space. For the rest, the darker and the lighter shades denote 1$\sigma$ and 2$\sigma$ level of the constrained parameter space, respectively. All the contours are smoothed for illustrative purposes. Our simulation input is based on the new Planck 2018 CMB primary results, hence the contours are centering at a different location.
\label{fig:constrain}
\end{figure}

We have quantitatively estimated the completeness of the expected cluster sample from the proposed \HUBS-DF survey, and its constraints on the cosmological parameters. Assuming an average CXB background derived from the \Chandra or \XMMNewton observations, the \HUBS 300 ks deep exposure will allow us to achieve the most complete X-ray sample of clusters within $z \lesssim 0.5$ with $M_{500} > 5\times10^{13} M_{\odot}$. Under such mass limit, HUBS detection will be 100\% complete within $z \lesssim 0.48$, 93.2\% complete at $z=0.5$, 30.5\% complete at $z=0.75$, and 10.6\% complete at $z=1$. Although HUBS will detect a much smaller total number of clusters than all-sky surveys, the expected detected number density will be one order of magnitude better than the eROSITA shallow survey. The depth of our survey guarantees that there will be $>2600$ photon counts for each target, which is crucial in measuring gas properties. Compared to the extremely low counts (50 counts per target used as detection limit in forecast \cite{Pillepich12}) of the \eROSITA detected clusters, the remarkably improved spectral quality will allow us to directly constrain the gas temperature in the spectral fittings, and estimate the cluster mass more accurately via the mass-temperature (\textit{M-T}) scaling relation. By applying the weak-lensing measurements and the excellent spectroscopic redshift measurements of the GAMA survey to, e.g.,  perform the mass calibration, we expect that our error budget can be notably reduced compared with that of the wide but shallow survey of eRASS. Finally, an early prediction of the cosmological constraints for $\Lambda$CDM models has been estimated using the cluster detection limit mentioned above and the Markov chain Monte Carlo (MCMC) exploration of the parameter space (\ref{fig:constrain}), which shows that \HUBS can greatly improve the constraints on cosmological model parameters as compared to the existing results of \ROSAT, \Chandra and the \eROSITA. 

\noindent{\bf B. Constraints from cluster gas fraction}

In contrast to the methods that use the abundance of galaxy clusters as a function of mass and redshift to constrain cosmological models, the method via calculating gas mass fraction ($f_{\rm gas}$) does not rely on the completeness of the cluster sample. This method focuses on the study of gas ratios and their dispersion, and the dependence of these quantities on the cluster mass, aperture and redshift. With the deduced gas mass fraction we can constrain $\Omega_{\rm m}$, using a combination of the Hubble parameter and cosmic baryon fraction as $h^{3/2} (\Omega_{b}/ \Omega{_m})$. The results obtained with this method, although to be improved, are competitive as well as consistent with those from recent CMB, Type Ia supernova and baryon acoustic oscillation data, and in the meantime to explain why $f_{\rm gas}$ is lower than expected in some low-temperature ($kT_{\rm 2500}< 5$ keV) systems.

The current relevant work from \Chandra \cite{Mantz14,Mantz22} is based on a morphological selection of relaxed clusters above 5 keV, with the study only confined within $r_{2500}$. However, at $r_{2500}$ the influence of various astrophysical processes, e.g. stellar winds from massive stars, AGN jets and supernova explosions, cannot be neglected. The study of \cite{Ettori06} indicated that even at $\sim r_{500}$ hot gas is still relatively significantly affected by various astrophysical processes, giving the dependence of the baryon fraction upon radiative cooling, star formation, feedback through galactic winds, conduction, and redshifts. As a result, extended investigation of $f_{\rm gas}$ until the virial radius (roughly $r_{200}$ where gravity dominates) is more cosmologically crucial. 

The large FoV of \HUBS brings a huge advantage to addressing this issue by allowing us to cover the entire virial region of a cluster with one pointing close as $z\sim0.02$. Moreover, its superior spectral resolution provides unique means to identify ICM bulk motions with radial velocities possibly from 90 to 6000 km s$^{-1}$, to crosscheck whether the cluster is truly relaxed. We expect to observe $\sim30-50$ relaxed brightest galaxy clusters to $r_{200}$ with $\sim 50$ ks single pointing for each. The selection should satisfy two criteria. Firstly, high-quality multi-band data are available for the target, for synergetic investigations to remove possible background pollution. Secondly, once weak lensing data is available, the total gravitating mass obtained in the X-ray will be calibrated, in order to guarantee high accuracy in the gas fraction calculation.

When the gas mass fractions of low-$z$ clusters are incorporated with baryon fraction measurements via the CMB, or with priors on the cosmic baryon density and the Hubble constant, the Hubble constant or the dark energy density as well as the equation of state can be deduced. Thus galaxy clusters in the low-$z$ region can not only help provide a more accurate sample for our measurements, but also have more important cosmological significance for they are in the dark-energy-dominated window of cosmic history. Finally, if the gas mass fraction of clusters is indeed constant, it can be used as a `standard ruler' to measure the space-time geometry of the Universe with the Hubble parameters determined by other measurements.

\section{Galactic Science} \label{sec:Galactic-Sciences}

Close to home, the Milky Way provides the nearest targets for detailed study of the ISM, energetic explosions, stars, compact objects, and so on. Hot gas is thought to permeate the ISM. It is most certainly related to supernovae, stellar wind, and the central supermassive black hole, and thus offers an excellent laboratory for studying the physics of feedback processes. There are a number of unresolved issues, including the origin of the soft X-ray background radiation, the origin of the \eROSITA Bubble (which might be related to the Fermi Bubble), and the properties of the hot halo, which likely bear relevance to the physics of CGM and feedback processes.

\subsection{The Cosmic X-ray background} \label{subsec:cosmic_xray}
The cosmic X-ray background (CXB) is one of the first discoveries of X-ray astronomy, along with the first extrasolar X-ray source Scorpius X-1 \cite{Giacconi:1962aa}. 
Later the flux level of the soft X-ray band ($44-70~$\AA) was successfully measured by \cite{Bowyer:1968aa} and \cite{Bunner:1969aa}, and interpreted as truly diffuse emission of hot plasma \cite{Weymann:1967aa, Henry:1968aa}. 
Our understanding of the soft X-ray background has progressed considerably in the ensuing more than 50 years, with generations of X-ray instruments.
Aside from the in-service all-purpose telescopes such as \Chandra X-ray observatory and \XMMNewton, as well as retired ones such as \ROSAT and \textit{Suzaku}, many kinds of space missions have been dedicated to probing the nature of the soft X-ray background. 
Recent missions include the space shuttle payload ``\textit{Diffuse X-ray Spectrometer}'' (\textit{DXS}, \cite{Sanders:2001aa}), dedicated explorer ``\textit{Cosmic Hot Interstellar Plasma Spectrometer}'' (\textit{CHIPS}, \cite{Hurwitz:2005aa}), sounding rocket mission ``\textit{Diffuse X-rays from the Local galaxy}'' (\textit{DXL}, \cite{Galeazzi:2011aa, Galeazzi:2014aa}), and the recent soft X-ray surveyor \textit{HaloSat} \cite{Kaaret2019}. 

In general, the CXB can be decomposed into two kinds of origins, galactic and extragalactic.
The galactic soft X-ray emission comes from three distinct components, the solar wind charge exchange (SWCX), the Local Hot Bubble (LHB), and the Galactic halo, while the extragalactic origin is dominated by AGNs.
Distinguishing these components from each other and quantifying their contribution to the soft CXB are limited by the spectral resolution of current space missions and model dependent.
For example, it is still an open question whether the observed soft X-ray emission at 1/4 keV is due to LHB or purely from SWCX (e.g., \cite{Lallement:2004aa, Koutroumpa:2009aa, Robertson:2009aa}). 

\HUBS can provide unprecedented line diagnostics to help to understand the origin of the CXB.
On the other hand, both in-service and past missions barely cover the $0.1-0.5$ keV band, which seems to be a turnover in the spectral energy distribution of cosmic UV/X-ray background (\cite{Haardt:2012aa, Khaire:2019aa, Faucher:2020aa}).
With \HUBS, we can obtain finer constraints on the UV/X-ray background modeling.

\subsubsection{Local hot bubble}
It has been identified that the solar system resides in a cavity of low-density and ionized gas, surrounded by a shell of cold neutral gas and dust. 
The existence of such a cavity was implied by the soft X-ray emission seen in the \ROSAT map at 1/4 keV \cite{Snowden:1997aa}, and was dubbed as the ``local hot bubble'' \cite{Sanders:1977aa, Cox:1987aa}.
Though debates on the model exist, the LHB still led its popularity and was strongly supported by recent studies \cite{Snowden:2015aa, Snowden:2015ab}.
Followup studies revealed its irregular shape and extent, suggesting a pathlength of order 100 pc \cite{Snowden:2015ab, Liu:2017aa, Zucker:2022aa}, possibly created and maintained by stellar winds or supernova explosions due to nearby star formation activities.

Despite the change in intensities, soft X-ray emission from the LHB can be characterized by a hot phase plasma with $k_{\rm B} T \sim 0.1$~keV \cite{Sanders:2001aa, Wulf:2019aa}.
However, possible contamination arises from foreground SWCX, and Galactic halo at intermediate and high latitudes.

X-ray shadowing method is invoked to divide the observed emission into the foreground and background components \cite{Snowden:2015aa, Uprety:2016aa}.
Based on the \textit{DXL} data, Liu et al. \cite{Liu:2017aa} removed the contribution by SWCX and reported a uniform temperature $k_{\rm B} T = 0.097 \pm 0.013$~keV, consistent with previous results. In addition, combining the
\textit{DXL} result and other measurements, Snowden et al. \cite{Snowden:2015ab} showed the total pressure in the LHB
is in pressure equilibrium with the local interstellar clouds, eliminating the long-standing pressure problem 
of the LHB \cite{Jenkins2009}.

\subsubsection{Galactic halo}
The hot gaseous halo was first predicted by \cite{Spitzer:1956aa}, while the first hint of the hot gaseous halo was observed by RASS \cite{Truemper:1982aa}. 
It was implied by the anti-correlation between the soft X-ray emission at 1/4 keV and the column densities of the neutral hydrogen (e.g. \cite{Snowden:1995aa}).
After \ROSAT, high spectral resolution spectra obtained by \textit{DXS} revealed that the soft X-ray emission was dominated by thermal emission of hot gas ($k_{\rm B} T \approx 0.1-0.2$ keV), which favored a galactic halo origin \cite{McCammon:2002aa}.
Furthermore, the launch of flagship telescopes \XMMNewton and \Chandra enabled high spatial resolution observations to decompose the diffuse contribution seen by \ROSAT into the extragalactic point sources (e.g., AGN; \cite{Hornschemeier:2000aa}) and truly diffuse emission (e.g., \cite{LiJ13a}).

In the past two decades, the understanding of the Galactic hot halo has been improved a lot by deep \XMMNewton and \Chandra observations in both emission and absorption. 
Particularly, the spatial distribution of the Milky Way hot halo has been established as the first-order approximation assuming the spherical symmetry (e.g. \cite{Henley:2012aa, Gupta:2012aa, Miller:2013aa}). 
Furthermore, the temperature distribution of the hot halo has been investigated, showing another extremely hot phase at $k_{\rm B} T \approx 0.7$ keV (e.g., \cite{Das:2019aa}). 
In the current decade, newly launched surveyors \eROSITA and \textit{HaloSat} also continuously provide new insights, such as the discovery of the soft X-ray bubbles on both sides of the Milky Way (i.e., \eROSITA bubbles; \cite{Predehl:2020aa}).

Although it has been astounding progress to understand the Galactic hot halo from decades ago, there are still fundamental open questions in the field. 
For instance, the metallicity of the hot halo is still controversial. 
On one hand, the continuum of the thermal emission due to the Galactic hot halo is hard to be decomposed from other contributors to the continuum (e.g., the CXB or the soft proton), limited by the relatively poor spectral resolution of the existing instruments. 
On the other hand, the SWCX contributes to the soft X-ray line emission, which is also blended with the emission of the Galactic hot halo. 
These difficulties make it a hard problem to determine the hot halo metallicity. 
The high resolution and spectral coverage down to 0.1 keV of \HUBS could bring new possibilities to determine the metallicity by clearly decomposing line emission and the continuum or determining the line ratios between forbidden and resonant lines.

Another intriguing question is about the potential extremely hot phase in the Galactic halo. 
Currently, detection of an extremely hot phase at $k_{\rm B} T \approx 0.7$ keV has been claimed in both emission and absorption. 
However, these pieces of evidence have limitations in different ways. 
The absorption line analyses rely on weak detection ($\approx 2-3\sigma$) of Ne {\sc ix} and Ne {\sc x} in two sight lines \cite{Das:2019aa, Das:2021aa}. 
The modeling of this absorption system requires both super high temperature and super solar neon abundance of [Ne/O] $\approx$ 0.7, which raises questions about its origin (e.g., in the Galactic disk or the Galactic halo). 
The emission evidence of the super hot component is mainly the unexpected enhanced feature at $0.8-0.9$ keV in the single-temperature hot halo model. 
However, as suggested in \cite{Wulf:2019aa}, the similar feature observed at low Galactic latitudes can be explained by the hot corona of M dwarf stars in the disk. 
Adopting the model in \cite{Wulf:2019aa}, M dwarf stars can contribute $\approx 2 - 4 \times 10^{-7}$ kpc cm$^{-3}$ at high latitudes, which can be 50 - 100\% of the claimed detection of extremely hot phase (e.g., \cite{Das:2019ab, Bluem:2022aa}). 
The high spectral resolution and relatively high spatial resolution of \HUBS could provide unique insights into the extreme hot phase by constraining line ratios (determining radiation mechanisms) and resolving possible M dwarfs in the field.
Therefore, although \HUBS is not a dedicated surveyor focusing on the diffuse emission of the Galactic hot halo, its unique combination of large FOV and high spectral resolution opens a special window to study the Galactic hot halo.

\subsubsection{Extragalacitc sources}
While the diffuse Galactic and the local emission dominate the CXB in the $0.5-1$ keV band, the majority of the X-ray background has been recognized as discrete extragalactic sources, mostly AGN and star-forming galaxies. 
According to the deepest observations by \Chandra and \XMMNewton, in the $0.5-2$ keV band, the resolved fraction of the extragalactic background reaches about $80-90$\% (e.g., \cite{Moretti:2003aa, Worsley:2005aa, Hickox:2006aa, Lehmer:2012aa, Xue:2012aa}).
As a consequence, insights into the extragalactic background may serve as a constraint on the integrated SMBH growth and the accretion physics of galaxies. 

However, there still remains unresolved CXB of unknown origin, for instance, about 10\% diffuse emission in the $1-2$ keV band \cite{Hickox:2006aa}. 
It may be from CGM of galaxies within their Virial radii \cite{Mineo:2012aa} or ``warm-hot'' intergalactic medium with the temperature of $10^{5-7}$ K (WHIM; \cite{Cen:1999aa}). 
On the other hand, with the current CCD energy resolution, some models for the components of CXB are oversimplified, for e.g., a single-T APEC model for describing the local thermal-like emission cannot interpret the emission excess of CXB below 0.5 keV \cite{Hickox:2006aa}. 
Consequently, the uncertainties of the AGN contribution are larger in the $1-2$ keV band, which is essential for disentangling the obscured and the unobscured AGNs (e.g., \cite{Treister:2009aa}).

\HUBS has a large FOV ($\sim 1$ deg$^2$), and therefore most observations will partly cover the region of CXB, with a remarkable effective area ($\sim$500 cm$^2$). 
The X-ray integral field units in \HUBS cover the $0.1-2$ keV band which is complementary to the bands of \Chandra or the future \textit{Athena}, and together can give better constraints on the composition of CXB. 
With the 2 eV high energy resolution, the local diffuse emission can be exclusively determined, and the obscured fraction of AGNs would be more precise. 
Taking that \HUBS is designed for observing CGM and WHIM (e.g., \cite{Zhang22} ), it will quantitatively predict their contributions in the CXB and fill the final gap of the unresolved CXB.

\subsubsection{Solar wind charge exchange}
SWCX is generated when the highly ionized solar wind ions interact with the neutral materials within the solar system, gaining an electron in a highly excited state which then decays emitting an X-ray or UV photon with the characteristic energy of the ion. 
It was first proposed to explain the cometary soft X-ray emission \cite{Cravens:1997aa}, and then identified as the source of the long-term enhancements observed in the \ROSAT All-Sky Survey (RASS; \cite{Snowden:1994aa, Snowden:1995aa}).

Based on the target neutrals, there are in general two kinds of SWCX, i.e., the geocoronal SWCX and the heliospheric SWCX. 
The former is due to the interaction between the compressed solar wind ions in the magnetosheath and the neutrals (mostly hydrogen) in the exosphere of the Earth.
Its strength and location depend strongly on the strength of the solar wind.
The latter, on the other hand, is due to the interaction between the free-flowing solar wind and the neutral ISM within the entire heliosphere (up to $\sim100$ AU).
Heliospheric SWCX shows direction dependence as a consequence of the structured solar wind and the neutral distribution in the heliosphere.

Due to its ubiquity, SWCX emission contaminates every X-ray observation of astrophysical objects.
In particular, the spectrum of SWCX contains rich lines, some of which are the same lines used for the diagnostics of astrophysical plasma. 
The inclusion of SWCX emission could significantly change the derived plasma temperature of the astrophysical object, and/or mimic a separate diffuse soft X-ray component. 

Despite the difficulties in separating SWCX emission from that of astrophysical plasma, different groups have developed models to calculate SWCX emission based on solar wind conditions, neutral distributions, and theoretical interaction cross-sections (e.g., \cite{Robertson2003,Koutroumpa2006,Koutroumpa2007,Bodewits2007}). However, there are, sometimes, large discrepancies between the model predictions and the observational results (e.g., \cite{Koutroumpa2009bb,Snowden09,Ringuette2021}). The largest uncertainties of these models are mainly due to the lack of detailed information about the solar wind abundance and ionization state, and the theoretical and experimental interaction cross-section.

Owing to its high spectral resolution, \HUBS will allow us to resolve most of the fine-structure lines, which suits well for learning SWCX.
In principle, line intensity ratios in triplets of the He-line ions (e.g., O {\sc vii}) from SWCX emission are different from those in astrophysical thermal emission (e.g., \cite{Lallement2009}), and
the \HUBS spectroscopy will help to distinguish and separate SWCX emission from the thermal components, e.g., from the LHB, the Galactic halo, and other distant components. 

\begin{figure}[H]
  \centering
  \includegraphics[width=0.8\columnwidth]{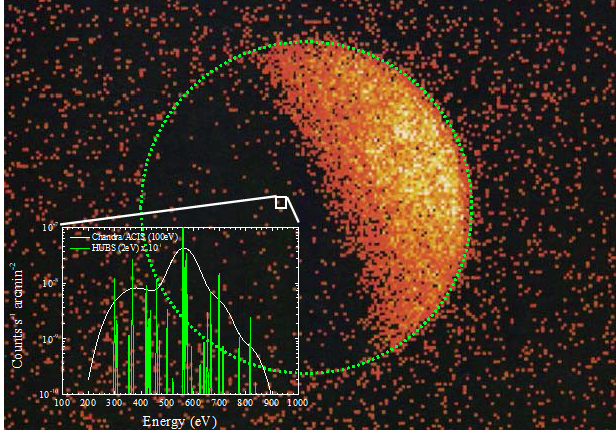}
  \caption{The \ROSAT PSPC observation for the Moon \cite[green dotted circle]{Schmitt1991}, inside panel shows the \HUBS simulation (2 eV, green curve) for an average slow solar wind ($v=$400~km/s) at one chip marked by open white box along with the comparison to \Chandra/ACIS simulation (100~eV, white curve). Here, the CX model uses the same exospheric hydrogen as in previous work \cite{Wargelin04}.}
  \label{fig:HUBS-moon}
\end{figure}

Due to the low-Earth orbit, \HUBS observation will be inevitably affected by the geocoronal SWCX. One strategy to study the SWCX in the near-Earth environment is through \HUBS observations of the Moon. \HUBS observation will cover the full Moon with its field of view and clearly resolve the SWCX lines with its superior energy resolution (see Fig.~\ref{fig:HUBS-moon}). On the bright side, strong fluorescence lines from O, Mg, Al, and Si can serve as a remote sensor of the element composition on the lunar surface. While the observation of the dark side of the Moon will maximize the SWCX signal by blocking the thermal emission from our galaxy and distant objects. 
The X-ray emission from the dark Moon mainly consists of two parts: the emission from the magnetosheath (the near-Earth environment $<10$ $R_{E}$) and from the region between the bow-shock ($\sim10$ $R_{E}$) and the Moon (60 $R_{E}$). For the near-Earth environment, the high variability of SWCX is complicated by the solar wind temporal variation. Real-time monitoring of the solar wind is necessary for accurate data analysis. In-situ measurements from ACE \footnote{https://solarsystem.nasa.gov/missions/ace/in-depth/} and/or the future Chinese space mission SMILE \footnote{http://english.cssar.cas.cn/smile/} will provide valuable data. Another important factor in the SWCX luminosity is the neutral distribution in the magnetosheath, which requires sophisticated magneto-hydrodynamic modeling for the solar wind interaction with Earth's atmosphere \cite{Sun2020}.
The high-resolution spectra obtained by \HUBS will precisely measure the SWCX contribution in the near-Earth environment and help to test the results of MHD models.
In addition, \HUBS data can be used to constrain the charge exchange cross-sections measured in the laboratory. 

\subsection{Supernova remnants} \label{subsec:snr}

Supernovae (SNe) are among the most violent explosions in the universe, which release a typical energy of $\sim 10^{51}$~erg in a rather short timescale. As an essential part of the galactic ecosystem, SNe play an important role in the baryon cycle and the energy feedback.
Supernova remnants (SNRs) are SNe interacting with the surrounding circumstellar material (CSM) and interstellar medium (ISM), which provide an important means to study the physics of both sides of the interaction.

SNRs are bright sources in the X-ray sky and the nearest targets to observationally constrain how SNe influences galactic ecosystems. Over a hundred X-ray-bright SNRs have been found thus far in our Galaxy, LMC, and SMC. These extended sources are actively heating the interstellar medium with fast shocks and enriching it with heavy elements. The X-ray observations in past decades have greatly
advanced our knowledge on SNRs \cite{2012A&ARv..20...49V}, but also post
some challenges that require X-ray observations with high spectral resolution.

Some crucial questions in SNRs are yet to be answered with future X-ray instruments with high spectroscopic capabilities: 1) What are the metal compositions in diverse SNRs and how do different supernovae contribute to producing heavy metals in our Universe? 2) How are the hot plasmas in non-equilibrium ionization produced? 3) How to constrain charge exchange and resonant scattering processes using emission lines?
Below we summarize why \HUBS will help us address these key questions.

\subsubsection{Associate SNRs with their progenitors}

One of the major challenges in the SNR study concerns the identification of the progenitor type. The two major types of SNe --- the core-collapse SNe and the Type Ia (thermonuclear) SNe --- can be well-defined and easily distinguished based on their optical spectrum around maximum light. However, it is not that straightforward to associate an evolved SNR with its original progenitor system, which needs a detailed investigation into the properties of the SN ejecta and the CSM.

Type Ia SNe represents the thermonuclear explosions of C/O white dwarfs. The nuclear burning in Type Ia SNe typically results in a large amount of iron-group elements (IGEs) such as Fe and Ni as well as intermediate-mass elements (IMEs) such as Si, S, Ar, and Ca \cite{2010ApJ...712..624M,2013MNRAS.429.1156S}. However, in the case of core-collapse SNe, one may expect oxygen as the major product of the nucleosynthesis \cite{2006NuPhA.777..424N,2016ApJ...821...38S}. Therefore, the SN ejecta metal abundances (or abundance ratios) can be used as diagnostics for typing their remnants \cite{2018A&A...615A.150Z}. SNRs showing evidence of enhanced oxygen abundances (so-called oxygen-rich SNRs) are commonly considered from the core-collapse explosions of the most massive stars, while SNRs dominated by IGEs and IMEs are more likely from Type Ia events. The X-ray spectra of SNRs contain most of the prominent emission lines from these metal species, which are essential for constraining the ejecta properties. However, the X-ray spectra of SNRs can always be a combination of the non-thermal emission from the accelerated particles and the thermal emission from both the shocked ejecta and CSM/ISM. Therefore, a precise measurement of the metal abundances relies on high-resolution X-ray spectroscopy that allows us to separate, identify, and measure the individual emission lines, and to distinguish the ejecta from other components. This can be challenging for the CCD instruments. For example, with a typical energy resolution $\Delta E\sim100$\,eV, CCD instruments can hardly resolve the Fe-L complex and the Ne He$\alpha$ lines around $\sim0.7$--$1.0$\,keV, which will be seen as a bump-like structure or a pseudo-continuum and lead to large uncertainties in the measured abundances. The current grating instruments such as \XMMNewton RGS and \Chandra LETG/HETG may partially solve this problem, but they are limited to those bright remnants with small angular sizes and the remnants with bright knot/filament structures. 

\begin{figure}[H]
    \centering
    \includegraphics[width=0.5\textwidth]{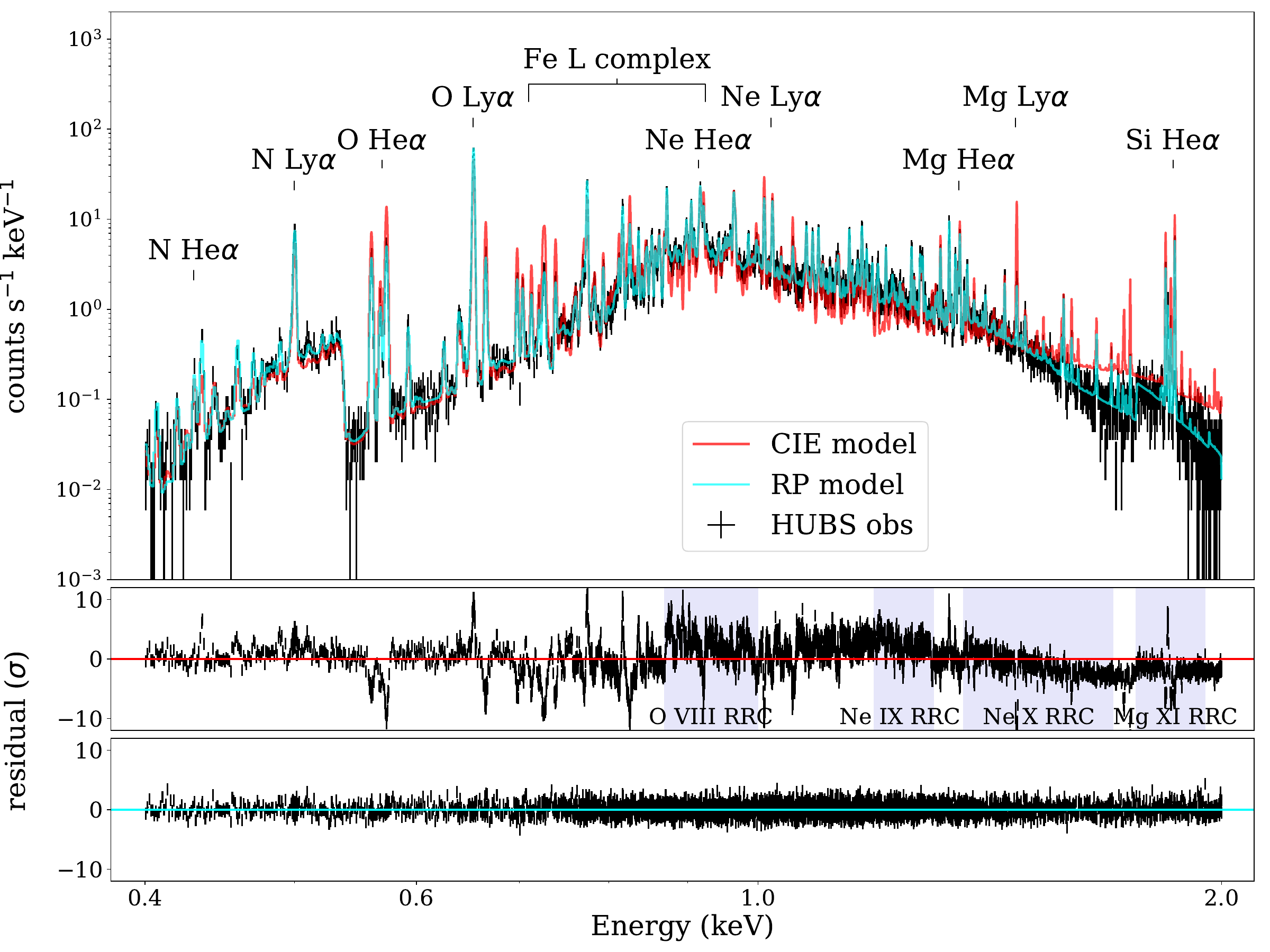}
    \caption{Simulated 100\,ks \HUBS spectrum of SNR 3C 400.2 (black data points), fitted with a collisional ionization equilibrium plasma model (CIE model, the red curve) and a recombining plasma model (RP model, the cyan curve), respectively. The lower panels show the residuals.}
    \label{fig:3C400}
\end{figure}

The constraint on the X-ray properties of the shocked plasmas in SNRs, and our understanding of the SN-SNR connection, will be greatly improved with the help of \HUBS. The \HUBS energy band ($0.1$--$2$\,keV) covers most of the He-like and H-like emission lines from C, N, O, Ne, Mg, Si, and L-shell emission from Fe and Ni. With an ultra-high energy resolution of $\sim2$\,eV of the main array and $\lesssim1$\,eV of the central sub-array, \HUBS is capable of resolving individual emission lines, especially the He-like triplets (i.e., the resonant lines, the forbidden lines, and the intercombination lines) and the Fe-L complex. 
Figure \ref{fig:3C400} shows the simulated 100\,ks \HUBS spectrum of a mix-morphology SNR 3C 400.2, which provides an illustration of its extraordinary capability on detecting and resolving diverse metal species under different ionization states in SNRs.
On the other hand, the spatial resolving ability and the large field of view may help to map the spatial distributions of the plasma parameters over the whole remnant.

\subsubsection{Constrain the origins of non-equilibrium ionization plasmas}

At the early phase of the SNR evolution, due to the low density of the shocked plasma, the ionization process may take a rather long timescale before reaching equilibrium ($n_{\rm e}t\sim10^{12}$\,cm$^{-3}$\,s). Therefore, the shocked plasma in young SNRs is expected to be in the non-equilibrium ionization (NEI) state, where the plasma is still under-ionized (ionizing plasma, IP), characterized by an ionization temperature $kT_{\rm i}$ which is lower than the electron temperature $kT_{\rm e}$. The observational evidence for this under-ionized NEI plasmas has been extensively established for a number of young SNRs such as Cas A, Kepler's SNR, SN 1006, SN 1987A, etc (e.g., \cite{1996A&A...307L..41V,2007ApJ...668L.135R,2019ApJ...872...45S,2015MNRAS.453.3953L,2021ApJ...916...41S}). However, recent X-ray spectroscopic studies have revealed the existence of over-ionized plasma (recombining plasma, RP) in several SNRs, where $kT_{\rm i}$ goes even higher than $kT_{\rm e}$ (e.g., IC 443, G359.1-0.5, W28, W44, etc., \cite{2002ApJ...572..897K,2009ApJ...705L...6Y,2011PASJ...63..527O,2012PASJ...64...81S,2012PASJ...64..141U}). So far, RPs have been found in over a dozen of SNRs, which may represent a new subclass of SNRs \cite{2020ApJ...893...90S}.

The physical origin of the RPs in SNRs has not yet been fully understood. Theoretically, there are two approaches to an over-ionization state of the plasma: increase of $kT_{\rm i}$ (extra ionization) or decrease of $kT_{\rm e}$ (electron cooling). The extra ionization can be caused by suprathermal electrons \cite{2011PASJ...63..527O}, high-energy photons \cite{2002ApJ...572..897K}, and low-energy cosmic ray protons \cite{2021PASJ...73..728Y}. On the other hand, the electron cooling scenario, which is considered to be better applied to the SNR evolution, may arise from adiabatic expansion \cite{1989MNRAS.236..885I} and thermal conduction \cite{2002ApJ...572..897K,2011MNRAS.415..244Z}. In addition, simulations indicate that various scenarios, such as the adiabatic expansion and the thermal conduction, may simultaneously contribute to the formation of RP \cite{2011MNRAS.415..244Z,2019ApJ...875...81Z}.

The X-ray emission of RPs is characterized by several distinct spectral features, including the radiative recombination continua (RRCs), enhanced Ly$\alpha$ to He$\alpha$ line ratios, and enhanced He-like ion G ratios (defined as $G=(f+i)/r$, where $r$, $f$, and $i$ stand for the resonant, forbidden, and intercombination line flux, respectively). Limited by the energy resolution of the current CCD instruments, the studies on RPs by far are mostly based on the RRCs and Ly$\alpha$ lines lying in the $\gtrsim2$\,keV band (covers mainly the heavier elements such as Si, S, and Fe), and thus may leave some bias. \HUBS will extend our study into lower energy band ($0.1$--$2$\,keV). 
SNR 3C400.2 is one of the few remnants in which people have detected recombining features in the $<2$\,keV band so far \cite{2015MNRAS.446.3885B} (another possible example could be SNR CTB 1 \cite{2018PASJ...70..110K}). In Figure \ref{fig:3C400}, we present a simulation of the 100\,ks \HUBS spectrum of 3C 400.2. A collisional ionization equilibrium (CIE) plasma model leaves significant residuals at the RRCs of O {\sc viii}, Ne {\sc ix}, Ne {\sc x}, and Mg {\sc xi}, as well as the He$\alpha$ and Ly$\alpha$ lines, which can be clearly identified with the help of \HUBS.
In addition, the spatial resolving ability of \HUBS can help to map the distribution of RPs in SNRs, which is crucial in determining their physical origins.

\subsubsection{Diagnose charge exchange and resonant scattering processes}

The excellent energy resolution of \HUBS is especially suitable for line-oriented studies. Here, we bring up two examples in SNR physics, concerning the charge exchange process and the resonant scattering effect.

Charge exchange (CX) takes place in various astrophysical environments where the hot ionized plasma interacts with the neutral gas, such as solar wind interacting with planet atmospheres, comets, and the heliosphere. The collisionless shocks in SNRs provide a promising site for the CX study. Right behind the SNR shock front, unshocked cold neutrals may collide with the shocked hot ions and go through the CX processes, resulting in a population of highly excited recombined ions (or neutrals) which then produce cascade emission lines. Observational evidence of CX emission has been obtained from the optical band in many SNRs for over 30 years \cite{1980ApJ...235..186C,2001ApJ...547..995G}. However, the study of CX-induced X-ray emission in SNRs is still limited. Possible evidence has been found for a number of SNRs, including Galactic remnants Cygnus Loop \cite{2011ApJ...730...24K,2015MNRAS.449.1340R,2019ApJ...871..234U}, Puppis A \cite{2012ApJ...756...49K}, and G296.1-0.5 \cite{2022ApJ...933..101T}, SMC remnant 1E0102.2-7219 \cite{2001A&A...365L.231R}, as well as LMC remnants N132D \cite{2020ApJ...900...39S} and J0453.6-6829 (SNR B0453-68.5) \cite{2022PASJ...74..757K}. These studies are mostly based on investigations into the O VII triplets: the CX emission could be indicated by an unusually high G ratio. However, the precise measurement of G ratios could still be challenging with current X-ray instruments. CCD cameras are not able to resolve the He-like triplets, which are shown as one single line in the spectrum. Thereby one can only roughly estimate the G ratio based on the line centroid energy, which may lead to large uncertainties. In addition, CCD observations may be contaminated by the emission from solar wind charge exchange (SWCX). Grating instruments can help to resolve the triplets and to improve the constraint on G ratios, but the energy resolution may still be affected by the angular size of the source (morphological broadening). On the other hand, an enhanced G ratio is not necessarily originated from CX, it can also be induced by other mechanisms such as resonant scattering and inner-shell ionization. One possible way to distinguish CX from other mechanisms is to look for enhanced high-level excitation lines (e.g., enhanced Ly$\gamma$/Ly$\beta$ line ratios), which is another prominent and unique feature of CX emission. However, these lines are usually too weak to be detected or blended with other emission lines. Taking together the spatial resolving ability, the high energy resolution, and the large effective area, \HUBS provides us with an unprecedented opportunity to study the CX phenomenon in SNRs.

Due to the rather low density, the hot X-ray plasma in SNRs can be safely assumed as optically thin in most cases. However, for some emission lines with large transition oscillator strengths, the resonant scattering (RS) effect may not be ignored when the remnant contains a large column density. The optical depth at the line centroid can be estimated following \cite{1995A&A...302L..13K}:
\begin{equation}
    \tau=\frac{4.24\times10^{26}fN_{\rm H}\left(n_{\rm i}/n_{\rm z}\right)\left(n_{\rm z}/n_{\rm H}\right)\left(M/T_{\rm keV}\right)^{1/2}}{E_{\rm eV}\left(1+0.0522Mv_{100}^2/T_{\rm keV}\right)^{1/2}}
\end{equation}
where $f$ is the oscillator strength of the line, $E_{\rm eV}$ the line centroid energy in eV, $N_{\rm H}$ the hydrogen column density in cm$^{-2}$, $n_{\rm i}$ the ion density, $n_{\rm z}$ the element density, $M$ the atomic weight, $T_{\rm keV}$ the plasma temperature in keV, and $v_{100}$ the turbulence velocity in 100\,km\,s$^{-1}$. Taking the O {\sc vii} resonant line ($f\sim0.72$) as an example, in a dense remnant like SN 1987A ($n_{\rm e}\sim2400$\,cm$^{-3}$ \cite{2021ApJ...916...41S}) or a large remnant like Cygnus Loop (diameter of $\sim2.8^\circ$ at a distance of $\sim540$\,pc \cite{1998ApJS..118..541L,2005AJ....129.2268B}), the column density may go to $N_{\rm H}\gtrsim10^{20}$\,cm$^{-2}$, resulting in an optical depth $\tau\sim1$. The RS process will scatter the incident photon into another random direction. For a non-uniform distribution of the plasma or an asymmetric remnant, it will then change the line flux and modify the surface brightness distribution. Therefore, similar to CX, RS may also be indicated by an enhanced G ratio --- in this case, it is due to the reduced resonant line flux rather than the enhanced forbidden line flux. The RS effect in X-rays has been extensively studied in diffuse hot plasma of massive elliptical galaxies, galactic bulges, and clusters of galaxies \cite{2002ApJ...579..600X,2018ApJ...861..138C,2018PASJ...70...10H}. The current study on the X-ray RS effect in SNRs is still quite limited. One possible observational evidence comes from the LMC remnant N49, for which people find enhanced O {\sc vii} G ratio as well as O {\sc viii} Ly$\beta$/$\alpha$ and Fe {\sc xvii} ($3s$--$2p$)/($3d$-$2p$) ratios, indicating RS effect on several resonant lines \cite{2020ApJ...897...12A}. \HUBS will be capable of resolving all of the bright resonant emission lines lying in the 0.1--2\,keV band. Taking advantage of its spatially-resolved high energy resolution and the large field of view, we will be able to map out the surface brightness distributions of individual emission lines for the whole remnant, which has never been done before and will certainly improve our insight into the RS effect in SNRs. 

\subsection{Stars and compact objects} \label{subsec:star_compact_obj}
As the fundamental units of galaxies, stars play a key role in the recycling of matter. X-ray observations associated with stars not only deepen understanding of a wealth of astronomical phenomena, but also contribute to the understanding of extreme physical processes.

Several crucial questions have to be answered with \HUBS:  1) Can we detect spectral features on the neutron star surface or the surrounding accretion disk to constrain the equation of the state of compact objects? 2) How does the hot plasma near the WD surface in the accretion column/boundary layer cool and how are the emitted X-ray photons absorbed by the accreted matter? 3) How to understand the X-ray flare mechanism and coronal heating process?

\HUBS, with its large area and high spectral resolution, will clarify these unanswered questions about neutron stars, white dwarfs and active stars. 
\subsubsection{Neutron Stars}

Neutron stars formed by supernova explosions are the most compact objects in the universe. The equation of state for cold and dense matter is still inconclusive with respect to the understanding of the non-perturbative nature of the fundamental interactions between quarks \cite{2007PhR...442..109L}. The equation of state of a neutron star and a strangeon star predicts different mass-radius relations \cite{2013ApJ...765L...5S,2017JPhCS.861a2027X}. The accurate measurements of neutron star mass and radius could put stringent constraints on the equation of state \cite{2016ARA&A..54..401O}. Mass measurements of massive neutron stars, $M>2M_\odot$, have already excluded a number of equations of state that predict the maximum mass smaller than $2M_\odot$ \cite{2010Natur.467.1081D}. Although a number of masses of neutron stars in compact binaries have been measured from radio observations with high precision, radius measurements are much more difficult to achieve with comparable precision.

Usually, neutron star mass and radius can be measured from type I X-ray bursts occurring in NS LMXBs, pulse profile modeling of X-ray pulsars and so on. NS low-mass X-ray binaries (NS LMXBs) are composed of NS and a main sequence donor orbiting each other.  The masses of the NS and its companion can be determined by kinetic methods, with the orbital motion of the star in the NS LMXBs causing its spectral lines to undergo periodic redshifts and blue shifts due to the Doppler effect. The optical and/or near-infrared (NIR) spectroscopic observations can determine the mass function (stellar apparent velocity profile) of the star. Over the past decades, optical/NIR observations have shown that this method has the potential to constrain the compact object mass of LMXBs. However, there are also some shortcomings, mainly in that (1) this method requires a relatively bright optical/infrared flux of stars in LMXBs with strong absorption or emission lines, which is difficult with current optical/NIR telescopes for optically faint LMXBs; (2) This method can only measure the stellar mass function, but not the dense stars. If we can use an X-ray telescope with high energy resolution and a large effective area, we will be able to measure the velocity profile of dense stars and obtain the mass function of dense stars, which can be combined with optical/NIR observations to measure the binary mass ratio. The masses of dense stars can be constrained more precisely if the stellar masses can be determined from optical observations. This has important implications for the mass spectrum of black holes and neutron stars, and for the solution of the ``mass gap" problem. Even if the stellar masses cannot be determined, the mass ratio of the two objects, combined with other measurements, can be used to constrain the binary masses very well. 

Zhang et al. \cite{2012MNRAS.421.3550Z} suggested that absorption lines from accretion disk winds are redshifted or blueshifted due to the Doppler effect of orbital motion. These spectral features are produced in the vicinity of compact objects and trace their motion,  which can constrain the mass of compact objects in LMXBs. This approach was subsequently applied to the eclipsing NS LMXB MXB 1659--298, but the uncertainties of the measured apparent velocities are large because the energy resolution of \Chandra and NuSTAR is not high enough \cite{2018MNRAS.481L..94P}. In general, the maximum apparent velocity of compact objects in X-ray binaries is in the order of 100 km/s, which causes a spectral shift of order $10^{-4}$. Therefore a high energy resolution of the detector is required. The energy resolution of \HUBS has the possibility to measure the Doppler effect of the spectral lines with high precision.

The LMXB 4U 1700+24 has a red giant companion, and the X-ray emission is dominated by wind accretion. In the X-ray spectrum of 4U 1700+24, the O {\sc viii} (hydrogen-like Ly-$\alpha$) emission line is found with a central energy of about 0.65 keV \cite{2014A&A...562A..55N}. The spectral line structure is corresponding to a gravitational redshift of 0.009, suggesting that 4U 1700+24 is a candidate for a low-mass neutron star, which has to be verified by \HUBS observations \cite{2014RAA....14..617X}.

Type-I X-ray bursts are the unstable thermonuclear burning of accreting matter on the NS surface. The unstable thermonuclear burning of the hydrogen and helium, also known as type I X-ray burst, usually has a duration of $\sim 10-100$ s with a typical energy release of $10^{39}$ erg, recurs from few hours to days, and ignites at a column depth of $\sim10^8~ \mathrm{g~cm^{-2}}$ \cite{2021ASSL..461..209G}.  In a rare case, superbursts, which are believed due to burning carbon, have been identified from the total energy release of $\sim 10^{42}$ erg and the duration of $>10^3$ s \cite{2001ApJ...559L.127C}. Cottam et al. \cite{2002Natur.420...51C} reported the identification of absorption lines by stacking of \XMMNewton spectra over dozens of type-I X-ray bursts from NS LMXB EXO 0748--676, which they claimed were gravitational redshifted Fe and O lines from the stellar surface. However, the spectral lines have not been confirmed by the following observations. This particular source is now believed to be rotating rapidly with the frequency of 552 Hz from its burst oscillation \cite{2010ApJ...711L.148G}, which makes it challenging to explain the relatively narrow spectral features. in't Zand et al. \cite{2017arXiv170307221I} also reported the none detection of spectral lines in Rapid Burster from \Chandra/HETG observations.  Rauch et al. \cite{2008A&A...490.1127R} calculated the possible spectral lines in the soft X-ray band, i.e., Fe and O, that can be generated during X-ray bursts. \HUBS provides a larger effective area accompanied with a high energy resolution than \XMMNewton and \Chandra, which could resolve spectra features with a high S/N ratio from NS in LMXBs also by adding many X-ray bursts, or from a single superburst. 

The X-ray dimmed isolated neutron stars (XDINS) mainly emit blackbody spectra in X-ray bands, and show optical/ultraviolet (UV) excesses \cite{2007Ap&SS.308..181H}.  All seven known XDINSs were found by soft X-ray detectors. RX J1856--3754 is the brightest and closest neutron star. \Chandra observations showed that RX J1856--3754 had an almost blackbody spectrum in the soft X-ray band, with no emission or absorption lines \cite{2001A&A...379L..35B}, but absorption lines may be present in other XDINS. It is generally believed that the soft X-ray thermal spectrum of XDINS comes from the surface of the star. The absorption lines in the XDINS spectrum are produced in the magnetic environment of the neutron star. Moreover, the structure of the absorption lines is related to the stellar surface properties. The temperature and stellar radius determined by the continuum spectrum depend on the equation of state of the compact object (see \cite{2005AAS...20719803H} for neutron stars; \cite{2017ApJ...837...81W} for strangeon stars).  Therefore, XDINS is also an excellent target to study the surface properties and the equation of the state of NSs. 

Besides measuring the NS mass and radius, \HUBS can also study the magnetic field of anomalous X-ray pulsars (AXPs) and soft-$\gamma$-ray repeaters (SGRs). AXPs and SGRs are slowly rotating, isolated and ultra-magnetized neutron stars \cite{2007Ap&SS.308....1K,2013BrJPh..43..356M}. Their X-ray activities, short bursts and outbursts, are powered by magnetic energy.  During outbursts, the X-ray spectra of AXPs and SGRs may show absorption lines, which is interpreted as a proton cyclotron feature. \HUBS could resolve the absorption features in 0.1--2 keV from AXPs and SGRS, and measure the magnetic field (see e.g., \cite{2013Natur.500..312T}).

\subsubsection{Cataclysmic Variables}

Cataclysmic variables (CVs) are binaries consisting of a white dwarf (WD) and a late-type main sequence or sub-giant star. CVs are the most populated binaries consisting of a compact star and their spatial density can reach up to $10^{-6}$ to $10^{-5}\,\mathrm{pc^{-3}}$ in the solar neighborhood \cite{2006A&A...450..117S}. The WD in a CV accretes matter from its companion and emits mostly in UV and X-ray energy range. CVs are not only laboratories of stellar evolution theory, but also related to other important astrophysical questions. For example, CVs collectively contribute up to 80\% of the Galactic Diffuse X-ray Background (GDXE). What's more, CVs are closely related to the progenitor of type Ia supernovae since the latter are supposed to be binaries harboring one or two WDs.

X-ray observations provide unique information to understand the accretion and emission process of CVs. the X-ray luminosity of CVs can reach $10^{33-34}$ erg s$^{-1}$, high enough to study the structure of the X-ray emitting region through X-ray spectroscopy. For example, the hard X-ray (around 2 to 50 keV) spectra of CVs have been well described by the multi-temperature thermal plasma model (mkcflow), and are used to constrain the maximum emission temperature and the mass of the WD. In contrast, the soft (0.1-2 keV) X-ray spectra of CVs are less well understood, and the usual characterization (the same mkcflow emission partially covered by the accreted matter) failed to explain the He-like and H-like lines from different elements (e.g., C, N, O) \cite{2003ApJ...586L..77M, 2005ApJ...626..396P}. Until now, there are only 15 CVs with high-resolution X-ray spectra at present, and about half of them are not well explained. Since the soft X-ray are supposed to be originated from the region fairly close to the surface of the WD, the failure of a widely-accepted model in this energy range leads to the lack of understanding of the accretion process near the WD itself. 

\HUBS provides a unique opportunity to explore the details of the emission region in CVs. The high-resolution spectra would certainly allow a detailed investigation of the distribution of the differential emission measure (deM) and the metallicity on a large sample of CVs. Combined with hard X-ray data, a thorough understanding of the structure and evolution of the accreted matter in CVs could be reachable.

A rough estimation of the exposures could be done. For CVs within 50 pc,  the typical $0.1-2$ keV X-ray flux is $\sim10^{-12}$ to $\sim10^{-11}$ erg s $^{-1}$ cm $^{-2}$. Simulation shows an \HUBS snapshot of 10 to 100 ks (depending on the flux of the target)  could provide a spectrum with sufficient photons for a targeted CV to identify the important emission lines (e.g., of the O and Ne elements) of one targeted CV for further investigation. A total sample of 30 bright CVs in the solar neighborhood requires 30 snapshots with a total exposure of $1.4\times10^{6}$ s (16 days).

With the large effective area and high spectral resolution, \HUBS can greatly improve our understanding of the accretion process in CVs.

\subsubsection{Stars}

Stars located across almost all regions of a Hertzsprung-Russell diagram have been identified as X-ray sources, although with different mechanisms. Stellar magnetic corona is the predominant origin of X-rays for late-type stars, while for massive and hot stars, the X-ray emission is from shocks forming in unstable winds. The X-ray radiation of pre-main sequence stars may originate both in hot coronal plasma or shocks \cite{2009A&ARv..17..309G}.

The stellar magnetic activity provides substantial information on the magnetic dynamo and the coronal heating process. It is also of great value for exploring the interaction between stars and their planets and determining the habitable zone of different stars \cite{2009A&ARv..17..309G, 2009PASJ...61S.115M}. Stellar magnetic activity is ubiquitous in late-type stars, which can be traced by various proxies, including spots and flares from the photosphere, emission lines from the chromosphere, and X-ray and radio emissions from the corona. 
The activity level strongly depends on stellar parameters (e.g., stellar mass, age).

X-ray astronomy has played a key role in stellar activity studies. The X-ray luminosity of active stars in the quiet state ranges from $10^{27}$ to $10^{31}\mathrm{erg\,s^{-1}}$, while it is $1-2$ orders of magnitude brighter during flares \cite{2004ApJ...617..508T, 2009A&ARv..17..309G, 2015A&A...578A.129J}.
It helps establish the famous activity-rotation relation (e.g., \cite{2003A&A...397..147P, 2011ApJ...743...48W}).
In the relation, the X-ray activity is described as the ratio between X-ray luminosity and bolometric luminosity, while the Rossby number is used to trace stellar rotation, which is defined as the ratio of the rotation period to the convective turnover time.
The relation is usually suggested to consist of two distinct sequences: the saturated region for rapidly rotating stars, in which the activity level keeps constant, and the power-law decay region for slowly rotating stars, where the activity level is rotation-dependent \cite{2011ApJ...743...48W}. 

X-ray spectral observations have yielded a typical temperature of about $0.1-1$ keV for the stellar corona, belonging to the soft X-ray band.
Previous studies using the low-resolution spectra (e.g., \Chandra/ACIS, \XMMNewton/MOS) have measured the coronal temperatures and discussed the distribution of differential emission measures (dEM) for some nearby active stars \cite{2004A&ARv..12...71G}.
High-resolution X-ray spectroscopy, on the other hand, is mainly done with \Chandra/HETG and \XMMNewton/RGS high-resolution spectrometers.
By using the He-like and H-like lines from different elements (e.g., C, N, O), the distributions of some physical parameters (e.g., coronal temperature and density, dEM, metallicity) during quiet states and flares have been well constrained for dozens of stars \cite{2004ApJ...617..508T}.
High-resolution spectra of active stars revealed a new trend that runs opposite to the solar FIP effect, called the ``inverse FIP (IFIP) effect" \cite{2008A&A...482..639N}.

Although previous studies provide a number of surprising findings, there are many key issues unresolved. The standard picture of the activity-rotation relation has been challenged by recent studies, such as the variable activity level in the saturation region \cite{2014ApJ...794..144R} and more sequences possibly divided in the relation \cite{2018A&A...618A..48M}.
It is also doubted that the distribution of coronal physical parameters is universal among stars due to the small and incomplete sample with high-resolution spectroscopic observations.
For example, more than 900 F/G/K-type stars are located within 30 pc of the solar system \cite{2017ApJ...848...34H}, but only about 40 ones were observed; most stars around the solar system are M-type dwarfs, but only a few were observed (e.g., Proxima Cen \cite{2004A&A...416..713G} and CN Leo \cite{2010A&A...514A..94L}).
More importantly, some basic physical questions including the mechanism of the saturation and the connection between the relation and magnetic dynamo are poorly understood.  
A large sample covering different types of stars, with well-measured activities and spectral parameters, can help investigate the physical properties of the stellar magnetic dynamo and provide potential diagnostics of heating mechanisms.

\HUBS can help establish a large high-resolution X-ray spectral sample for stars with different spectral types, rotation periods, ages, and metallicities. 
For single stars, detailed diagnostics of the coronal temperature and density can be done with the emission lines from different elements. With further investigation of the dEM, the FIP and IFIP effects, and the area of the active region, a comparison with the Sun can help explore the flaring mechanism and heating process.  
On the other hand, by using the large sample, the distribution of these parameters and their relationships with different stellar parameters (e.g., mass, age, rotation) will help understand the structure and evolution of stars.

For typical active stars, an exposure of 100 ks by \HUBS can obtain a spectrum with a sufficiently high signal-to-noise ratio for the following studies; for close stars, the exposure time can be reduced to around 10--30 ks. Therefore, the exposure time of 100 stars is about 10$^3$ to 10$^4$ ks.  Taking Proxima Cen as an example, the simulation shows that \HUBS can clearly distinguish emission lines in its spectrum (typical for M-type active stars) compared with \XMMNewton/RGS observations with an exposure time of $\approx$800 s.

With the large effective area and high spectral resolution, it can be predicted that \HUBS will provide a valuable opportunity to advance stellar magnetic activity studies.

\section{Exploiting the Capabilities of \HUBS}
\label{sec:unlocking}
As shown in previous sections, \HUBS will observe various types of warm and hot plasmas across more than ten orders of magnitude in size, such as stellar coronae, supernova remnants, AGN winds, hot plasmas around individual galaxies and galaxy assemblies, and cosmic web filaments. Characteristic emission and absorption lines in the high-resolution X-ray spectra will enable us to measure various physical properties of these astrophysical plasmas, including but not limited to temperature, density, elemental abundances, and kinematics \cite{Paerels2003, Porquet2010,Mao2017b}. These fundamental parameters are essential to fill the gaps in our understanding of the role of warm and hot plasmas in the formation and evolution of the hot Universe. These astrophysical plasmas, play an important role in the galactic ecosystem \cite{Crenshaw2003,Veilleux2005,Vink2012,Tumlinson2017,Werner2019,Laha2021,Faucher2023}. 

As we have experienced with the era of diffractive grating spectrometers aboard \XMMNewton and \Chandra \cite{Brinkman2000,denHerder2001,Canizares2005}, the next-generation high-resolution X-ray spectroscopy will offer both an opportunity and a challenge. On one hand, they will greatly advance our knowledge of the Universe more than what we have learned from \XMMNewton and \Chandra. On the other hand, they will also challenge us on how to quantify key observables precisely and efficiently. To better prepare us for the upcoming new era, we need to improve the status quo in the following three aspects: atomic data, plasma models, and spectral analysis techniques.
  
\subsection{Atomic data}
\label{subsec:atomic_data}
Various types of microscopic atomic processes give rise to continuum and line features in the observed spectra. Generally speaking, the interactions between electrons, ions, and photons can be divided into collision, ionization, and recombination \cite{Kaastra2008}. Each category can be further divided into several sub-classes. For instance, radiative, di-electronic, and multi-electron recombination all contribute to the continuum and line emission in the observed spectrum. Even if we are limited to the simplest radiative recombination rates of H- to Na-like ions with $Z\le30$, there are $3\times10^4$ levels to consider \cite{Mao2016}. Each level-resolved rate is provided either on a few temperature grids or described with a few parameters \cite{Mao2016}. The entire atomic database can easily grow to a significant size.

The associated large amount of atomic data is the building block of astrophysical plasma codes widely used in the community: APEC \cite{Smith2001,Foster2012}/ACX \cite{Smith2012}/NEI \cite{Smith2010}, CHIANTI \cite{Dere1997,DelZanna2021}, Cloudy \cite{Ferland1998,Ferland2017}, SPEX\cite{Kaastra1996,Kaastra2020}, SASAL\cite{LiangGY2014a,LiangGY2014b} and XSTAR \cite{Kallman2001,Mendoza2021}. Caution that the underlying atomic databases are not perfect (e.g., \cite{Mao2019d,Mao2022c}). Continuous developments including both theoretical calculations and lab measurements are required \cite{Badnell2016,BMartinez2019,Smith2020}. 

In 2016, we had a test of the next generation of high-resolution X-ray spectroscopy with \textit{Hitomi} \cite{Hitomi2016}. While the statistical uncertainty of the observed spectrum is less than 1\%, the Fe abundance measured with APEC and SPEX differ by 16\% \cite{Hitomi2018atom}. This is mostly attributed to the different atomic data used by these two plasma models \cite{Hitomi2018atom}. The Fe abundance is measured from H- and He-like lines, but their transition rates (i.e., A-values) and electron-impact excitation rates can differ up to 40\% \cite{Hitomi2018atom}. That is to say, the accuracy of the atomic data is not adequately converged to match the accuracy of the observed data. 

When the mysterious 3.5 keV line was in the spotlight \cite{Bulbul2014,Boyarsky2014}, it was unclear whether natural atomic processes like di-electronic recombination and charge exchange process can account for this instead of the dark matter decay process. This was largely due to the incompleteness of the atomic database. New theoretical calculations and lab measurements were then pursued to quantify the role of these two recombination processes \cite{GuLY2015,GuLY2016,Shah2016,Bulbul2019}. Ar {\sc xvii} di-electronic recombination line is at 3.62 keV, while the S {\sc xvi} charge exchange recombination line is at $3.47\pm0.06$ keV. Due to insufficient energy resolution of CCD instruments, the line center of the 3.5 keV line is not tightly constrained: $3.57\pm0.02$ keV by \cite{Bulbul2014} and $3.52\pm0.02 keV$ by \cite{Boyarsky2014}. On the other hand, while the 3.5 keV line is found in some mega-second CCD observations, it is absent in the $\sim300$ ks microcalorimeter (Hitomi/SXS) observation. Deeper microcalorimeter observations with fine energy resolution (e.g. HUBS) are certainly required.

\subsection{Plasma diagnostics}
\label{subsec:plasma_diag}
Plasma diagnostics play a crucial role when interpreting characteristic continuum and line features in the observed high-resolution spectra. Fundamental physical properties of the observing target are measured by matching the data and model.

In the $0.1-2$~keV soft X-ray bandpass covered by \HUBS, H-like Lyman series and He-like triplets are the most prominent emission line features. In a low-density CIE plasma, such as the majority of hot gas in individual galaxies and galaxy assemblies, the Ly$\alpha$ lines should have the highest intensity among the Lyman series. Ly$\alpha$ might be optically thick in some astrophysical environments so that the intensity of Ly$\alpha$ will be reduced by resonance scattering (a fraction of Ly$\alpha$ photons are scattered out of our line-of-sight). Other Lyman series lines with smaller oscillator strength suffer less from this issue, leading to larger ratios of Ly$\beta$/Ly$\alpha$, Ly$\gamma$/Ly$\alpha$, and Ly$\delta$/Ly$\alpha$. Furthermore, at the interface between the hot plasma and cold media (e.g., comets), the charge-exchange process can selectively increase the intensity of e.g., Ly$\gamma$ or Ly$\delta$ \cite{GuLY2016}. 

The He-like triplet consists of the resonance ($w$), inter-combination ($x$ and $y$), and forbidden ($z$) lines. The line ratio among the three is rather sensitive to a wide range of plasma temperature, density, and the astrophysical environment of the plasma \cite{Porquet2010}. In a CIE plasma, the $G=(x+y+z)/w$ ratio decreases with an increasing plasma temperature, while the $R=z/(x+y)$ ratio decreases with an increasing plasma density \cite{Porquet2010}. In photoionized plasmas (e.g., the X-ray narrow line region of AGN), the external radiation field can boost both the $G$- and $R$-ratios \cite{Porquet2010,Mao2019c}. The charge exchange process can also increase the $G$-ratio \cite{BRaymont2007,ZhangSN2014,GuLY2016}. The optical depth effect also applies to He-like resonance lines as well \cite{Chakraborty2020a,Chakraborty2020b,Chakraborty2021,Chakraborty2022}.

Apart from H- and He-like lines, the Fe-L complex is also prominent \cite{Phillips1996,Behar2001,XuH2002,dePlaa2012,Pinto2016,Ogorzalek2017}. These $n\ge3$ to $n=2$ (i.e. L-shell) transitions of Fe {\sc xvi} to Fe {\sc xxiv} are susceptible to a wide range of atomic processes: direct and resonance excitation, radiative and di-electronic recombination, inner-shell ionization. Consequently, they are notorious for spectral modeling. The line ratios among the Fe {\sc xvii} 15.01~\AA\ (3C), 15.26~\AA\ (3D), 17.05~\AA\ (3G), and 17.09~\AA\ (M2) lines have been the hot topic for both theoretical calculations and lab measurements for decades \cite{Brown1998,ChenGX2002,Beiersdorfer2002,Beiersdorfer2004,Brown2006,LiangGY2010,DelZanna2011,Bernitt2012,GuLY2019,GuLY2020}.

Thanks to the fine energy resolution and large effective area of \HUBS (Figure~\ref{fig:FoM}), some weak line diagnostics become possible and effective. The width of radiative recombination continua (RRC) is an effective measure of plasma temperature. These RRC can be found in the hot recombining plasma of supernova remnants \cite{Ozawa2009,Yamaguchi2009} or warm photoionized gas of X-ray binary \cite{Paerels2000} or AGN \cite{Kinkhabwala2002,Mao2018}. Di-electronic recombination satellite lines of He-like ions can effectively verify the presence of non-Maxwellian electrons, such as those supra-thermal electrons behind the shock of merging galaxy clusters \cite{Kaastra2009}. Meta-stable absorption lines of Be-like to F-like ions can probe a wide range of number density for AGN winds \cite{Mao2017b}.

All these diagnostics have been implemented in the astrophysical plasma codes widely used for X-ray spectral analysis: APEC/ACX/NEI, CHIANTI, Cloudy, SPEX, and XSTAR. Continuous developments of these plasma codes are still required. For instance, pre-calculated charge-state distribution tables \cite{Bryans2009,Urdampilleta2017} are not applicable to high-density plasma \cite{Dufresne2019,Dufresne2020}. Self-consistent charge-state distribution calculations involving excitation, recombination, and ionization from and to meta-stable levels are required. Radiation transfer for the high-density plasma is also required. On one hand, this calls for a large amount of atomic data that is not yet available. On the other hand, as the complexity grows, computational efficiency needs to be improved. 

\subsubsection{Laboratory benchmark required by \HUBS}
Plasma diagnostics for various objects are strongly dependent on the above-listed models, including SPEX/CX \cite{GuLY2016} and ACX \cite{Smith2012} for charge-exchange emissions in SWCX foreground and SNRs. However, these two models are not perfect. To obtain high-resolution spectra, both models use some approximations to redistribute total or $n-$ or $nl-$resolved cross-sections. For the $n-$resolved cross-section, limited experimental data are available in collisions with different neutrals at some energies \cite{xu21}. For the $nl$-resolved cross-section, only theoretical calculations are available while experiments are lacking. In the high-resolution spectra obtained with \HUBS ($\Delta E\le2$~eV), most of the lines are fine-structure levels ($nLSJ$) resolved. However, the present CX models including ACX and SPEX-CX, have rather large uncertainties \cite{GuLY2022}. This calls for the laboratory benchmark of the CX model.

By comparison of the resultant spectra from both the experimental and theoretical cross-sections, the accuracy of the CX model will be examined at given collision energies. In some cases, CX high-resolution spectrum can be measured directly in the laboratory, which can be used to fit the \HUBS observation. Besides the charge-exchange data, laboratory measurements on other atomic data including ionization, di-electronic/radiative recombination, excitation as well as spectra, will improve our interpretation of the \HUBS observations.

In turn, the \HUBS spectroscopy will prompt the progress of the collision theories in atomic physics, including the $nl$-resolved charge-exchange cross-section. Generally, the $n$-resolved CX cross-sections can be obtained in a heavy ion source by the cold target recoil ion momentum spectroscopy (COLTRIM) apparatus with the electron energy resolution of $\sim$10~eV \cite{xu21}. The $nl$-resolved cross-section of He-like captured ions has never been obtained by experiments. For the collision of O$^{7+}$ with H, the dominant capture channel by O$^{7+}$ projectile of the bound electron from the H donor is $n=4$ captured ion states by accurate close-coupling calculations. The radiative decay rates of the dipole transitions of O {\sc vii} have an accuracy of $\le$5\%. The O {\sc vii} resonance, intercombination, and forbidden lines at the rest-frame energy of 561~eV, 569~eV, and 574~eV are well resolved in the \HUBS observation. The observed data can be used to determine the $l-$distribution in the $n=4$ channel with an accuracy better than 10\% by an iterative algorithm. 

In summary, the \HUBS spectroscopy with high resolution requires laboratory measurements to benchmark the CX model, it also gives some constraints for the $nl$-distribution of the cross-sections measured in the laboratory. Both of them complement each other.

\subsection{Spectral analysis techniques}
\label{subsec:spectral_analysis}
With diffractive grating spectrometers, we typically obtain one high-resolution X-ray spectrum for each observation. It might take weeks or months for experts to finish a thorough spectral analysis. With X-ray integral field units like those on \HUBS, we might get up to thousands of high-resolution X-ray spectra in one single observation. We need to quantify key observables precisely and efficiently with limited manpower and computation resources. 

For observations targeting point-like sources, an efficient and automated line detection algorithm without any prior knowledge of the targets is required as the first step \cite{Mao2019c}. If the spectrum is not featureless, we need to identify these lines and extract preliminary information such as the line center, velocity shift, line broadening, equivalent width, and plasma types according to the intensities of characteristic lines. This might call for a machine-learning approach. For observations targeting extended sources, imaging spectroscopic approaches including but not limited to Weighted Voronoi Tessellations \cite{Diehl2006} and smoothed particle inference \cite{Peterson2007} are to be pursued to get a comprehensive and self-consistent view of the observing target.

\section{Status of \HUBS}
\label{sec:status}
The \HUBS project is being funded by the \textit{China National Space Administration} for key technology development (which corresponds roughly to Phase A, in terms of NASA project cycles). The critical technologies identified include superconducting microcalorimeter (detector), wide field-of-view X-ray focusing optics (telescope), multiplexing signal readout electronics, mechanical cooler and adiabatic demagnetization refrigerator. The goal is to advance the technical readiness levels (TRLs) of those technologies sufficiently by the end of 2023, before the project can enter the next phase. Looking ahead, the important milestones will include the completion of technology development and payload design, the construction of the \HUBS satellite, and the launch and operation of the satellite (around 2030 and beyond). 

Mock observations have been made to assess the scientific capabilities of \HUBS and also to help formulate observing strategies \cite{Zhang22}. The results suggest that CGM studies require deep exposures on carefully-selected targets, while group or cluster observations are likely quite efficient at low redshifts, thanks to the large field of view. For IGM studies, on the other hand, medium-exposure mosaic observations will be necessary to acquire sufficient spatial coverage, so the total exposure time is also expected to be long for each selected field. It is, therefore, clear that target selection is critical to the success of \HUBS. Discussion is ongoing on the scientific values (vs resource investment) of an all-sky survey in the extended mission period.

\section{Summary}
\label{sec:summary}

The \textit{Hot Universe Baryon Surveyor} mission aims at studying the hot gas in
the universe with unprecedented sensitivity and spatial resolution in X-rays.
Among the core sciences for which the \HUBS is tailored, the feedback in
the galactic ecosystem and the cosmic baryon budget are of particular
importance. The \HUBS will provide a unique opportunity to study the hot
gas in the ISM, the CGM and the ICM by resolving the X-ray spectrum in both emission
and absorption, from which the spatial distribution and the kinematics of the
hot gas can be confidently obtained. Since the thermal and kinematic status of
the hot gas is closely related to the star formation and the AGN activity, the
\HUBS mission will be a huge leap forward in our understanding of 
galaxy formation and evolution. Moreover, with its high sensitivity and large
field of view, the \HUBS is highly capable of searching for multi-phase hot
gas in galaxy groups, clusters and the cosmic web, which will pave the way for
the future study of the cosmic baryon budget.

The \HUBS may also extend its application to other X-ray-related observatory
sciences. For instance, \HUBS will be able to directly constrain the gas number
densities and opening angles of the AGN-driven outflows. The \HUBS can also
probe the nearest X-ray sources within our Galaxy, such as cosmic X-ray
background, supernova remnants, activities of stars and compact objects, and the
emission from the Solar system.

The capability of the \HUBS can be exploited further by renovating the knowledge
of atomic data, plasma models and the techniques of spectral analysis. The
laboratory measurements on atomic data will be used to benchmark the CX model
and to improve the interpretation of the \HUBS observations. In addition, since
the \HUBS will be able to obtain thousands of high-resolution X-ray spectra in
one single observation, the spectral analysis techniques will be developed to
quantify key observables precisely and efficiently with limited manpower and
computation resources.

A staged construction plan has been carefully designed for the \HUBS mission.
With the support from the \textit{China National Space Administration} for key
technology development, a number of critical technologies have been identified,
and the current goal is to sufficiently enhance the technical readiness levels
of those technologies by the end of 2023 before entering the next phase.
Besides, mock observations have also been developed to test the feasibility of
observing targets and strategies.

As we celebrate the 60th anniversary of X-ray astronomy, the field is about to enter a new era, in which spatially-resolved, high-resolution spectroscopy is expected to become increasingly exquisite and routine, thanks to the advancement of new detector technologies. The imminent launch of \textit{XRISM} \cite{xrism_2022} is highly anticipated, as the first mission employing microcalorimeters for spectroscopy observations. The scientific potential of such a spectrometer has been well illustrated by sounding-rocket experiment \cite{McCammonetal_2002} and the \textit{Hitomi} satellite mission \cite{Hitomi2016}, so breakthroughs are expected of \textit{XRISM}, especially in the studies of ICM and AGN, as it is optimized to detect emission lines at higher energies than \HUBS. With the new generation of microcalorimeters, \HUBS, as well as \textit{Athena} \cite{PeillAthena_2020SPIE}, will not only provide higher spectral resolution, but significantly improve the detection sensitivity at energies where the emission lines associated with hot CGM/IGM are expected to lie and thus provide new avenues to exploring baryonic processes in the cosmos. These improvements are expected to significantly advance our understanding of many important astrophysical fields, as we have stated in detail in the present paper.

\Acknowledgements{This work is supported by the National Natural Science Foundation of China (Grant Nos. 11721303, 11821303, 11825303, 11873029, 11890693, 11973033, 11991052, 12025303, 12033004, 12041301, 12121003, 12133008, 12173018, 12192220, 12192223, 12221003, 12233001, 12233005, 12273010, 12273030, 12273057, 12011540375, U1931140), the China Manned Space Project (Grant Nos. CMS-CSST-2021-A04, CMS-CSST-2021-A06, CMS-CSST-2021-A10, CMS-CSST-2021-B02), the Ministry of Science and Technology of China through its National Key R\&D Program (Grant No. 2018YFA0404502), the National SKA Program of China (Grant No. 2020SKA0120300), the National Key Research and Development Program of China (Grant No. 2022YFA1602903), the Outstanding Young and Middle-aged Science and Technology Innovation Teams from Hubei colleges and universities (Grant No. T2021026), the Young Top-notch Talent Cultivation Program of Hubei Province, the National Science Foundation (Grant Nos. AST-2107735 and AST-2219686), and NASA (Grant No. 80NSSC22K0668). Mr. Yongkai Zhu (Shanghai Jiao Tong University) provided useful comments on the manuscript.}

\InterestConflict{The authors declare that they have no conflict of interest.}


\AuthorContributions{This paper was organized and structured by Wei Cui and Feng Yuan, and was primarily contributed by each author as follows: \S\ref{sec:intro} (Wei Cui, Suoqing Ji, Feng Yuan), \S\ref{subsec:AGN_feedback} (Suoqing Ji, Junjie Mao, Feng Yuan), \S\ref{subsec:stellar_feedback} (Hui Li, Miao Li), \S\ref{subsec:CGM_physics} (Suoqing Ji, Jiangtao Li, Miao Li), \S\ref{subsec:CGM_physics_additional} (Suoqing Ji), \S\ref{subsec:potential_case_studies} (Jiangtao Li), \S\ref{subsec:Multi-phaseHotGas} (Dandan Xu, Haiguang Xu), \S\ref{subsec:StackCosmicWeb} (Dandan Xu), \S\ref{subsec:cosmology} (Zhongli Zhang, Haiguang Xu), \S\ref{subsec:cosmic_xray} (Guiyun Liang, Wenhao Liu, Zhijie Qu, Hang Yang, Shuinai Zhang), \S\ref{subsec:snr} (Lei Sun, Ping Zhou, Yang Chen), \S\ref{subsec:star_compact_obj} (Zhaosheng Li, Song Wang, Xiaojie Xu), \S\ref{subsec:atomic_data} (Junjie Mao), \S\ref{subsec:plasma_diag} (Guiyun Liang, Junjie Mao), \S\ref{subsec:spectral_analysis} (Junjie Mao, Ping Zhou, Shuinai Zhang), \S\ref{sec:status} (Wei Cui), and \S\ref{sec:summary} (Wei Cui, Suoqing Ji). All authors critically reviewed and made contributions to the manuscript.}

\bibliographystyle{unsrt}
\bibliography{HUBS}

\end{multicols}
\end{document}